\documentclass[journal]{IEEEtran}
\usepackage{caption} 
\usepackage{soul}
\usepackage{lipsum}
\usepackage{indentfirst}
\usepackage{booktabs}
\usepackage{enumerate}
\usepackage{tabu}
\usepackage{tabularx}
\usepackage{environ}         
\usepackage{etoolbox}        

\usepackage{cite}

\usepackage{graphicx}
\usepackage{wrapfig}
\usepackage[labelformat=simple]{subcaption}

\usepackage{float}

\usepackage{amsfonts, eqnarray, amsmath, amssymb}
\usepackage{mathptmx}
\usepackage{mathtools}
\usepackage{xfrac}
\usepackage[cmintegrals]{newtxmath}
\usepackage{breqn}
\usepackage{bm}

\makeatletter
\def\@begintheorem#1#2{%
    \parskip 0pt 
    \trivlist
    \item[%
        \hskip 10\p@
        \hskip \labelsep
        {{\bfseries #1\hskip 5\p@\relax#2.}}%
    ]
    \it
}
\def\@opargbegintheorem#1#2#3{%
    \parskip 0pt 
    \trivlist
    \item[%
        \hskip 10\p@
        \hskip \labelsep
        {\bfseries #1\ #2\       
   \setbox\@tempboxa\hbox{(#3)}  
        \ifdim \wd\@tempboxa>\z@ 
            \hskip 5\p@\relax    
            \box\@tempboxa       
        \fi.}%
    ]
    \it
}

\makeatother
\newtheorem{theorem}{Theorem}[section]
\newtheorem{proposition}[theorem]{Proposition}

\newtheorem{lemma}[theorem]{Lemma}

\usepackage{algorithm}
\usepackage[noend]{algpseudocode}

\usepackage{array}

\usepackage{url}

\NewEnviron{resizealign}{\sbox0{
    $\begin{matrix}\displaystyle\BODY\end{matrix}$}%
  \sbox1{$(\theequation)$}%
  \sbox2{\parbox{\dimexpr \wd0 + 2\wd1}%
    {\begin{align}\BODY\end{align}}}
  \noindent\resizebox{\columnwidth}{!}{\usebox2}%
}

\begin{document}
\title{Centralized and Decentralized Non-Cooperative Load-Balancing Games among Federated Cloudlets}
\author{Sourav~Mondal,~\IEEEmembership{Student~Member,~IEEE,}
        Goutam~Das,~\IEEEmembership{Member,~IEEE,}
        and~Elaine~Wong,~\IEEEmembership{Senior~Member,~IEEE}
\thanks{S. Mondal and E. Wong are with the Department of Electrical and Electronic Engineering, The University of Melbourne, VIC 3010, Australia. (e-mail: smondal@student.unimelb.edu.au, ewon@unimelb.edu.au).}
\thanks{G. Das is with the G. S. Sanyal School of Telecommunications, Indian Institute of Technology Kharagpur, 721302, India. (e-mail: gdas@gssst.iitkgp.ac.in).}}

\maketitle

\begin{abstract}
Edge computing servers like cloudlets from different service providers compensate scarce computational, memory, and energy resources of mobile devices, are distributed across access networks.  However, depending on the mobility pattern and dynamically varying computational requirements of associated mobile devices, cloudlets at different parts of the network become either overloaded or under-loaded.  Hence, load balancing among neighboring cloudlets appears to be an essential research problem.  Nonetheless, the existing load balancing frameworks are unsuitable for low-latency applications.  Thus, in this paper, we propose an economic and non-cooperative load balancing game for low-latency applications among federated neighboring cloudlets from the same as well as different service providers and heterogeneous classes of job requests. Firstly, we propose a centralized incentive mechanism to compute the pure strategy Nash equilibrium load balancing strategies of the cloudlets under the supervision of a neutral mediator.  With this mechanism, we ensure that the truthful revelation of private information to the mediator is a weakly-dominant strategy for all the federated cloudlets.  Secondly, we propose a continuous-action reinforcement learning automata-based algorithm, which allows each cloudlet to independently compute the Nash equilibrium in a completely distributed network setting.  We critically study the convergence properties of the designed learning algorithm, scaffolding our understanding of the underlying load balancing game for faster convergence. Furthermore, through extensive simulations, we study the impacts of exploration and exploitation on learning accuracy. This is the first study to show the effectiveness of reinforcement learning algorithms for load balancing games among neighboring cloudlets.
\end{abstract}
\vspace{-\baselineskip}

\IEEEpeerreviewmaketitle

\section{Introduction}\label{sec1}
\IEEEPARstart{T}{he} next-generation Internet is not only expected to route data, but also to store and process data, generated from a large number of pervasive mobile devices like smartphones and Internet-of-Thing devices. In state-of-the-art cloud-computing networks, mobile devices can offload data to remote cloud servers for storage and processing to compensate for their computation, memory, and energy resource poverty \cite{migrate}. With the recent emergence of ultra-reliable and low-latency communication (uRLLC) applications such as virtual/augmented reality, automotive, and teleoperation as part of the Tactile Internet \cite{Imali,9139304}, the long communication-latency between mobile devices and remote cloud server appears to be a major bottleneck to satisfy the low-latency requirements of 10-100 ms \cite{edge}. To overcome this hurdle, researchers from both industry and academia proposed edge computing solutions like \emph{multi-access edge computing}, \emph{fog computing}, and \emph{cloudlet computing} \cite{SurveyEls}. For sixth generation (6G) networks, edge computing nodes are also being used to implement artificial intelligence-based protocols, e.g., \emph{application layer prediction} and \emph{network layer prediction} that can facilitate various uRLLC applications across long-distance networks. In application layer prediction, different application-specific data (e.g., sensor data from a teleoperator to forecast haptic feedback in Tactile Internet) is used and in network layer prediction, various network parameter data (e.g., statistical parameters of various network traffic) is used for network load prediction to reduce the decision making latency \cite{6G1,EW_learn}.\par
Edge computing servers like \emph{cloudlets} are fundamentally a computer or cluster of computers installed in the proximity of mobile device users and distributed across access networks \cite{Satyanarayanan}. Thus, the authors of \cite{Ceselli,efficient,AnsariPlace,SouravJOCN,Martin2,SouravEls} proposed efficient \emph{cloudlet placement frameworks} over wireless and fiber-wireless access networks. As cloudlet computing systems are essentially distributed computing systems, the authors of \cite{offload2,offload3,GameML,price1,price2,game_offload,LB_IoT} addresses the \emph{optimal job request allocation} problem from mobile devices to cloudlets while meeting computation and communication constraints.  Although job allocation frameworks allocate job requests to the most favorable cloudlets, due to the dynamic nature of job request arrival process, cloudlets at different parts of a large network become overloaded and under-loaded at different times.  Thus, the authors of \cite{Jiaload,iotj_load,TONload,Rev13,sourav_CL} designed efficient \emph{load balancing frameworks} among neighboring cloudlets.\par
In this paper, we focus on the load balancing problem among neighboring cloudlets with heterogeneous job requests. We critically observe that the existing literature mainly stressed on minimizing the overall latency of cloudlets while addressing load balancing problems and considered only one job class. However, in practice, job requests can be heterogeneous and users are indifferent if the cloudlets can process the incoming job requests within the requested quality-of-service (QoS) latency, i.e., with a QoS latency target of 10 ms, users do not differentiate among job request processing times 4 ms, 8 ms, or 10 ms. Nonetheless, failing to meet the QoS latency target should incur a significant penalty on the cloudlets. With this realization, we propose a novel game-theoretic utility function which is maximum when the end-to-end latency is equal to the QoS latency target for each job class. In turn, this objective function makes each cloudlet interested in receiving some extra job requests from their neighboring cloudlets and gain some economic benefit, whenever the respective cloudlet is meeting the desired QoS latency target.\par
For load balancing among neighboring cloudlets from the same service provider (SP), network optimization based frameworks proposed in \cite{Jiaload,iotj_load,TONload,Rev13} performs very efficiently.  Nonetheless, in a real heterogeneous deployment scenario, usually multiple cloud SPs install cloudlets over the same customer base and a game-theoretic framework is required as different SPs are non-cooperative in general. Thus, to capture this multi-party economic interaction among heterogeneous neighboring cloudlets, our problem formulation acts as an optimization problem among cloudlets from the same SP and acts like a non-cooperative game among cloudlets from different SPs. In addition, \emph{processor slicing} technique is used by the cloudlets to handle heterogeneous job classes within this framework. As the existing load balancing frameworks make the load balancing decisions after the actual job request arrival to cloudlets, the overhead time of the load balancing algorithms makes these frameworks highly unfit, especially for low-latency applications. To deal with such scenarios, we make the cloudlets predict the job request arrival rates and make the load balancing decisions beforehand, so that immediate processing after actual job request arrival is possible, which is also in line with 6G network vision \cite{6G1}.\par
To compute the Nash equilibrium (NE) load balancing strategies of the cloudlets, firstly we propose a centralized framework where all the federated cloudlets send their predicted job request arrival rates to a mediator.  The mediator computes the NE load balancing strategies for the cloudlets and broadcasts to them before the actual job request arrival.  It is important to note that federated cloudlets are \emph{not always necessarily truthful} while revealing private information e.g., total incoming job requests, and may adopt strategies to gain some additional economic benefit from the market. Thus, we propose a scheme where the neutral mediator is present in the system to impose dominant strategy incentive compatible mechanisms such that \emph{revelation of truthful information is ensured} \cite{groves}. To the best of our knowledge, none of the existing game-theoretic frameworks on load balancing among cloudlets designed any truthful mechanisms.\par
Secondly, we propose a distributed framework to compute NE load balancing strategies among federated cloudlets, which is independent of the truthfulness of cloudlets.  Although distributed frameworks are more robust than centralized frameworks, all the cloudlets need to exchange extensive control information among themselves \cite{survey_load_bal}. This issue can be resolved by using various artificial intelligence-based schemes to learn network conditions and make load balancing decisions. However, the job request arrival process sometimes may vary rapidly and for the sake of robustness against dynamic network scenarios, where there is a very low correlation between the trained data and real-time data, we avoid artificial neural networks that heavily rely on historical data. Therefore, we propose a \emph{reinforcement learning automata-based algorithm} such that quick convergence is ensured. This empowers the cloudlets to make load balancing decisions independently, with a minimal exchange of control information among themselves \cite{Schaerf}. Recently, in \cite{8985529} we proposed a centralized game-theoretic load balancing framework among heterogeneous cloudlets and a single class of job requests. However, the handling of heterogeneous job classes, the incentive-compatible mechanism design, and the reinforcement learning automata-based algorithm were not part of this framework.\par
Our primary contributions in this paper are as follows:
\begin{enumerate}[(i)]
\item We formulate the load balancing problem among federated cloudlets (i.e., cloudlets from the same as well as different SPs) with heterogeneous classes of job requests as a novel economic and non-cooperative game-theoretic problem.  We prove the existence of NE of this game formulation and show that each cloudlet is able to maximize their respective utilities by participating in the load balancing game.
\item We propose a centralized scheme for computation of NE by a neutral mediator that supervises the load balancing game among the cloudlets to ensure fairness in the market competition.  Hence, we design an efficient direct revelation incentive compatible mechanism that ensures that the truthful revelation of private information to the mediator is always a weakly-dominant strategy for the federated cloudlets.
\item We design a distributed continuous-action reinforcement learning automata-based algorithm such that neighboring cloudlets can independently compute the NE load balancing strategy, with a minimal exchange of control information among themselves. We further scaffold the learning algorithm with the particular characteristics of the underlying load balancing game for faster convergence.  We critically study the impacts of exploration and exploitation on the accuracy of the NE learning.
\item Finally, we show that any participating cloudlet can achieve better utilities by mutual computation offloading under different network load conditions following our proposed NE strategies than any of the recent game-theoretical load balancing models. In terms of average end-to-end latency and utility values, the performance of our proposed model is also better than those models, particularly under highly overloaded conditions.
\end{enumerate}
\par The rest of this paper is organized as follows.  Section \ref{sec2} reviews some recent related works.  In Section \ref{sec3}, the details of the system model are presented.  In Section \ref{sec4}, a non-cooperative game-theoretic problem among federated cloudlets for computation offloading is formulated.  In Section \ref{sec5}, a dominant strategy incentive compatible mechanism is designed. In Section \ref{sec6}, a distributed continuous-action reinforcement learning automata-based algorithm is proposed.  In Section \ref{sec7}, the proposed load balancing framework is evaluated.  Finally, in Section \ref{sec8}, our primary achievements by using the game-theoretic framework are summarized.\par
\section{Related Works}\label{sec2}
Load balancing among edge computing nodes such as cloudlets is a major research issue and some researchers have recently proposed load balancing models based on optimization and game-theoretical methods. Primarily, centralized and decentralized control mechanisms are used in existing literature to address load balancing problems \cite{survey_load_bal}.  A common objective function and a series of constraints are formulated in \emph{centralized optimization models} to determine the optimal load balancing strategies for all cloudlets. A centralized problem of latency minimization is formulated by the authors of \cite{Jiaload} and proposed a network-flow based heuristic algorithm for solving it.  These models can provide quick and efficient solutions to the problem, but they are hard to implement on a realistic network situation where cloudlets from various SPs coexist. On the contrary, in \emph{decentralized control models}, all distributed nodes exchange their local control information among themselves and determine the load balancing strategies without any central controller node. Although for large networks such models are more robust, they cause inefficient sharing of control messages and computational burden on the network. Furthermore, reinforcement learning algorithms also seem to be a valuable approach to solving load balance problems but can present different complexity and convergence issues in real-time \cite{iwqos}. In \cite{Res1}, the authors proposed an efficient reinforcement learning algorithm to find the optimal load balancing decision for fog nodes with unknown reward and transition functions and the authors of \cite{Res2} proposed a deep recurrent Q-network approach to approximate the optimal joint task offloading and resource allocation for heterogeneous service tasks in multi-fog node systems.\par
Recently, there is a growing interest in applying cooperative and non-cooperative game-theoretical models to various network-related issues, as game theory offers many effective tools for evaluating and researching the relationship between distributed agents in conflict and cooperation \cite{game_load}. We observed that two recent non-cooperative load balancing frameworks published in \cite{TONload,Rev13} are close to our present work. The authors of \cite{TONload} proposed a distributed non-cooperative load balancing game in small cell networks among the neighboring cloudlets, and compared its findings with a centralized load balancing system that leverage the Lyapunov-drift technique to maximize the long-term system performance. Each cloudlet tries to minimize end-to-end latency costs under specific energy and latency constraints in this formulation. This model, therefore, works very well if the network is loaded moderately, but under very high load conditions it performs very poorly because under very high load conditions, some of the cloudlets start to violate the latency constraints and the NE solution becomes infeasible. By identifying the estimated latency as the dis-utility function of every cloudlet, the authors of \cite{Rev13} devised a non-cooperative load balancing game where cloudlets try to minimize its disutility and proposed an iterative proximal algorithm to compute the NE solution. In this framework, none of the cloudlets is allowed to offload until their incoming job requests reach a certain threshold. Nonetheless, this algorithm tends to assign a large number of job requests to the under-loaded cloudlets and hence, the end-to-end latency overshoots under very high load conditions.\par
To prevent the aforementioned issues, we tactfully integrate the QoS latency target into our game design so that the game does not become infeasible under any circumstances, even under very high load condition. Although the overloaded cloudlets can not offload their entire extra loads to their under-loaded neighbors, our game formulation allows them to offload job requests to the maximum extent. In such cases, the overloaded cloudlets fail to meet the QoS latency target, but their penalties are minimized to some extent. Moerover, the under-loaded cloudlets will still meet the QoS latency target, and hence, all the cloudlets will be able to maximize their individual utilities. The utility of each cloudlet consists of the revenue earned from all the incoming job requests and the penalty for failing to satisfy the QoS latency requirements.

\vspace{-\baselineskip}
\section{System Model}\label{sec3}
In this section, we discuss the considered system model and the primary assumptions made. We consider a general heterogeneous deployment scenario for federated cloudlets over access networks. The total number of federated cloudlets in the network is $N\geq 2$ and $\mathcal{C}=\{1,2,\dots,N\}$ denotes the set of all federated cloudlets.\par
\textbf{(a) Job request arrival process:} We consider multiple classes of job requests with heterogenous QoS requirements and denote the average job request arrival rate of class-$m\in \mathcal{J} = \{1,2,\dots,M\}$ to the $i^{\text{th}}$ cloudlet by $\lambda_i^m$. The total job request arrival rate at the $i^{\text{th}}$ cloudlet is given by $\lambda_i = \sum_{m=1}^M \lambda_i^m$. Each cloudlet relies on the access network SPs to successfully deliver the job requests from the associated mobile devices to them and pays for the bandwidth consumed. However, if the network fails to deliver some of the job requests to the cloudlets due to bandwidth constraints, then the network SPs pay a penalty in proportion to the undelivered job requests. As the average job request arrival varies from time instance to time instance, we assume that each $\lambda_i^m$ is \emph{independently distributed} over the support $\Lambda_i^m = [0,\lambda_i^{m,max}],\forall i\in\mathcal{C}$. Therefore, the \emph{computation job request profile} or \emph{true type} of all the federated cloudlets is represented as $\bm{\lambda}^m = (\lambda_1^m,\lambda_2^m,\dots,\lambda_N^m)\in \bm{\Lambda}^m = (\Lambda_1^m\times \Lambda_2^m\times\dots\times \Lambda_N^m)$. In practice, the job request arrival process to cloudlets is \emph{self-similar} and \emph{non-stationary} \cite{trace}. Therefore, the cloudlets predict the incoming job request arrival rates by employing long short-term memory (LSTM) networks \cite{lstm}. The transmission latency of the incoming job requests and the intermediate transmission latencies with the neighboring cloudlets are also estimated by each cloudlet.\par
\textbf{(b) Job request service process:} We assume that each processor in a cloudlet has similar job processing capabilities and the average service rate of incoming job requests of class-$m$ is $\mu_i^m$ (jobs/s). Therefore, $\mu_i^m$ indicates a parametric description of the job requests arrived at the $i^{\text{th}}$ cloudlet. We consider that each $i^{\text{th}}$ cloudlet has total $n_i\in\mathbb{Z}_{\geq 1}$ processors and dedicates $n_i^m\in\mathbb{R}_{\geq 1}$ processors for each job class-$m$. By using Google cluster-usage traces, the authors of \cite{poisson} showed that \emph{exponential distribution} fits perfectly with the service times of the job requests. In practice, some of the jobs are completely parallelized and some cannot be parallelized at all. Hence, to guarantee the worst-case performance of the cloudlets, we model the cloudlets as $M/M/c$ queuing systems \cite{MJia}. \par
\textbf{(c) QoS latency requirements of job requests:} The individual job requests from mobile devices demand a certain number of CPU cycles to process the jobs within a predefined QoS latency target $D_{Qm}$ \cite{GameML}. However, in this paper we are considering a batch of incoming job requests rather than individual job requests to cloudlets. Thus, we denote the computational and latency requirements of the job requests of class-$m$ to $i^{\text{th}}$ cloudlet by the consolidated tuple $(\mu_i^m,\lambda_i^m,D_{Qm})$. To handle the processing of different job classes, we use \emph{processor slicing} technique \cite{Res2} to slice the total $n_i$ processors into $M$ slices of $n_i^m$ processors for each job class-$m$. Moreover, each cloudlet uses a \emph{timeslotted model} to ensure the QoS latency target $D_{Qm}$ for the job requests of each job class-$m$. The duration of each fundamental timeslot is chosen to be equal to $D_Q$ and without loss of generality, we assume that $D_{Qm}$ values are integer multiples of $D_Q$, i.e., $D_{Q1} = I_1\times D_Q$, $D_{Q2} = I_2\times D_Q$, $D_{Q3} = I_3\times D_Q$, and so on ($I_1,I_2, I_3,\dots, I_M$ are integer values). Depending on the stationarity of the incoming job request traffic, we choose a bigger time interval $\tau_x$ which is an integer multiple of all of $I_1,I_2, I_3,\dots, I_M$. Each cloudlet uses a prediction algorithm a few timeslots before the beginning of $\tau_x$ interval to predict the job request arrival rates for all the $M$ job classes. Based on this predicted job request arrival rates, the processor slicing and the NE load balancing strategies for each cloudlet are computed. These values remain unchanged over each interval $\tau_x$. As the job requests arrive within each timeslot, they are marked with an integer $(I_m-1)$. If some jobs could not be processed within that timeslot, they are rolled over to the next timeslot by decreasing the marking by 1. This can continue until those jobs are processed and the jobs can be deleted after the marking becomes 0. Therefore, the job request arrival queue of the cloudlets can maintain a steady-state unless they are extremely overloaded. As the $M/M/c$ queue provides the worst-case processing latency of the cloudlets, we are also ensuring that all the incoming job requests are processed when the average latency of each cloudlet is $\leq D_{Qm}$. Please refer to Appendix-A for more discussions on timeslots and modelling of cloudlets as $M/M/c$ queues.\par
\begin{table}[!t]
\centering
\caption{Symbol definitions}
\label{table1}
\resizebox{\columnwidth}{!}{%
\begin{tabular}{ll}
\toprule \toprule
\multicolumn{1}{c}{\textbf{Symbol}} & \multicolumn{1}{c}{\textbf{Definition}}                                                     \\ \hline
$N$                                 & Total number of federated cloudlets in the network                                          \\
$M$                                 & Total number of different job classes generated from mobile devices                   \\
$\mu_i^m$                             & Average service rate of job class-$m$ at $i^{\text{th}}$ cloudlet                                            \\
$\lambda_i^m$                       & Average job request arrival rate of job class-$m$ to $i^{\text{th}}$ cloudlet               \\
$\hat{\lambda}_i^m$                 & Revealed average job request arrival rate of job class-$m$ to $i^{\text{th}}$ cloudlet      \\
$D_{Qm}$                            & QoS latency target of job class-$m$                                                         \\
$n_i$                               & Total number of processors installed in $i^{\text{th}}$ cloudlet                            \\
$n_i^m$                             & Number of processors for job class-$m$ of $i^{\text{th}}$ cloudlet                          \\
$\varphi_{ij}^m$                    & The fraction of jobs $i^{\text{th}}$ cloudlet offloads to $j^{\text{th}}$ cloudlet          \\
$\Omega_{i,1}^m$                    & Revenue earned by $i^{\text{th}}$ cloudlet per workload of class-$m$                        \\
$\Omega_{ij,2}^m$                   & Incentive paid for offloading class-$m$ jobs by $i^{\text{th}}$ to $j^{\text{th}}$ cloudlet \\
$\Omega_{i,3}^m$                    & Penalty paid by $i^{\text{th}}$ cloudlet for violating QoS target latency                   \\
$\Omega_{i,4}^m$                    & Penalty paid to the market regulator for not processing received jobs        \\
$t_{ui}^m$                          & Data transmission latency between mobile devices and $i^{\text{th}}$ cloudlet               \\
$t_{ij}^m$                          & Data transmission latency between $i^{\text{th}}$ and $j^{\text{th}}$ cloudlets             \\ \hline
\end{tabular}
}
\end{table}
\textbf{(d) User mobility model:} We assume that the mobile users cannot move beyond the coverage area of a cloudlet within a few milliseconds, thus consider the \emph{quasi-static mobility model} for mobile users. This means that mobile users can be considered almost stationary to the corresponding cloudlets during computation offloading period, but may move on later \cite{GameML}. Each cloudlet prioritizes the processing of the incoming job requests internally or offloads to a neighboring cloudlet to satisfy the QoS latency target $D_{Qm}$ through some internal scheduling algorithm (beyond the scope of this paper).\par

\section{Economic and Non-cooperative Load Balancing Game among Cloudlets}\label{sec4}
In this section, we formulate the load balancing problem among $N\geq 2$ neighboring federated cloudlets as a continuous-kernel non-cooperative game.  In a practical deployment scenario, overloaded cloudlets intend to offload a fraction of its job requests to its under-loaded neighboring cloudlets. We denote the fraction of class-$m$ job requests $i$\textsuperscript{th} cloudlet offloads to its $j^{\text{th}}$ neighboring cloudlet by $\varphi_{ij}^m$. The complete job request offloading strategy space of all cloudlets for each job class-$m$ is defined as a matrix $\Phi^m =(({\Phi_1^m})^T,(\Phi_2^{m})^T, \dots,(\Phi_N^{m})^T)^T \subset \mathbb{R}^{N\times N}$, where $\bm{\varphi}_i^m = (\varphi_{i1}^m,\varphi_{i2}^m,\dots,\varphi_{iN}^m)\in \Phi_i^m \subset \mathbb{R}^{N}$, $\varphi_{ij}\in \Phi_{ij}^m = [0,1]\subset\mathbb{R}$, $\sum_{j=1}^{N}\varphi_{ij}^m =1,\forall i\in \mathcal{C}$. In a stable market scenario, all the SPs tend to install cloudlets with similar processing capacity (i.e., $\mu_i^m = \mu_j^m, \forall i, j$) to meet a standard QoS for the same customer base. Hence, the total processing latency of the class-$m$ job requests at $i^{\text{th}}$ cloudlet with $\bar{\lambda}_i^m(\bm{\varphi}_i^m,\bm{\varphi}_{-i}^m) = \{(1-\sum_{j\neq i} \varphi_{ij}^m)\lambda_i^m +\sum_{j\neq i} \varphi_{ji}^m\lambda_j^m\}$ and $\bar{\rho}_i^m = \bar{\lambda}_i^m/(n_i^m\mu_i^m)$ can be derived as follows:
\begin{align}
\mathbb{T}_i^m (\mu_i^m,\bar{\lambda}_i^m) = \frac{1}{\mu_i^m}+ \frac{\mathcal{E}_C(n_i^m,\bar{\lambda}_i^m/\mu_i^m)}{n_i^m\mu_i^m-\bar{\lambda}_i^m}, \label{eq01}
\end{align}
where, $\mathcal{E}_C(n_i^m,\bar{\lambda}_i^m/\mu_i^m)$ is the \emph{Erlang-C formula}, given by:
\begin{align}
\mathcal{E}_C(n_i^m,\bar{\lambda}_i^m/\mu_i^m) = \frac{1}{1+(1-\bar{\rho}_i^m)\left(\frac{n_i^m !}{(n_i^m \bar{\rho}_i^m)^{n_i^m}}\right)\sum\limits_{k=0}^{n_i^m-1} \frac{(n_i^m \bar{\rho}_i^m)^k}{k!}}. \label{eq02}
\end{align}
\par Each $i^{\text{th}}$ cloudlet makes an optimal processor slicing by observing their load conditions and by solving the following optimization problem:
\begin{align}
\mathcal{P}_i^m: \quad \quad \min_{(n_i^1,n_i^2,\dots,n_i^M)} \quad & \max_{m\in\mathcal{J}} \left\{t_{ui}^m+\mathbb{T}_i^m(\mu_i^m,\lambda_i^m;n_i^m) - D_{Qm}\right\} \nonumber\\
\text{subject to}\quad &1 \leq n_i^m \leq n_i, \sum_{m=1}^{M} n_i^m = n_i, \forall m \in \mathcal{J}.  \nonumber
\end{align}
\par We consider that soft processor slicing is available and $n_i^m$ can take any real and $\geq 1$ value. Hence, $\mathcal{P}_i^m$ is a continuous convex optimization problem and can be solved by gradient projection algorithm (refer to Appendix-B). 

\vspace{-\baselineskip}
\subsection{Economic and Non-cooperative Game Formulation}
In this paper, we consider the most commonly used pricing schemes e.g., \emph{pay-as-you-go} policy, where users pay a fixed price per job request without any long-term commitments \cite{IBM_pay}. For the total amount of incoming class-$m$ job requests from all the connected mobile devices, each $i^{\text{th}}$ cloudlet earns a linearly proportional revenue ($\Omega_{i,1}^m$) per workload. Each $i^{\text{th}}$ cloudlet also pays a linearly proportional price per workload ($\Omega_{ij,2}^m$) for offloading job requests to a neighboring $j^{\text{th}}$ cloudlet from a different SP and also, receives a linearly proportional price for executing its neighbor's offloaded jobs. The cloudlets can also use cooperative or bargaining strategies among themselves to decide the value of $\Omega_{ij,2}^m$. We define a parameter $\gamma_{ij}$ to distinguish the price for offloading a job request to neighboring cloudlets as follows:
\begin{equation*} 
\gamma_{ij}=\begin{cases}1; & \text{if neighboring cloudlet belongs to different SP }\\ 0 ; & \text{if neighboring cloudlet belongs to the same SP} \end{cases}
\end{equation*}
\par This means that each $i^{\text{th}}$ cloudlet pays a price to $j^{\text{th}}$ cloudlet to offload any job requests when it belongs to another SP, i.e. $\gamma_{ij}=1$. In addition to these, each $i^{\text{th}}$ cloudlet pays a penalty price with a proportionality cost factor ($\Omega_{i,3}^m$) for exceeding the QoS target latency $D_{Qm}$. Note that if an overloaded cloudlet offloads some job requests to a neighboring cloudlet and it fails to process them for some reason, then the penalty is actually paid by the neighboring cloudlet. In this work, we consider a linear penalty price similar to the linear latency cost designed in \cite{TONload}. Therefore, all the federated cloudlets with \emph{utility functions} $\mathcal{U}^N_i(\bm{\varphi}_i,\bm{\varphi}_{-i}),\forall i\in\mathcal{C}$, where $\bm{\varphi}_{-i} = (\bm{\varphi}_1,\dots,\bm{\varphi}_{i-1},\bm{\varphi}_{i+1},\dots,\bm{\varphi}_N)$, in the load balancing game intend to solve the following maximization problem:\par
\begin{resizealign} 
&\max_{\bm{\varphi}_i \in \Phi_i}\mathcal{U}^{N}_i(\bm{\varphi}_i,\bm{\varphi}_{-i}) = \sum_{m=1}^M \Omega_{i,1}^m \frac{\lambda_i^m}{n_i^m\mu_i^m} \nonumber\\
& +\sum_{m=1}^M \sum_{j=1,j\neq i}^{N} \Omega_{ji,2}^m \gamma_{ji} \frac{\varphi_{ji}^m\lambda_j^m}{n_i^m\mu_i^m}- \sum_{m=1}^M \sum_{j=1,j\neq i}^{N} \Omega_{ij,2}^m \gamma_{ij}\frac{\varphi_{ij}^m\lambda_i^m}{n_j^m\mu_j^m}\nonumber\\
&-\sum_{m=1}^M \Omega_{i,3}^m \left[\left(1-\sum_{j=1,j\neq i}^{N}\varphi_{ij}^m\right)\frac{\lambda_i^m}{n_i^m \mu_i^m} \max\left\{0,\left(t_{ui}^m+\mathbb{T}_i^m(\mu_i^m,\bar{\lambda}_i^m)-D_{Qm}\right)\right\}\right.\nonumber\\
& \left.+\sum_{j=1,j\neq i}^{N} \frac{\varphi_{ji}^m\lambda_j^m}{n_i^m\mu_i^m} \max\left\{0,\left(t_{uj}^m+\mathbb{T}_i^m(\mu_i^m,\bar{\lambda}_i^m)+t_{ji}^m-D_{Qm} \right)\right\} \right] \label{eq03}
\end{resizealign}\par \vspace{-0.5\baselineskip} 
\begin{align}
\text{subject to}\quad &0 \leq \varphi_{ij}^m \leq 1, \forall m \in \mathcal{J} \nonumber\\
& \sum\nolimits_{m=1}^M b_i^m\varphi_{ij}^m\lambda_i^m \leq B_{ij}, \forall j\neq i \in \mathcal{C}. \label{eq04}
\end{align} 
\par The first term in (\ref{eq03}) denotes the total payment received by the cloudlet from mobile users and is linearly proportional to the average workload.  The second term denotes the payment $i^{\text{th}}$ cloudlet receives from $j^{\text{th}}$ cloudlet to execute its offloaded job requests and the third term denotes the payment $i^{\text{th}}$ cloudlet makes to $j^{\text{th}}$ cloudlet for offloading job requests.  The fourth term denotes the penalty $i^{\text{th}}$ cloudlet pays for overall latency (sum of transmission, processing, and queuing latencies) if it exceeds $D_{Qm}$ against the total incoming class-$m$ job requests, otherwise no penalty is applied. We denote the average round-trip data transmission latency among mobile devices and the corresponding $i^{\text{th}}$ cloudlet by $t_{ui}$ and the inter-cloudlet round-trip data transmission latency by $t_{ij},\forall i,j\neq i\in\mathcal{C}$. We also consider that overloaded cloudlets may face network bandwidth constraint while offloading job requests as $\sum_{m=1}^M b_i^m \varphi_{ij}^m\lambda_i^m \leq B_{ij}$, where $b_i^m$ denotes the average number of bits/job request of class-$m$ and $B_{ij}$ denotes the bandwidth available in the link between $i^{\text{th}}$ and $j^{\text{th}}$ cloudlets. Each cloudlet needs to pay a price to network SPs for the bandwidth consumed by the offloaded job requests to neighboring cloudlets, but this price is paid separately and over a longer period of time. We assume that the cloudlets operate under the \emph{condition of stable operation}, i.e., $\{(1-\sum_{j\neq i}\varphi_{ij}^m)\lambda_i^m+\sum_{j\neq i}\varphi_{ji}^m\lambda_j^m\}/(n_i^m \mu_i) < 1, \forall i,j\neq i \in\mathcal{C}, \forall m \in \mathcal{J}$. The utility of each cloudlet in this load balancing game is an affine function when the total latency is within $D_{Qm}$, otherwise, it becomes a non-linear function whose maximum value is achieved at total end-to-end latency equal to $D_{Qm}$. Hence, the cloudlets are always interested in gaining some incentives by receiving some extra job requests from neighboring cloudlets without exceeding $D_{Qm}$ but, the utility starts to decrease beyond this point. Moreover, the \emph{individual rationality} of each federated cloudlet is maintained due to the \emph{default utility}, $\mathcal{U}_i^0, \forall i\in \mathcal{C}$.\par
Furthermore, due to the utility function (\ref{eq03}) and constraints (\ref{eq04}), which does not provide an explicit latency bound on the participating cloudlets, even highly over-loaded cloudlets can participate in the game and can offload some of the job requests to the relatively under-loaded neighboring cloudlets. This leads to a utility higher than the utility achieved without participating in the game. Note that under such network load conditions, the game formulation in \cite{TONload} that has explicit delay bound on participating cloudlets becomes infeasible and a valid NE solution can not be computed. We prefer to investigate the NE of the game $\Gamma$, because none of the federated cloudlets find it beneficial to deviate unilaterally from the NE computational offload strategy $\bm{\varphi}^{*}=(\bm{\varphi}^*_{1},\bm{\varphi}^*_{2},\dots,\bm{\varphi}^*_{N})$.
\begin{lemma}\label{lm1}
\textit{The utility functions $\mathcal{U}^N_i(\bm{\varphi}_i,\bm{\varphi}_{-i}),\forall i\in\mathcal{C}$ of the under-loaded cloudlets are affine and the overloaded cloudlets are quasi-concave functions of $\bm{\varphi}_i$.}
\end{lemma}
\begin{IEEEproof}
Please refer to Appendix C.
\end{IEEEproof}
\begin{theorem}\label{th1}
\textit{At least one pure strategy NE exists for the game $\Gamma=\langle \mathcal{C},(\Phi_i^m)_{i\in \mathcal{C},m\in\mathcal{J}},(\mathcal{U}^N_i(\bm{\varphi}_i,\bm{\varphi}_{-i}))_{i\in \mathcal{C}} \rangle$.}
\end{theorem}
\begin{IEEEproof}
In the game $\Gamma$, the strategy spaces of all the federated cloudlets $\Phi_i$ are \emph{compact and convex} in nature. The utility functions $\mathcal{U}^N_i(\bm{\varphi}_i,\bm{\varphi}_{-i}), \forall i\in \mathcal{C}$ are continuous functions of $(\bm{\varphi}_i,\bm{\varphi}_{-i})$ with the condition of stable operation, $\{(1-\sum_{j\neq i}\varphi_{ij}^m)\lambda_i^m+\sum_{j\neq i}\varphi_{ji}^m\lambda_j^m\}/(n_i^m \mu_i^m) < 1, \forall i,j\neq i \in\mathcal{C}, \forall m \in \mathcal{J}$.  Moreover, we showed in Lemma \ref{lm1} that $\mathcal{U}_i^N (\bm{\varphi}_i,\bm{\varphi}_{-i}), \forall i \in \mathcal{C}$ are affine or quasi-concave functions of $\bm{\varphi}_i$. These are the sufficient conditions to ensure the existence of a pure strategy NE for the non-cooperative load balancing game $\Gamma$.
\end{IEEEproof}
\subsection{Computation of the Pure-Strategy Nash Equilibrium}
We observe that the utility functions $\mathcal{U}^N_i(\bm{\varphi}_i,\bm{\varphi}_{-i}), \forall i\in \mathcal{C}$ are non-differentiable in nature due to the presence of $\max\{0,x\}$ function. Hence, we cannot derive the \emph{best response functions} of the cloudlets by directly differentiating the utility functions. At first, we identify whether the cloudlets are under-loaded or overloaded and organize their utilities accordingly. As different job classes are processed independent of each other through processor slicing, each cloudlet can simultaneously be under-loaded for one job class while overloaded for some other job class. After this, we proceed to use the \emph{necessary conditions}, i.e., the first-order KKT conditions (FOC) to compute the pure-strategy NE load balancing strategies \cite{political}. Intuitively, three cases can arise in practice and the pure strategy NE solutions are described as follows:\par
\textbf{Case-1:} $\left[t_{ui}^m+ \mathbb{T}_i^m(\mu_i^m,\lambda_i^m) \right] < D_{Qm}$, $\left[t_{uj}^m+ \mathbb{T}_j^m(\mu_j^m,\lambda_j^m) \right] < D_{Qm},\forall j\neq i\in\mathcal{C},\forall m\in \mathcal{J}$. In this case, all the cloudlets are under-loaded, i.e., they have sufficient computational resources to meet the QoS latency target $D_{Qm}$. Hence, the \emph{unique NE solution} is $\varphi_{ij}^{m*} = \varphi_{ji}^{m*} = 0$. This implies that both the cloudlets can achieve their maximum utilities as the total revenue earned without offloading any job requests to each other.\par
\textbf{Case-2:} $\left[t_{ui}^m+ \mathbb{T}_i^m(\mu_i^m,\lambda_i^m) \right] \geq D_{Qm}$, $\left[t_{uj}^m+ \mathbb{T}_j^m(\mu_j^m,\lambda_j^m) \right] \geq D_{Qm}, \forall j\neq i\in\mathcal{C},\forall m\in \mathcal{J}$. In this case, all the cloudlets are overloaded and they can not reduce their individual latencies by offloading any job requests to each other. Thus, it is obvious that the \emph{unique NE strategy} for both the cloudlets is not to offload any job requests to each other, i.e., $\varphi_{ij}^{m*} = \varphi_{ji}^{m*} = 0$.\par
\textbf{Case-3:} $\left[t_{ui}^m+ \mathbb{T}_i^m(\mu_i^m,\lambda_i^m) \right] \geq D_{Qm}$, $\left[t_{uj}^m+ \mathbb{T}_j^m(\mu_j^m,\lambda_j^m) \right] < D_{Qm}$, $\forall i\in\mathcal{C}_o, j\in \mathcal{C}_u,m\in \mathcal{J}$, where $\mathcal{C}_o \subset \mathcal{C}$ is the set of overloaded cloudlets and $\mathcal{C}_u = \mathcal{C}\setminus \mathcal{C}_o$ is the set of under-loaded cloudlets. Thus, $j^{\text{th}}$ cloudlets do not need to offload anything, i.e., $\varphi_{ji}^{m*} = 0$ but $i^{\text{th}}$ cloudlet needs to offload their excess job requests to $j^{\text{th}}$ cloudlets to meet the QoS target latency $D_Q$, as long as overloaded cloudlets do not exceed $D_{Qm}$. Hence, the following \emph{NE solution} is $\varphi_{ij}^{m*} > 0$ and $\varphi_{ji}^{m*} = 0$. As we can not solve this game analytically, we can verify this solution by a suitable algorithmic solution. As long as the under-loaded $j^{\text{th}}$ cloudlets can process the entire extra load from $i^{\text{th}}$ cloudlet, they accept the entire workload. However, when overloaded cloudlets cannot process the entire extra load, then they allow the $i^{\text{th}}$ only to offload job requests partially such that they do not exceed their QoS target latency $D_{Qm}$.\par
In this case, we realized that the NE load balancing strategies are not entirely controlled by the overloaded cloudlets but also by the under-loaded cloudlets. Hence, we introduce a new set of decision variables for each $i^{\text{th}}$ cloudlet denoting the fraction of job requests received from all $j^{\text{th}}$ cloudlets i.e., $\bm{\psi}_i^m = (\psi_{1i}^m,\psi_{2i}^m,\dots,\psi_{Ni}^m)^T$ with the jointly shared equality constraints, $\varphi_{ji}^m = \psi_{ji}^m, \forall i,j\neq i \in\mathcal{C}, m \in \mathcal{J}$. We can mark the $N_u$ under-loaded cloudlets and $N_o$ overloaded cloudlets by observing $\lambda_i^m$ such that $N_u+N_o = N$. Moreover, the latency constraints and FOC from (\ref{eq03})-(\ref{eq04}) for under-loaded cloudlets with $\tilde{\lambda}_i^m = \{(1-\sum_{j\neq i} \varphi_{ij}^m)\lambda_i^m +\sum_{j\neq i} \psi_{ji}^m\lambda_j^m\}$ and vectors of Lagrange multipliers $\bm{\alpha}_i^m$, $\bm{\beta}_i^m$, $\bm{\xi}_i^m$, and $\bm{\eta}_i$ are written as:
\begin{gather} 
\left(t_{uj}^m  + \mathbb{T}_i^m(\mu_i^m,\tilde{\lambda}_i^m)+t_{ij}^m - D_{Qm}\right) \leq 0, \forall i \in \mathcal{C},m\in\mathcal{J}, \label{eq05}\\ 
\nabla_{\varphi_i^m} \mathcal{U}_i^N  +  \nabla_{\varphi_i^m} \left[(\bm {\alpha}_i^m)^T \bm {\varphi}_i^m + (\bm {\beta}_i^m)^T \left(\bm {1}-\bm {\varphi}_i^m \right) \right. \nonumber\\
 - (\bm {\xi}_i^m)^T \left(t_{uj}^m+ \mathbb{T}_i^m(\mu_i^m,\tilde{\lambda}_i^m) +t_{ji}^m-D_{Qm} \right) \nonumber\\
\left. +\eta_i \left(B_{ij}-\sum\nolimits_{m=1}^M b_i^m\varphi_{ij}^m\lambda_i^m\right) \right] = 0, \forall i \in \mathcal{C},m\in\mathcal{J}. \label{eq06}
\end{gather}
\par Similarly, the latency constraints and FOC from (\ref{eq03})-(\ref{eq04}) for the overloaded cloudlets respectively are written as follows:
\begin{gather} 
\left(t_{ui}^m+\mathbb{T}_i^m(\mu_i^m,\tilde{\lambda}_i^m)-D_{Qm}\right) \geq 0, \forall i \in \mathcal{C},m\in\mathcal{J}, \label{eq07}\\
\nabla_{\varphi_i^m} \mathcal{U}_i^N + \nabla_{\varphi_i^m} \left[(\bm {\alpha}_i^m)^T \bm {\varphi}_i^m + (\bm {\beta}_i^m)^T \left(\bm{1}-\bm {\varphi}_i^m \right) \right. \nonumber\\
 + (\bm {\xi}_i^m)^T \left(t_{ui}^m+\mathbb{T}_i^m(\mu_i^m,\tilde{\lambda}_i^m)-D_{Qm} \right) \nonumber\\
\left.+\eta_i \left(B_{ij}-\sum\nolimits_{m=1}^M b_i^m\varphi_{ij}^m\lambda_i^m \right)\right] = 0, \forall i \in \mathcal{C},m\in\mathcal{J}. \label{eq08}
\end{gather} 

\begin{theorem}\label{th2}
\textit{The pure strategy NE of the load balancing game $\Gamma=\langle \mathcal{C},(\Phi_i^m)_{i\in \mathcal{C},m\in\mathcal{J}},(\mathcal{U}^N_i(\bm{\varphi}_i,\bm{\varphi}_{-i}))_{i\in \mathcal{C}} \rangle$ can be computed as the solution of the equivalent constrained optimization problem $\bm{\mathcal{P}_{\Gamma}^m}$.}
\end{theorem}\par \vspace{-\baselineskip}
\begin{resizealign} 
\mathcal{P}_{\Gamma}^m: & \quad \quad \quad  \text{Minimize} \sum_{m=1}^M \sum_{i=1}^{N} \left[(\bm{\alpha}_i^m)^T \bm{\varphi}_i^m + (\bm{\beta}_i^m)^T \left(1-\bm{\varphi}_i^m\right) \right] \nonumber\\
& -\sum_{m=1}^M \sum_{i=1}^{N_u} (\bm{\xi}_i^m)^T \left(t_{uj}^m+\mathbb{T}_i^m(\mu_i^m,\tilde{\lambda}_i^m) +t_{ji}^m - D_{Qm} \right)\nonumber\\
& +\sum_{m=1}^M \sum_{i=1}^{N_o} (\bm{\xi}_i^m)^T \left(t_{ui}^m+ \mathbb{T}_i^m(\mu_i^m,\tilde{\lambda}_i^m) - D_{Qm}\right) \nonumber\\
& +\sum_{i=1}^N \sum_{j=1,j\neq i}^N \bm{\eta}_i^T \left(B_{ij}-\sum\nolimits_{m=1}^M b_i^m\varphi_{ij}^m\lambda_i^m \right)
\end{resizealign} \vspace{-\baselineskip}
\begin{align}
\text{subject to}\quad &\bm{\alpha}_i^m, \bm{\beta}_i^m, \bm{\xi}_i^m, \bm{\eta}_i \succeq  \bm{0}, \forall i \in \mathcal{C},m\in\mathcal{J} \nonumber \\
&0 \leq \varphi_{ij}^m \leq 1, \forall i,j\neq i \in \mathcal{C},m\in\mathcal{J} \nonumber \\
&\varphi_{ji}^m = \psi_{ji}^m, \forall i,j\neq i \in \mathcal{C},m\in\mathcal{J} \nonumber\\
& \text{constraints (\ref{eq05})-(\ref{eq08})}, \forall i \in \mathcal{C}. \nonumber
\end{align} \vspace{-\baselineskip}
\begin{algorithm}[b!]
\caption{Projection Algorithm with Constant Step Size}\label{alg1}
\begin{algorithmic}[1]
\State \textbf{Input:} Network parameters $n_i^m$, $\mu_i^m$, $\lambda_i^m$, $t_{ui}^m$, $t_{ij}^m, \forall i,j\neq i \in\mathcal{C}, m \in \mathcal{J}$.
\State \textbf{Output:} The pure-strategy NE of the non-cooperative load balancing game.
\State \textbf{Initialization:} Choose any Lagrange multipliers $\bm{\alpha}^{m,(0)}$, $\bm{\beta}^{m,(0)}$, $\bm{\xi}^{m,(0)}$, $\bm{\eta}^{(0)} \succeq 0$, step size $\omega > 0$, and tolerance limit $\epsilon > 0$.  Set the iteration index $t=0$.
\If {all cloudlets are under-loaded for job class-$m$, i.e., $\left[t_{ui}^m+ \mathbb{T}_i^m(\mu_i^m,\lambda_i^m) \right] < D_{Qm}$, or all cloudlets are overloaded for job class-$m$, i.e., $\left[t_{ui}^m+ \mathbb{T}_i^m(\mu_i^m,\lambda_i^m) \right] \geq D_{Qm}$}
choose not to offload, i.e., set $\bm{\varphi}^{m*}=\bm{I}_N$: STOP;
\Else { identify the FOC of the under-loaded cloudlets as (\ref{eq05})-(\ref{eq06}) and the FOC of the overloaded cloudlets as (\ref{eq07})-(\ref{eq08}).  Use the corresponding CSC to formulate the objective function of $\mathcal{P}_{\Gamma}^m$.}
\EndIf
\If {$\bm{\varphi}^{m,(t)}(\bm{\chi}^{m,(t)})$, $\bm{\alpha}^{m,(t)}(\bm{\chi}^{m,(t)})$, $\bm{\beta}^{m(t)}(\bm{\chi}^{m(t)})$, $\bm{\xi}^{m,(t)}(\bm{\chi}^{m,(t)})$, $\bm{\eta}^{(t)}(\bm{\chi}^{m,(t)})$ and the objective satisfies the desirable tolerance limit $\epsilon$} STOP;
\EndIf
\State With given $\bm{\chi}^{m,(t)}$, compute $\bm{\varphi}^{m,(t)}(\bm{\chi}^{m,(t)})$, $\bm{\alpha}^{m,(t)}(\bm{\chi}^{m,(t)})$, $\bm{\beta}^{m,(t)}(\bm{\chi}^{m,(t)})$, $\bm{\xi}^{m,(t)}(\bm{\chi}^{m,(t)})$, $\bm{\eta}^{(t)}(\bm{\chi}^{m,(t)})$ as the solution of the reformulated optimization problem $\mathcal{P}_{\Gamma}^m$;
\State Update the Lagrange multipliers $\bm{\chi}^m$ corresponding to the constraints of $\mathcal{P}_{\Gamma}^m$ by gradient projection as follows:
\begin{align} 
\bm{\chi}^{m,(t+1)} = [\bm{\chi}^{m,(t)} + \omega \Theta(\bm{\chi}^{m,(t)})]^+, \label{eq09}
\end{align} 
where $\Theta$ is the feasible set of $\mathcal{P}_{\Gamma}^m$.
\State Set $t \gets t+1$; go to Step 6.
\end{algorithmic}
\end{algorithm} 
\par Please refer to Appendix D for the proof. Nonetheless, it is noteworthy that the problem $\mathcal{P}_{\Gamma}^m$ is not a convex optimization problem. Hence, we can \emph{reformulate the problem $\mathcal{P}_{\Gamma}^m$ with a convex feasible set and a continuously differentiable objective function} by adding the squares of left hand sides of (\ref{eq06}) and (\ref{eq08}) to the objective \cite{game_opt}.  This problem now can be solved by using a \emph{gradient-projection algorithm} or any commercially available solver package. The global optimal solution is achieved only when the value of the objective is 0 (or approximately 0 within a desired tolerance limit) and this corresponds to the NE of the game $\Gamma$. Therefore, an algorithm to solve the game $\Gamma$ is summarized in Algorithm \ref{alg1}.\par
\begin{figure}[t!]
\centering
\includegraphics[width=\columnwidth,height=5.9cm]{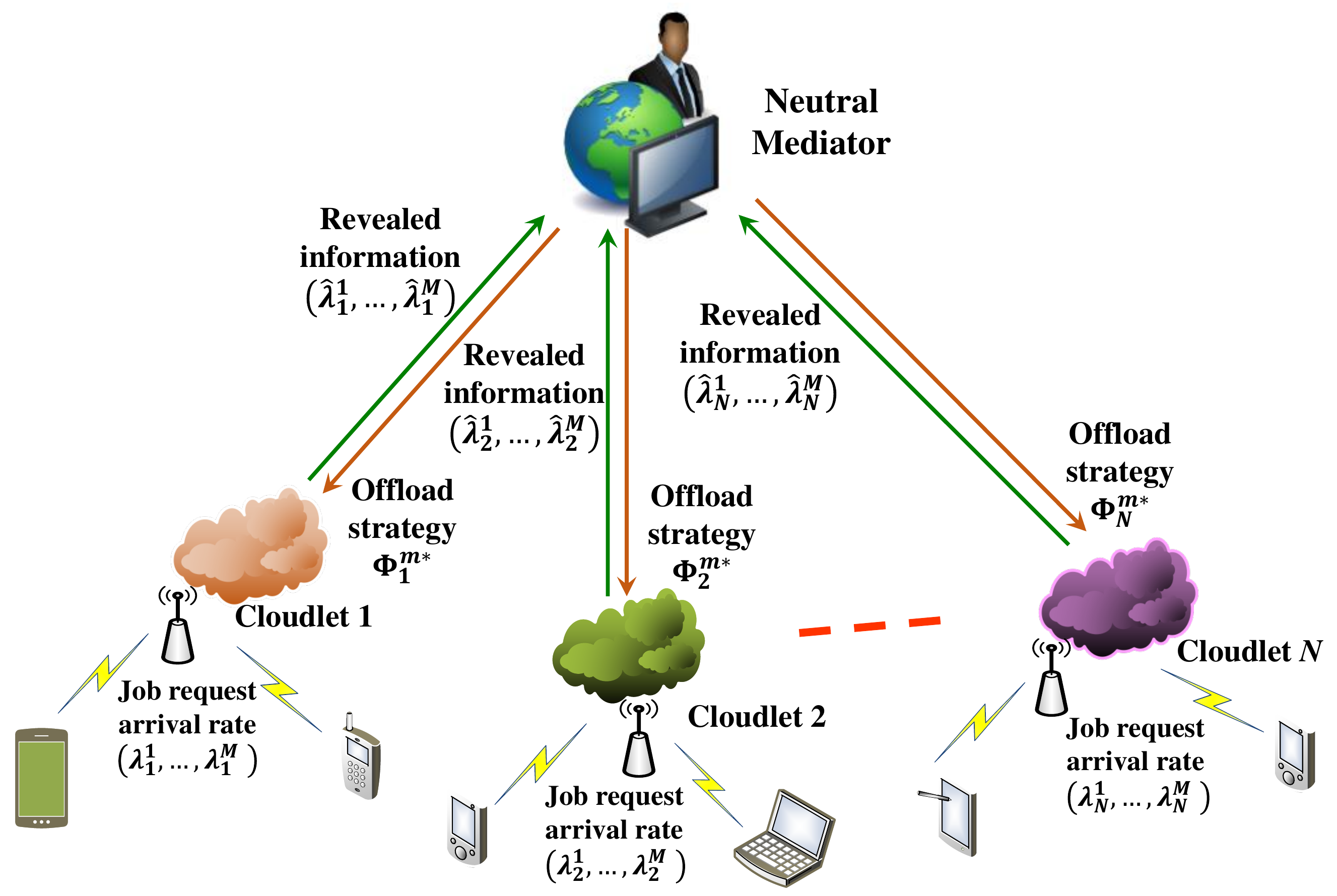}
\caption{A simplified schematic diagram to show the information exchange among $N$ federated cloudlets and a neutral mediator.}
\label{game_cloudlet}
\end{figure}
\setlength{\textfloatsep}{5pt}

\section{Centralized Load Balancing Framework} \label{sec5}
In this section, we propose a centralized load balancing framework by using our game formulation in Section \ref{sec4}. Private information of a cloudlet is defined as the information that is known only to that cloudlet and no other entity in the network \cite{Narahari}. As the federated cloudlets are \emph{non-cooperative} and \emph{rational utility maximizers}, they are not interested in sharing their private information like incoming job request arrival rates with their neighboring cloudlets. On the other hand, a neutral mediator does not have any utility associated with the incoming job requests. The mediator acts like a supervisor that ensures fair participation of all the federated cloudlets in the market competition and installs a computational facility in the proximity of the federated cloudlets to compute the NE for all the cloudlets by using a centralized algorithm, as shown in Fig. \ref{game_cloudlet}. In a centralized framework, the mediator firstly addresses the \emph{preference elicitation problem} to elicit truthful information from the federated cloudlets by imposing efficient mechanisms on the cloudlets who periodically reveal their private information about incoming job requests. We represent the \emph{revealed type profile} for job class-$m$ of the federated cloudlets as $\hat{\bm{\lambda}}^m = (\hat{\lambda}_1^m,\hat{\lambda}_2^m,\dots,\hat{\lambda}_N^m)\in \bm{\Lambda}^m$.
However, we assume that cloudlets usually truthfully reveal other network parameters like $\mu_i^m$, $t_{ui}^m$, and $t_{ij}^m$, because $\mu_i^m$ is dependent on the type of job requests and usually do not change frequently, and $t_{ui}^m$, $t_{ij}^m$ can be cross-verified as they are reported by multiple cloudlets. This mechanism provides us rules/guidelines on how the cloudlets and the mediator should communicate with each other. Secondly, the mediator addresses the \emph{preference aggregation problem} to compute the NE load balancing strategies for each cloudlet based on the communicated information, to induce the desired strategic behavior from the cloudlets \cite{Narahari}. The fundamental stages of the overall control design are summarized below:\par 
\begin{enumerate}[(a)]
\item Each cloudlet continuously observes a set of historical data samples and uses a \emph{load-predictive learning algorithm} to predict the average incoming class-$m$ job request arrival rate $\lambda_i^m$ for the interval $\tau_x$ at the interval $\tau_{x-1}$.

\item Each cloudlet also estimates the transmission latency of the incoming job requests by using the given stochastic parameters of the wireless and optical interfaces between mobile devices and cloudlets. Each cloudlet also estimates the intermediate transmission latencies with its neighbors.

\item Each cloudlet communicates their latest predictions on incoming job request arrival rate for each class and the transmission latencies to the centralized computational facility installed by the mediator.

\item Based on the communicated information, firstly, the mediator computes the processor slicing in each cloudlet by solving $\mathcal{P}_i^m$. Then employs a centralized algorithm to compute the NE computation offloading strategies and broadcasts to all the federated cloudlets before the starting of time interval $\tau_x$.

\item Accordingly, the cloudlets offload some fraction of their total incoming job requests to their neighboring cloudlets during all the timeslots of duration $D_Q$ within the time interval $\tau_x$ such that $D_{Qm}$ is satisfied for all the job classes.
\end{enumerate}

\par To ensure the truthfulness of the federated cloudlets, the mediator imposes another penalty based on the revealed $\hat{\lambda}_i^m$ after each interval $\tau_x$ if the $i^{\text{th}}$ cloudlet fails to meet $D_{Qm}$ due to overload. A proportionality cost factor $\Omega_{i,4}^m$ is multiplied with the load of class-$m$ job requests at the $i^{\text{th}}$ cloudlet and this penalty is calculated as $\sum_{m=1}^M \Omega_{i,4}^m \left\{\frac{\hat{\lambda}_i^m+\sum_{j\neq i}\hat{\varphi}_{ji}^m \hat{\lambda}_j^m}{n_i^m\mu_i^m} \right\}$. However, in practice, each cloudlet will operate at a stable load condition most of the time and will try to avoid dropping of job requests to the best extent possible. Furthermore, if the $i^{\text{th}}$ cloudlet is under-loaded, then the mediator will pass some additional job requests $\aleph_i^m$ to force its overall latency at $D_{Qm}$ and prevent under-utilization of its resources.
\vspace{-\baselineskip}
\subsection{Incentive Compatible Mechanism Design}
In this section, we show that our proposed load balancing game is incentive compatible and ensures the truthful revelation of private information from the federated cloudlets. We consider a generic multi-cloudlet ($N\geq 2$) and multi-job class ($M \geq 1$) scenario and show that truthful revelation of private information to the mediator is a weakly-dominant strategy for $i^{\text{th}}$ cloudlet irrespective of the information shared by all other $j^{\text{th}}$ cloudlets. Recall that the true type of each $i^{\text{th}}$ cloudlet for all job classes is $(\lambda_i^1,\dots,\lambda_i^M)$ and its revealed type to the mediator is $(\hat{\lambda}_i^1,\dots,\hat{\lambda}_i^M)$. With the revealed information, the mediator solves the load balancing game and broadcasts the NE load balancing strategies $\hat{\varphi}_{ij}^{m*}$ and $\hat{\varphi}_{ji}^{m*}$ to the federated cloudlets. The utility of the $i^{\text{th}}$ cloudlet based on its true type is $\mathcal{U}_i^N$ and the utility computed by each cloudlet based on their revealed type is denoted by $\hat{\mathcal{U}}_i^N$.  The incentive mechanism design is truthful if $(\hat{\mathcal{U}}_i^N - \mathcal{U}_i^N) \leq 0$ always holds true.\par
\textbf{Case-1: [All cloudlets are under-loaded]}\par
\textbf{Sub-case-1A:} \emph{A truly under-loaded cloudlet cannot improve its utility by revealing a job request arrival rate such that it is still under-loaded}. In this case, $i^{\text{th}}$ and all other $j^{\text{th}}$ cloudlets are under-loaded but, they are not aware of each other's true load conditions. Mathematically, we express this scenario from $i^{\text{th}}$ cloudlet's perspective as $(t_{ui}^m+ \mathbb{T}_i^m(\mu_i^m,\lambda_i^m)) < D_{Qm}$, $(t_{uj}^m+ \mathbb{T}_j^m(\mu_j^m,\hat{\lambda}_j^m)) < D_{Qm},\forall j\neq i \in \mathcal{C}$. Now, if the revealed type $\hat{\lambda}_i^m < \lambda_i^m$ or, $\hat{\lambda}_i^m \geq \lambda_i^m$ such that $(t_{ui}^m+\mathbb{T}_i^m(\mu_i^m,\hat{\lambda}_i^m)) < D_{Qm}$, then the NE does not change from the true NE, i.e., $\hat{\varphi}_{ij}^{m*} = 0$ and $\hat{\varphi}_{ji}^{m*} = 0$. Hence, the utility of $i^{\text{th}}$ cloudlet does not improve, because $(\hat{\mathcal{U}}_i^N - \mathcal{U}_i^N) = \sum_m \Omega_{i,1}^m (\lambda_i^m/(n_i^m\mu_i^m))- \sum_m \Omega_{i,1}^m(\lambda_i^m/(n_i^m\mu_i^m)) = 0$. This means that if the revealed $\hat{\lambda}_i^m$ is either less or greater than $\lambda_i^m$ such that the $i^{\text{th}}$ cloudlet is still under-loaded, then the NE does not change and the utility remains the same, because the revenue earned from mobile users remains unchanged.\par
\textbf{Sub-case-1B:} \emph{The utility of a truly under-loaded cloudlet decreases if it reveals itself as overloaded when neighboring cloudlets are under-loaded}.
If $\hat{\lambda}_i^m \geq \lambda_i^m$ such that $(t_{ui}^m+ \mathbb{T}_i^m(\mu_i^m,\hat{\lambda}_i^m)) \geq D_{Qm}$, then NE based on revealed this information is that the $i^{\text{th}}$ cloudlet needs to offload $\hat{\varphi}_{ij}^{m*} \hat{\lambda}_i^m$ job requests to some under-loaded $j^{\text{th}}$ cloudlets. This decreases its utility of $i^{\text{th}}$ cloudlet due to the associated payment. The NE with the revealed information is $\hat{\varphi}_{ij}^{*m} > 0$ and $\hat{\varphi}_{ji}^* = 0$. Clearly, the utility of the $i^{\text{th}}$ cloudlet decreases as follows:\par \vspace{-0.5\baselineskip}
\begin{resizealign}
(\hat{\mathcal{U}}_i^N &- \mathcal{U}_i^N) = \sum_{m=1}^M \Omega_{i,1}^m \frac{\lambda_i^m}{n_i^m\mu_i^m} - \sum_{m=1}^M\sum_{j=1,j\neq i}^N \Omega_{ij,2}^m \gamma_{ij} \frac{\hat{\varphi}_{ij}^{m*}\hat{\lambda}_i^m}{n_j^m \mu_j^m} \nonumber\\ 
&- \sum_{m=1}^M \Omega_{i,1}^m \frac{\lambda_i^m}{n_i^m \mu_i^m}
= - \sum_{m=1}^M\sum_{j=1,j\neq i}^N \Omega_{ij,2}^m \gamma_{ij}\frac{\hat{\varphi}_{ij}^{m*} \hat{\lambda}_i^m}{n_j^m \mu_j^m} <0. \label{eq10}
\end{resizealign}

\par \textbf{Case-2: [Under-loaded and overloaded cloudlets]}\par
\textbf{Sub-case-2A:} \emph{The utility of a truly overloaded cloudlet decreases or remains same if it reveals itself as more or less overloaded or under-loaded when some neighboring cloudlets are under-loaded}. \textbf{(i)} We consider $(t_{ui}^m+ \mathbb{T}_i^m(\mu_i^m,\lambda_i^m)) \geq D_{Qm}$ and $(t_{uj}^m+ \mathbb{T}_j^m(\mu_j^m,\hat{\lambda}_j^m)) < D_{Qm}$ $\forall j\in \mathcal{C}_u,m\in \mathcal{J}$ i.e., the $i^{\text{th}}$ cloudlet is overloaded and all $j^{\text{th}}$ cloudlets are under-loaded.  If $\hat{\lambda}_i^m \geq \lambda_i^m$, then $i^{\text{th}}$ cloudlet offloads more job requests if $j^{\text{th}}$ cloudlet can process the excess jobs completely, i.e., $\hat{\varphi}_{ij}^{m*}\hat{\lambda}_i^m \geq \varphi_{ij}^{m*}\lambda_i^m$. The values of $\hat{\varphi}_{ij}^{m*}, \varphi_{ij}^{m*}$ can be computed by Algorithm \ref{alg1} with $\hat{\varphi}_{ji}^{m*}=0$. Therefore, an overloaded cloudlet revealing itself as more overloaded, needs to offload more than actually needed to meet $D_{Qm}$, and hence the utility of the $i^{\text{th}}$ cloudlet decreases as: $(\hat{\mathcal{U}}_i^N - \mathcal{U}_i^N) = -\sum_m \sum_{j\neq i} (\Omega_{ij,2}^m \gamma_{ij}/(n_j^m\mu_j^m))(\hat{\varphi}_{ij}^{m*}\hat{\lambda}_i^m-\varphi_{ij}^{m*}\lambda_i^m) < 0$. However, if $\hat{\lambda}_i^m \geq \lambda_i^m$ and $\hat{\lambda}_i^m \geq \hat{\lambda}_j^m$, then the $i^{\text{th}}$ cloudlet may occasionally gain a scope to offload more workloads than its overloaded neighbors when the under-loaded neighbors accept their workload partially because more overloaded cloudlet gets to offload more job requests. Nonetheless, as the $i^{\text{th}}$ cloudlet still remains overloaded, it needs to pay more penalty because the revealed $\hat{\lambda}_i^m \geq \lambda_i^m$ and its utility decreases as follows:\par
\begin{resizealign} 
&(\hat{\mathcal{U}}_i^N -\mathcal{U}_i^N) = -\sum_{m=1}^M\sum_{j=1,j\neq i}^N \Omega_{ij,2}^m \frac{\gamma_{ij}}{n_j^m\mu_j^m}(\hat{\varphi}_{ij}^{m*} \hat{\lambda}_i^m - \varphi_{ij}^{m*} \lambda_i^m)\nonumber\\
&- \sum_{m=1}^M \left(\frac{\lambda_i^m- \sum_{j\neq i}\hat{\varphi}_{ij}^{m*} \hat{\lambda}_i^m}{n_i^m\mu_i^m}\right) \left\{\Omega_{i,3}^m \left[t_{ui}^m+ \mathbb{T}_i^m(\mu_i^m,(\lambda_i^m- \sum_{j\neq i}\hat{\varphi}_{ij}^{m*}\hat{\lambda}_i^m))-D_{Qm} \right]\right\}\nonumber\\
&+\sum_{m=1}^M \left(\frac{\lambda_i^m- \sum_{j\neq i}\varphi_{ij}^{m*} \lambda_i^m}{n_i^m\mu_i^m}\right) \left\{\Omega_{i,3}^m \left[t_{ui}^m+ \mathbb{T}_i^m(\mu_i^m,(\lambda_i^m- \sum_{j\neq i}\varphi_{ij}^{m*}\lambda_i^m))-D_{Qm} \right]\right\} \nonumber\\
&- \sum_{m=1}^M \Omega_{i,4}^m \left(\frac{\hat{\lambda}_i^m}{n_i^m\mu_i^m}\right) + \sum_{m=1}^M \Omega_{i,4}^m \left(\frac{\lambda_i^m}{n_i^m\mu_i^m}\right) \leq 0. \label{eq11}
\end{resizealign}
\par We shall soon derive a necessary condition between $\Omega_{ij,2}^m$, $\Omega_{i,3}^m$ and $\Omega_{i,4}^m$ that ensured that $(\hat{\mathcal{U}}_i^N - \mathcal{U}_i^N) \leq 0$ for expressions like (\ref{eq11}). \textbf{(ii)} If $\hat{\lambda}_i^m < \lambda_i^m$ but $(t_{ui}^m+ \mathbb{T}_i^m(\mu_i^m,\hat{\lambda}_i^m)) \geq D_{Qm}$, then $i^{\text{th}}$ cloudlet offloads less than that required to meet the QoS target latency $D_{Qm}$, i.e., $\hat{\varphi}_{ij}^{m*}\hat{\lambda}_i^m \leq \varphi_{ij}^{m*} \lambda_i^m$ and $\hat{\varphi}_{ij}^{m*}, \varphi_{ij}^{m*}$ can be computed by Algorithm \ref{alg1} with $\hat{\varphi}_{ji}^{m*}=0$. This implies that $i^{\text{th}}$ cloudlet needs to pay lesser incentives with the revealed $\hat\lambda_i^m$ but, due to the partial job request offloading, the penalty for latency decreases its utility value as follows:\par
\begin{resizealign} 
&(\hat{\mathcal{U}}_i^N -\mathcal{U}_i^N) = -\sum_{m=1}^M\sum_{j=1,j\neq i}^N \Omega_{ij,2}^m \frac{\gamma_{ij}}{n_j^m\mu_j^m}(\hat{\varphi}_{ij}^{m*} \hat{\lambda}_i^m - \varphi_{ij}^{m*} \lambda_i^m) - \sum_{m=1}^M \Omega_{i,4}^m \left(\frac{\hat{\lambda}_i^m}{n_i^m\mu_i^m}\right)\nonumber\\
& - \sum_{m=1}^M \left(\frac{\lambda_i^m- \sum_{j\neq i}\hat{\varphi}_{ij}^{m*} \hat{\lambda}_i^m}{n_i^m\mu_i^m}\right) \left\{\Omega_{i,3}^m \left[t_{ui}^m+ \mathbb{T}_i^m(\mu_i^m,(\lambda_i^m- \sum_{j\neq i}\hat{\varphi}_{ij}^{m*}\hat{\lambda}_i^m))-D_{Qm} \right]\right\} \leq 0. \label{eq12}
\end{resizealign} 
\par \textbf{(iii)} Again, if $\hat{\lambda}_i^m < \lambda_i^m$ such that $(t_{ui}^m+ \mathbb{T}_i^m(\mu_i^m,\hat{\lambda}_i^m)) < D_{Qm}$, then $i^{\text{th}}$ cloudlet is treated as under-loaded and hence, it is not allowed to offload any job requests but receives some job requests from some overloaded neighbors and the mediator. Therefore, the NE solution with the revealed information is $\hat{\varphi}_{ij}^{m*} = 0$, $\hat{\varphi}_{ji}^{m*} > 0$ for some $j\neq i \in \mathcal{C}$ and the utility of $i^{\text{th}}$ cloudlet decreases due to the latency penalty:\par
\begin{resizealign}
&(\hat{\mathcal{U}}_i^N - \mathcal{U}_i^N) = \sum_{m=1}^M \sum_{j=1,j\neq i}^N \Omega_{ji,2}^m \gamma_{ji} \frac{\hat{\varphi}_{ji}^{m*}\hat{\lambda}_j^m}{n_i^m\mu_i^m} \nonumber\\
&+ \sum_{m=1}^M \sum_{j=1,j\neq i}^N \Omega_{ij,2}^m \gamma_{ij}  \frac{\varphi_{ij}^{m*} \lambda_i^m}{n_j^m\mu_j^m} - \sum_{m=1}^M \Omega_{i,4}^m \left\{\frac{\hat{\lambda}_i^m+\sum_{j\neq i}\hat{\varphi}_{ji}^{m*} \hat{\lambda}_j^m+\aleph_i^m}{n_i^m\mu_i^m} \right\} \nonumber\\
& - \sum_{m=1}^M \Omega_{i,3}^m \left[\frac{\lambda_i^m}{n_i^m\mu_i^m}\left(t_{ui}^m+ \mathbb{T}_i^m(\mu_i^m,\lambda_i^m+\sum_{j\neq i} \hat{\varphi}_{ji}^{m*} \hat{\lambda}_j^m+\aleph_i^m) -D_{Qm} \right) \right.\nonumber\\
& \left. +\sum_{j=1,j\neq i}^{N}\frac{\hat{\varphi}_{ji}^{m*} \hat{\lambda}_j^m}{n_i^m\mu_i^m} \left(t_{uj}^m+\mathbb{T}_i^m(\mu_i^m,\lambda_i^m+\sum_{j\neq i} \hat{\varphi}_{ji}^{m*} \hat{\lambda}_j^m+\aleph_i^m)+t_{ji}^m-D_{Qm} \right)\right] \leq 0. \label{eq13}
\end{resizealign} 
\par \textbf{Sub-case-2B:} \emph{The utility of a truly under-loaded cloudlet remains unchanged or decreases by revealing itself as more or less under-loaded when some of the neighboring cloudlets are overloaded}. \textbf{(i)} We consider $(t_{ui}^m+ \mathbb{T}_i^m(\mu_i^m,\lambda_i^m)) < D_{Qm}$, $(t_{uj}^m+ \mathbb{T}_j^m(\mu_j^m,\hat{\lambda}_j^m)) \geq D_{Qm}$, i.e., the $i^{\text{th}}$ cloudlet is under-loaded and some $j^{\text{th}}$ cloudlets are overloaded. In this case, if $\hat{\lambda}_i^m < \lambda_i^m$ or, $\hat{\lambda}_i^m \geq \lambda_i^m$ such that $(t_{ui}^m+ \mathbb{T}_i^m(\mu_i^m,\hat{\lambda}_i^m)) < D_{Qm}$ and can process the entire job request offload requests $\hat{\varphi}_{ji}^{m*} \hat{\lambda}_j^m$ from the $j^{\text{th}}$ cloudlet, then the NE solution $\hat{\varphi}_{ij}^{m*} = 0$ and $ \hat{\varphi}_{ji}^{m*} > 0$ does not change. Thus, the utility of the $i^{\text{th}}$ cloudlet remains unchanged because the same amount of job requests are offloaded irrespective of the falsely revealed information. However, if $\hat{\lambda}_i^m < \lambda_i^m$ and $\hat{\lambda}_i^m < \hat{\lambda}_j^m$, then the $i^{\text{th}}$ cloudlet may occasionally gain more workload and incentives than its under-loaded neighbors. However, it will fail to meet $D_{Qm}$ as the mediator has also offloaded some jobs at a rate of $\aleph_i^m$ to force its overall latency at $D_{Qm}$. Therefore, the mediator will be aware of the falsely revealed information and will apply the extra penalties, which will decrease the utility as follows:\par
\begin{resizealign}
&(\hat{\mathcal{U}}_i^N - \mathcal{U}_i^N) = \sum_{m=1}^M \sum_{j=1,j\neq i}^N \Omega_{ji,2}^m \frac{\gamma_{ji}}{n_i^m\mu_i^m}\left(\hat{\varphi}_{ji}^{m*}\hat{\lambda}_j^m-\varphi_{ji}^{m*}\lambda_j^m\right) \nonumber\\
& - \sum_{m=1}^M \Omega_{i,4}^m \left\{\frac{\hat{\lambda}_i^m+\sum_{j\neq i}\hat{\varphi}_{ji}^{m*} \hat{\lambda}_j^m+\aleph_i^m}{n_i^m\mu_i^m} \right\} \nonumber\\
& - \sum_{m=1}^M \Omega_{i,3}^m \left[\frac{\lambda_i^m}{n_i^m\mu_i^m}\left(t_{ui}^m+ \mathbb{T}_i^m(\mu_i^m,\lambda_i^m+\sum_{j\neq i} \hat{\varphi}_{ji}^{m*} \hat{\lambda}_j^m+\aleph_i^m) -D_{Qm} \right) \right.\nonumber\\
& \left. +\sum_{j=1,j\neq i}^{N}\frac{\varphi_{ji}^{m*}\lambda_j^m}{n_i^m\mu_i} \left(t_{uj}^m+\mathbb{T}_i^m(\mu_i^m,\lambda_i^m+\sum_{j\neq i} \hat{\varphi}_{ji}^{m*} \hat{\lambda}_j^m+\aleph_i^m)+t_{ji}^m-D_{Qm} \right)\right] \leq 0. \label{eq14}
\end{resizealign} 
\par \textbf{(ii)} Again, if $\hat{\lambda}_i^m \geq \lambda_i^m$ such that $(t_{ui}^m+ \mathbb{T}_i^m(\mu_i^m,\hat{\lambda}_i^m)) < D_{Qm}$ but cannot process the entire job request offload requests $\hat{\varphi}_{ji}^{m*} \hat{\lambda}_j^m$ from some of the overloaded $j^{\text{th}}$ cloudlets, then lesser amount of job requests are offloaded, i.e., $\hat{\varphi}_{ji}^{m*} \hat{\lambda}_j^m \leq \varphi_{ji}^{m*} \hat{\lambda}_j^m$ and $\hat{\varphi}_{ji}^{m*}, \varphi_{ji}^{m*}$ can be computed by Algorithm \ref{alg1} with $\hat{\varphi}_{ij}^{m*} = 0$. Thus, the utility of the $i^{\text{th}}$ cloudlet decreases because $(\hat{\mathcal{U}}_i^N - \mathcal{U}_i^N) = -\sum_m\sum_{j\neq i} (\Omega_{ji,2}^m \gamma_{ji}/(n_i^m\mu_i^m))(\varphi_{ji}^{m*} \hat{\lambda}_j^m - \hat{\varphi}_{ji}^{m*} \hat{\lambda}_j^m) < 0$.\par
\textbf{(iii)} Finally, if $\hat{\lambda}_i^m < \lambda_i^m$ but the $i^{\text{th}}$ cloudlet actually cannot process the entire extra load offloaded by the overloaded $j^{\text{th}}$ cloudlets, then more jobs are offloaded to the $i^{\text{th}}$ cloudlet but it fails to meet the QoS latency target. Therefore, $\hat{\varphi}_{ji}^{m*}\hat{\lambda}_j^m \geq \varphi_{ji}^{m*}\hat{\lambda}_j^m$ and $\hat{\varphi}_{ji}^{m*}, \varphi_{ji}^{m*}$ can be computed by Algorithm \ref{alg1} with $\hat{\varphi}_{ij}^{m*} = 0$. However, the latency penalty decreases the utility in spite of getting some extra incentives as shown in (\ref{eq14}).\par
\par \textbf{Case-3: [All cloudlets are overloaded]}\par
\textbf{Sub-case-3A:} \emph{A truly overloaded cloudlet cannot improve its utility by revealing a job request arrival rate such that it is still overloaded}. In this case, $(t_{ui}^m+ \mathbb{T}_i^m(\mu_i^m,\lambda_i^m)) \geq D_{Qm}$, $(t_{uj}^m+ \mathbb{T}_j^m(\mu_j^m,\hat{\lambda}_j^m) \geq D_{Qm}$. Now, $\hat{\lambda}_i^m \geq \lambda_i^m$ or, $\hat{\lambda}_i^m < \lambda_i^m$ such that $(t_{ui}^m+ \mathbb{T}_i^m(\mu_i^m,\lambda_i^m)) \geq D_{Qm}$, then the NE is same as the true NE, i.e., $\hat{\varphi}_{ij}^{m*} = 0$ and $\hat{\varphi}_{ji}^{m*} = 0$. Hence, the utility of $i^{\text{th}}$ cloudlet does not improve and $(\hat{\mathcal{U}}_i^N - \mathcal{U}_i^N)=0$.\par
%
\textbf{Sub-case-3B:} \emph{The utility of a truly overloaded cloudlet decreases if it reveals itself as under-loaded when all the neighboring cloudlets are overloaded}.
If $\hat{\lambda}_i^m < \lambda_i^m$ such that $(t_{ui}^m+ \mathbb{T}_i^m(\mu_i^m,\hat{\lambda}_i^m)) < D_{Qm}$, then the $i^{\text{th}}$ cloudlet is considered as under-loaded and overloaded $j^{\text{th}}$ cloudlets offload some job requests to it. The NE based on the revealed information is $\hat{\varphi}_{ij}^{m*} = 0$ and $\hat{\varphi}_{ji}^{m*} > 0$. This helps the $i^{\text{th}}$ cloudlet to earn some extra incentives but the latency penalty decreases the utility:\par
\begin{resizealign}
&(\hat{\mathcal{U}}_i^N - \mathcal{U}_i^N) = \sum_{j=1,j\neq i}^N \Omega_{ji,2}^m \gamma_{ji} \frac{\hat{\varphi}_{ji}^{m*} \hat{\lambda}_j^m}{n_i^m\mu_i^m}- \sum_{m=1}^M \Omega_{i,4}^m \left\{\frac{\hat{\lambda}_i^m+\sum_{j\neq i}\hat{\varphi}_{ji}^{m*} \hat{\lambda}_j^m+\aleph_i^m}{n_i^m\mu_i^m} \right\}\nonumber\\
& + \sum_{m=1}^M \Omega_{i,3}^m \frac{\lambda_i^m}{n_i^m\mu_i}\left[t_{ui}^m+ \mathbb{T}_i^m(\mu_i^m,\lambda_i^m)-D_{Qm} \right]\nonumber\\
& - \sum_{m=1}^M \Omega_{i,3}^m \frac{\lambda_i^m}{n_i^m\mu_i^m}\left[t_{ui}^m+ \mathbb{T}_i^m(\mu_i^m,\lambda_i^m+\sum_{j\neq i} \hat{\varphi}_{ji}^{m*} \hat{\lambda}_j^m+\aleph_i^m)-D_{Qm} \right]\nonumber\\
& - \sum_{m=1}^M \sum_{j=1,j\neq i}^N \Omega_{i,3} \frac{\hat{\varphi}_{ji}^{m*}\hat{\lambda}_j^m}{n_i^m\mu_i^m}\left[t_{uj}^m+ \mathbb{T}_i^m(\mu_i^m,\lambda_i^m+\sum_{j\neq i} \hat{\varphi}_{ji}^{m*} \hat{\lambda}_j^m+\aleph_i^m)+t_{ji}^m-D_{Qm} \right] \leq 0. \label{eq15}
\end{resizealign} 
\par Clearly, the above analysis shows that we ensured that any cloudlet can not achieve a better utility by revealing false job request arrival rate and the truthful revelation of all the cloudlets is a weakly-dominant strategy, irrespective of the information shared by the neighboring cloudlets, when the following proposition holds true.
\begin{proposition}
To implement the proposed incentive mechanism among federated cloudlets, $\Omega_{ij,2}^m \geq \Omega_{i,3}^m [t_{ui}^m+\mathbb{T}_i^m(\mu_i^m,\lambda_i^{m,max}) ]+ \Omega_{i,4}^m$ and $n_i\Omega_{ij,2}^m \leq (\Omega_{i,3}^m+\Omega_{i,4}^m), \forall i,j \in\mathcal{C},m \in \mathcal{J}$ are the necessary conditions.
\end{proposition} 
\par Please refer to Appendix E for the proof. It is very interesting to observe that each cloudlet may reveal slightly erroneous job request arrival rates due to prediction error. This error has no impact on system performance for Sub-case-1A and Sub-case-3A as the utilities do not change. Nonetheless, there is slight decrease of utilities in Sub-case-1B, Sub-case-2B(ii) (linear decrease) and Sub-case-2A(i)/(ii)/(iii), Sub-case-2B(i)/(iii), Sub-case-3B (just at the verge of $D_{Qm}$ violation). However, as the state-of-the-art traffic prediction algorithms can achieve reasonably high accuracy, this decrease of utilities of cloudlets is expected to be very minor in practice.

\section{Decentralized Load Balancing Framework}\label{sec6}
In a distributed cloudlet network, the incoming class-$m$ job request arrival rate $\lambda_i^m$ to each cloudlet $i\in\mathcal{C}$ is known only to that cloudlet and no other entity in the network. We still assume that the federated cloudlets are \emph{non-cooperative} and \emph{rational utility maximizers} and hence, they do not share this \emph{private information} with each other.  Therefore, a reinforcement learning automata-based algorithm helps the cloudlets to independently make load balancing decisions only from their private information and to aid this decision-making process, we use the economic and non-cooperative game formulated in Section \ref{sec4}. Moreover, some particular characteristics of the underlying game formulation is used to reduce the search space of the reinforcement learning algorithm and greatly improve the convergence rate of the algorithm. The fundamental stages of the overall control design are summarized below:\par
\begin{enumerate}[(a)]
\item A \emph{load-predictive learning algorithm} is executed by each cloudlet just before every time interval $\tau_{x-1}$ by using historical data to predict the incoming job request arrival rate of the next time interval $\tau_x$. Each cloudlet uses their predicted job request arrival rates to perform processor slicing for the time interval $\tau_x$.

\item Each cloudlet also estimates the transmission latency of the incoming job requests by using the given stochastic parameters of the access network between mobile devices and cloudlets. Each cloudlet also estimates the intermediate transmission latencies with its neighboring cloudlets.

\item Using this learned information, each cloudlet shares a random amount of its job requests to the neighboring cloudlets, depending on the latest probability distribution over its strategy space and observes the utility and rewards received.  Based on the reward values received at $n^{\text{th}}$ and $(n-1)^{\text{th}}$ time-slots, each cloudlet updates the probability distribution over its strategy space for $(n+1)^{\text{th}}$ time-slot.
\end{enumerate}

\vspace{-\baselineskip}
\subsection{Distributed Reinforcement Learning Algorithm}
In this subsection, we design a continuous-action reinforcement learning automata-based algorithm for learning the NE of the continuous-kernel non-cooperative load balancing game formulated in Section \ref{sec4}.  At first, we define the \emph{mixed-strategy} of each $i^{\text{th}}$ cloudlet as continuous probability density function (PDF) $\bm{f}_i(\bm{\varphi}_i)$ over its pure-strategy space $\Phi_i$. Therefore, the probability of randomly choosing an action within a close neighborhood of $\varphi_{ij}^m$ by $i^{\text{th}}$ cloudlet can be determined from the corresponding PDF $f_{ij}^m(\varphi_{ij}^m)$. The complete mixed-strategy of all the cloudlets is defined as $\mathcal{F} := \bm{f}_1\times ... \times\bm{f}_N$ over the complete pure-strategy space $\Phi$. When each $i^{\text{th}}$ cloudlet chooses an action $\varphi_{ij}^m$, i.e., offloads $\varphi_{ij}^m\lambda_i^m$ job requests to neighboring $j^{\text{th}}$ cloudlets, then the \emph{environment responds with a random reward} $\mathcal{R}_i (\bm{\varphi}_i,\bm{\varphi}_{-i}) \in [0,1]$, which is defined as:
\begin{align} 
\mathcal{R}_i (\bm{\varphi}_i,\bm{\varphi}_{-i}) = \frac{\mathcal{U}^{N}_i(\bm{\lambda}_i,\bm{\varphi}_i,\bm{\varphi}_{-i})}{\max\{\Omega_{i,1}^m,\Omega_{ij,2}^m,\Omega_{i,3}^m\}},\forall i\in \mathcal{C},m\in\mathcal{J}. \label{eq16}
\end{align} 
\par In the load balancing game, with a continuous-action reinforcement learning automata-based algorithm, each $i^{\text{th}}$ cloudlet starts with \emph{uniform probability distributions} as their mixed-strategies over their individual pure-strategy action spaces and keeps on updating the PDFs based on the received rewards in the following time-slots to ultimately \emph{find their pure-strategy NE} \cite{carla}.  After exploring an action $\bm{\varphi}_{i}^{(n)} \in \Phi_i$ during $n^{\text{th}}$ time-slot, the PDFs are updated for $(n+1)^{\text{th}}$ time-slot by the following update rule:\par
\begin{resizealign}
\bm{f}_i^{(n+1)}&(\bm{\varphi}_i) = \chi^{(n)} \left[\bm{f}_i^{(n)}(\bm{\varphi}_i) +\Theta^{(n)} \left(\mathcal{R}_i^{(n)}-\mathcal{R}_i^{(n-1)}\right) \exp\left\{-\frac{1}{2}\left(\frac{\bm{\varphi}_i -\bm{\varphi}_i^{(n)}}{\sigma^{(n)}}\right )^2\right\} \right], \label{eq17}
\end{resizealign} 
\par where, $\Theta$ is the \emph{learning rate parameter}, $\sigma$ is the \emph{spreading rate parameter} and $\chi$ is a \emph{normalization factor} so that $\int_{-\infty}^{+\infty} f^{(n+1)} dz = 1$, for any $z$.  Note that, our proposed reinforcement learning automaton (\ref{eq17}) operates as \emph{gradient bandit} algorithm, based on the idea of \emph{stochastic gradient ascent} algorithms \cite{Sutton}.  Moreover, the term $(\mathcal{R}_i^{(n)}-\mathcal{R}_i^{(n-1)})$ used in this model makes this algorithm \emph{highly robust in tracking a non-stationary job request arrival process}. The PDFs are continuously updated by the cloudlets based on their private information and rewards received at every time-slot to learn the pure-strategy NE of the non-cooperative load balancing game and the space complexity of this algorithm is $\mathcal{O}((N\times M)^2\times L)$, where $L$ is the length of memory required for storing the discrete version of $f_{ij}^m(\varphi_{ij}^m)$ \cite{Phansalkar1994}. \par

\begin{algorithm}[t!]
\caption{Distributed Reinforcement Learning Automata-based Algorithm for learning NE of the Load Balancing Game}\label{alg2}
\begin{algorithmic}[1]
\State \textbf{Initialization:} Set the iteration index $n=0$ and the PDFs $f_{ij}^{m,(n)}(\varphi_{ij}^m) = \text{uniform}(\Phi_{ij}^m), \forall i,j \in \mathcal{C},m\in\mathcal{J}$.
\State \textbf{Output:} The pure-strategy NE of the non-cooperative load balancing game.
\If {$i^{\text{th}}$ cloudlet is under-loaded, i.e., $(t_{ui}^m+ \mathbb{T}_i^m(\mu_i^m,\lambda_i^m)) < D_{Qm}$}
choose not to offload, i.e., $\bm{\varphi}^{m,(n)}_i = \bm{0}$;
\Else { $i^{\text{th}}$ cloudlet randomly choose an action $\bm{\varphi}_i^{(n)}$ based on its latest mixed-strategy $\bm{f}_i^{(n)}(\bm{\varphi}_i)$.}
\EndIf
\If {$j^{\text{th}}$ cloudlet is overloaded or partially processes the jobs received from all $i^{\text{th}}$ cloudlets} 
indicate all $i^{\text{th}}$ cloudlets that only $\hat{\psi}_{ij}^m\lambda_i^m$ jobs are processed ($0\leq \hat{\psi}_{ij}^m < \varphi_{ij}^m$), where $\hat{\psi}_{ij}^m\lambda_i^m$ are chosen according to the ratio of $\varphi_{ij}^m\lambda_i^m$.
\EndIf
\State At the end of $n^{\text{th}}$ time-slot, each $i^{\text{th}}$ cloudlet receives a reward $\mathcal{R}_i^{(n)}$ from the environment.
\State Each $i^{\text{th}}$ cloudlet update their mixed-strategy $\bm{f}_i^{(n+1)}(\bm{\varphi}_i)$ by the reinforcement learning automaton for $\bm{\varphi}_i^{(n)} \in \Phi_i$:\par
\begin{resizealign} 
\bm{f}_i^{(n+1)}&(\bm{\varphi}_i) = \chi^{(n)} \left[\bm{f}_i^{(n)}(\bm{\varphi}_i) +\Theta^{(n)} \left(\mathcal{R}_i^{(n)}-\mathcal{R}_i^{(n-1)}\right) \exp\left\{-\frac{1}{2}\left(\frac{\bm{\varphi}_i-\bm{\varphi}_i^{(n)}}{\sigma^{(n)}}\right )^2\right\} \right] \nonumber
\end{resizealign} 
\State Set $n \gets n+1$; go to Step 3.
\end{algorithmic}
\end{algorithm}
\begin{theorem}\label{th3} 
The continuous-action reinforcement learning automata-based algorithm with update rule (\ref{eq17}) converges to the pure-strategy Nash equilibrium of the non-cooperative load balancing game.
\end{theorem}
\par Please refer to Appendix F for the proof. Although, the convergence of the proposed algorithm is guaranteed but, we can \emph{speed up the convergence rate of the algorithm} several times more by scaffolding our understanding about the underlying load balancing game. We observe that whenever some cloudlet is in under-loaded condition, i.e., $(t_{ui}^m+ \mathbb{T}_i^m(\mu_i^m,\lambda_i^m)) < D_{Qm}$, its NE strategy is not to offload any job requests to its neighboring cloudlets. Thus, all cloudlets can update their PDFs accordingly without exploring many job request offloading strategies as long as the under-load condition persists. In addition to this, during overload condition, i.e., $(t_{ui}^m+ \mathbb{T}_i^m(\mu_i^m,\lambda_i^m)) \geq D_{Qm}$, each cloudlet will offload only a portion of the received job requests and try to shift the peak of the PDF $f_{ij}^m(\varphi_{ij}^m)$ around the pure strategy NE solution $\varphi^{m*}_{ij}$, such that it can meet $D_{Qm}$ with the rest of the job requests by itself. Hence, the corresponding search spaces can be reduced accordingly. However, as each cloudlet is unaware of the load condition of its neighboring cloudlets, an occasional feedback mechanism is required from the neighboring cloudlets when the offloaded jobs are partially processed, i.e., $\hat{\psi}_{ij}^m\lambda_i^m$. This implies that when $i^{\text{th}}$ cloudlet offloads $\varphi_{ij}^m\lambda_i^m$ job requests but the $j^{\text{th}}$ cloudlet sends feedback that only $\hat{\psi}_{ij}^m\lambda_i^m$ jobs are processed, where $0\leq \hat{\psi}_{ij}^m < \varphi_{ij}^m$. Therefore, in such cases, the $i^{\text{th}}$ cloudlet updates the PDF by using $\hat{\psi}_{ij}^m$ in (\ref{eq17}) instead of $\varphi_{ij}^m$. Furthermore, when $i^{\text{th}}$ cloudlet is under-loaded and receives job requests from multiple cloudlets but can partially process the job requests i.e., $(t_{ui}^m+ \mathbb{T}_i^m(\mu_i^m,\lambda_i^m)) < D_{Qm}$ but $(t_{uj}^m+ \mathbb{T}_i^m(\mu_i^m,(\lambda_i^m+\sum_{i\neq j}\varphi_{ji}^m\lambda_j^m))+t_{ji}^m) \geq D_{Qm}$ the residual job processing capacity of the $i^{\text{th}}$ cloudlet is distributed according to the ratio of $\varphi_{ji}^m\lambda_j^m$. The proposed algorithm is summarized in Algorithm \ref{alg2}.

\section{Results and Discussions}\label{sec7}
In this section, we investigate various behavioural aspects of the proposed load balancing strategy through numerical evaluations. For this purpose, we consider a set of 10 neighboring federated cloudlets. At first, we consider a single job class-$m$ to compare system performance with a few existing frameworks and the average processing rate of each of the cloudlets with multiple processors as $n_i^m\mu_i^m = 1000$ jobs/s and incoming job request rates to each cloudlet $\lambda_i^m$ varies within 0-1500 jobs/s. We consider the duration of each timeslot as $D_Q= 5$ msec, the average value of latency between mobile users and cloudlets $t_{ui}$ as 2 msec, the intermediate transmission latency between neighboring cloudlets $t_{ij}$ varies within 0.5-1 msec, the number of bits/job request $b_{ij}^m$ varies within 100-200 KB \cite{SouravEls}, and the available bandwidth $B_{ij}$ lies within 0.5-1 Gbps \cite{Imali}.  The optimal values of proportionality price factors $\Omega_{i,1}^m$, $\Omega_{ij,2}^m$, $\Omega_{i,3}^m$, and $\Omega_{i,4}^m$ can be determined by studying the market equilibrium conditions for providing cloud-based services \cite{net_price}. In actual practice, sometimes the proper price factors are also determined by applying the \emph{multiple criteria decision-making theory} \cite{decision}.  However, in this work we arbitrarily choose $\Omega_{i,1}^m= 5\times 10^3$, $\Omega_{ij,2}^m = 3\times 10^4$, $\Omega_{i,3}^m = 9\times 10^4$, and $\Omega_{i,4}^m = 6\times 10^3$ cost/unit load such that our necessary game design conditions $\Omega_{ij,2}^m \geq \Omega_{i,3}^m [t_{ui}^m+\mathbb{T}_i^m(\mu_i^m,\lambda_i^{m,max})]+ \Omega_{i,4}^m$ and $n_i \Omega_{ij,2}^m \leq (\Omega_{i,3}^m+\Omega_{i,4}^m), \forall i,j \in\mathcal{C},m \in \mathcal{J}$ are satisfied. Moreover, for our gradient projection algorithm to compute NE of the load balancing game, we choose a step size $\omega =0.1$, and a tolerance limit $\epsilon = 10^{-4}$.\par
\begin{figure}[t!]
  \centering

  \begin{subfigure}[t]{\columnwidth}
  \centering
  \includegraphics[width=\textwidth,height=4.5cm]{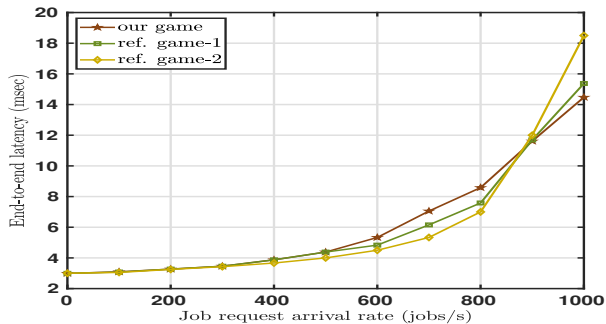}       
  \caption{Average end-to-end latency of cloudlets with $n_i^m\mu_i^m = 1000$ jobs/s.}
  \label{latency1}
  \end{subfigure}

  \begin{subfigure}[t]{\columnwidth}
  \centering
  \includegraphics[width=\textwidth,height=4.5cm]{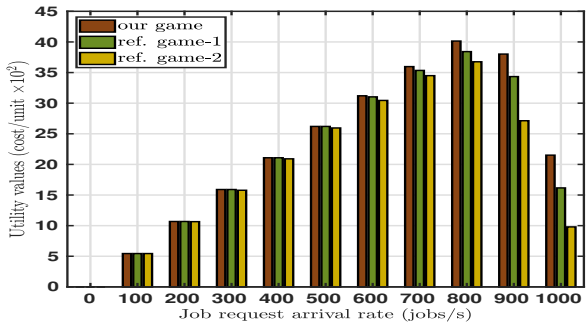}
  \caption{Average utility values of cloudlets with $n_i^m\mu_i^m = 1000$ jobs/s.}
  \label{utility1}
  \end{subfigure}

  \caption{Comparison of performance between our proposed game and other games proposed in \cite{TONload,Rev13} in terms of end-to-end latency and utility values with high difference ([0-400] jobs/s) among incoming job request arrival rates of neighboring cloudlets.}
\end{figure}
In Fig. \ref{latency1}, we compare average end-to-end latency of all the participating cloudlets against job request arrival rate with our currently proposed game and games proposed in \cite{TONload} (labeled as ``ref. game-1") and \cite{Rev13} (labeled as ``ref. game-2"), respectively. To make a fair comparison, we consider only a single class of job requests (as \cite{TONload} and \cite{Rev13} deal with a single job class), a \emph{high difference in job request arrival rates among under and overloaded cloudlets} (within 0-400 jobs/s), and the service rates of all the cloudlets are $n_i^m\mu_i^m = 1000$ jobs/s. As per the characteristics of our proposed game explained in Appendix D, we see that the ref. game-1 performs best when the load condition is low or moderate as it always tries to minimize end-to-end latency. Under these conditions, our proposed game as well as ref. game-2 performs slightly poorer as these models do not require the cloudlets to offload anything. However, ref. game-2 allows the cloudlets to offload job requests after reaching a certain threshold in incoming job requests and their latency performance begins to improve.\par
Nonetheless, under high load conditions, when all the cloudlets become sufficiently overloaded, our game also allows the cloudlets to strategically offload some job requests and latency performance becomes relatively better than ref. game-1 and ref. game-2. This is because it is ensured in our game that all the under-loaded cloudlets meet the QoS latency target $D_{Q1} = 10$ msec. The over-loaded cloudlets may exceed $D_{Q1}$, but they are allowed to offload job requests to the maximum extent possible. On the contrary, ref. game-1 becomes infeasible in high load conditions as some of the cloudlets start to violate explicit latency constraints and ref. game-2 appears to overload the under-loaded cloudlets by uncontrolled offloading of the job requests. Next, Fig. \ref{utility1} shows a comparison among average utility values of all the participating cloudlets against job request arrival rate with our game, ref. game-1, and ref. game-2. It is clear that with our game, the average economic utility values of the cloudlets are relatively better than both ref. game-1 and ref. game-2 under all the network scenario.\par
\begin{figure}[t!]
  \centering

  \begin{subfigure}[t]{\columnwidth}
  \centering
  \includegraphics[width=\textwidth,height=4.5cm]{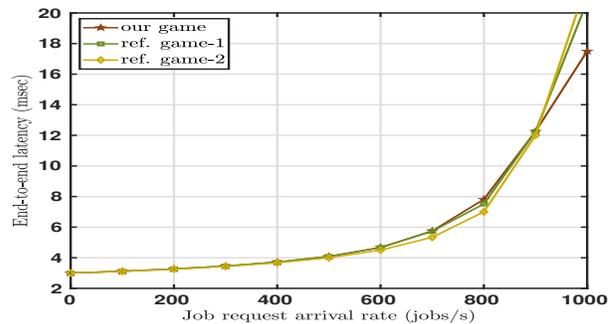}       
  \caption{Average end-to-end latency of cloudlets with $n_i^m\mu_i^m = 1000$ jobs/s.}
  \label{latency2}
  \end{subfigure}

  \begin{subfigure}[t]{\columnwidth}
  \centering
  \includegraphics[width=\textwidth,height=4.5cm]{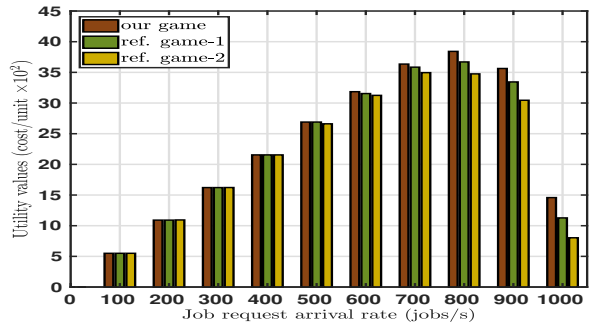}
  \caption{Average utility values of cloudlets with $n_i^m\mu_i^m = 1000$ jobs/s.}
  \label{utility2}
  \end{subfigure}

  \caption{Comparison of performance between our proposed game and other games proposed in \cite{TONload,Rev13} in terms of end-to-end latency and utility values with moderate difference ([0-200] jobs/s) among incoming job request arrival rates of neighboring cloudlets.}
\end{figure}
Similarly, Fig. \ref{latency2} shows a comparison of average end-to-end latency performance and Fig. \ref{utility2} shows a comparison among average utility values of all the participating cloudlets against job request arrival rate with our game, ref. game-1, and ref. game-2.  In this case, we consider a \emph{moderate difference in job request arrival rates} (within 0-200 jobs/s) among under-loaded and overloaded cloudlets and the service rates of all the cloudlets are $n_i^m\mu_i^m = 1000$ jobs/s.  Note that both the plots show similar behavior as in Fig. \ref{latency1} and Fig. \ref{utility1}.  However, since the difference between under-loaded and overloaded cloudlets in job request arrival rates is lower than the previous case, the scope for overloaded cloudlets to offload job requests is also lower.  Hence, the average end-to-end latency overshoots to a relatively higher value in overload conditions and the average utility gained are also lower.\par
\begin{figure*}[!t] 
\centering 
\includegraphics[width=\textwidth,height=4.5cm]{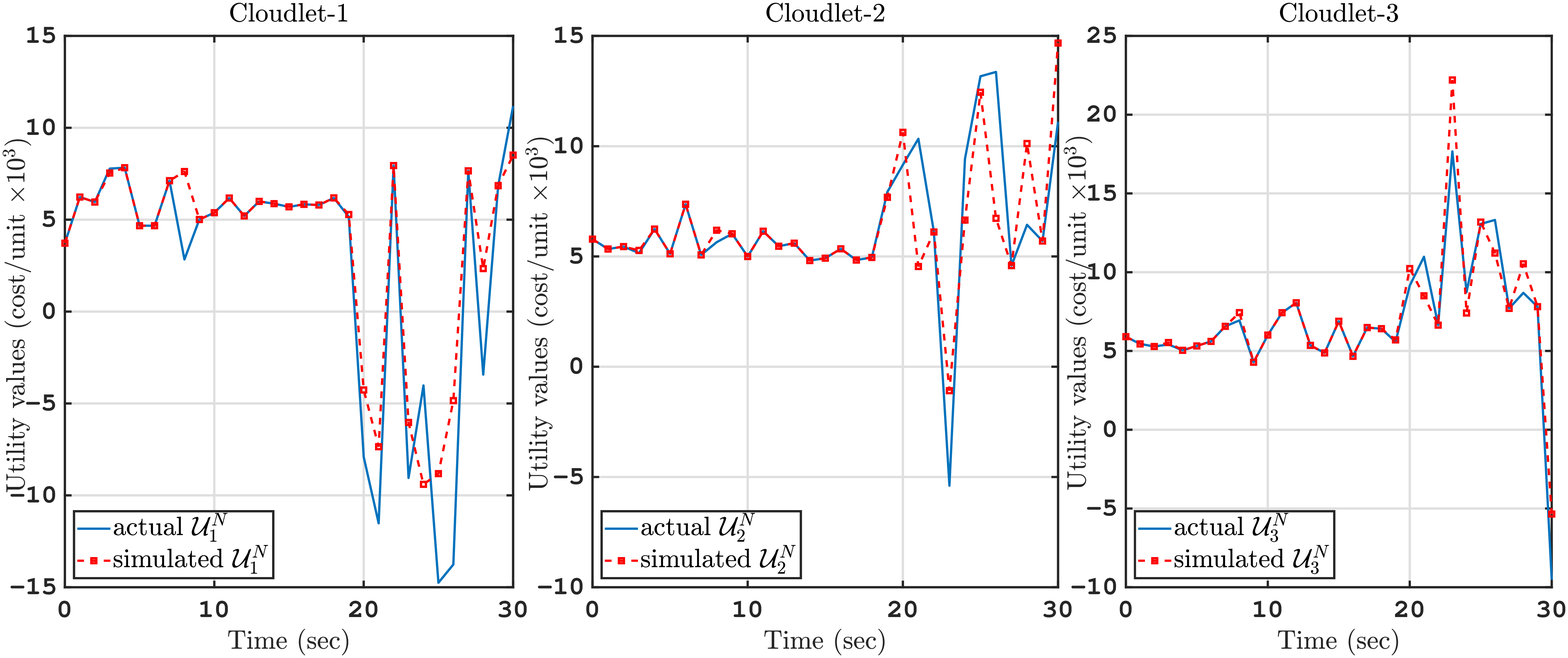}
\caption{Theoretical and simulated NE utility values ($10\%$ mean error) of three cloudlets with $D_{Q1} = 10$ msec and $D_{Q2} = 20$ msec.} 
\label{predict_compare}
\end{figure*} 
\setlength{\textfloatsep}{1\baselineskip plus 0.2\baselineskip minus 0.5\baselineskip}
%
\begin{figure}[!b]
\centering
\includegraphics[width=\columnwidth,height=4.5cm]{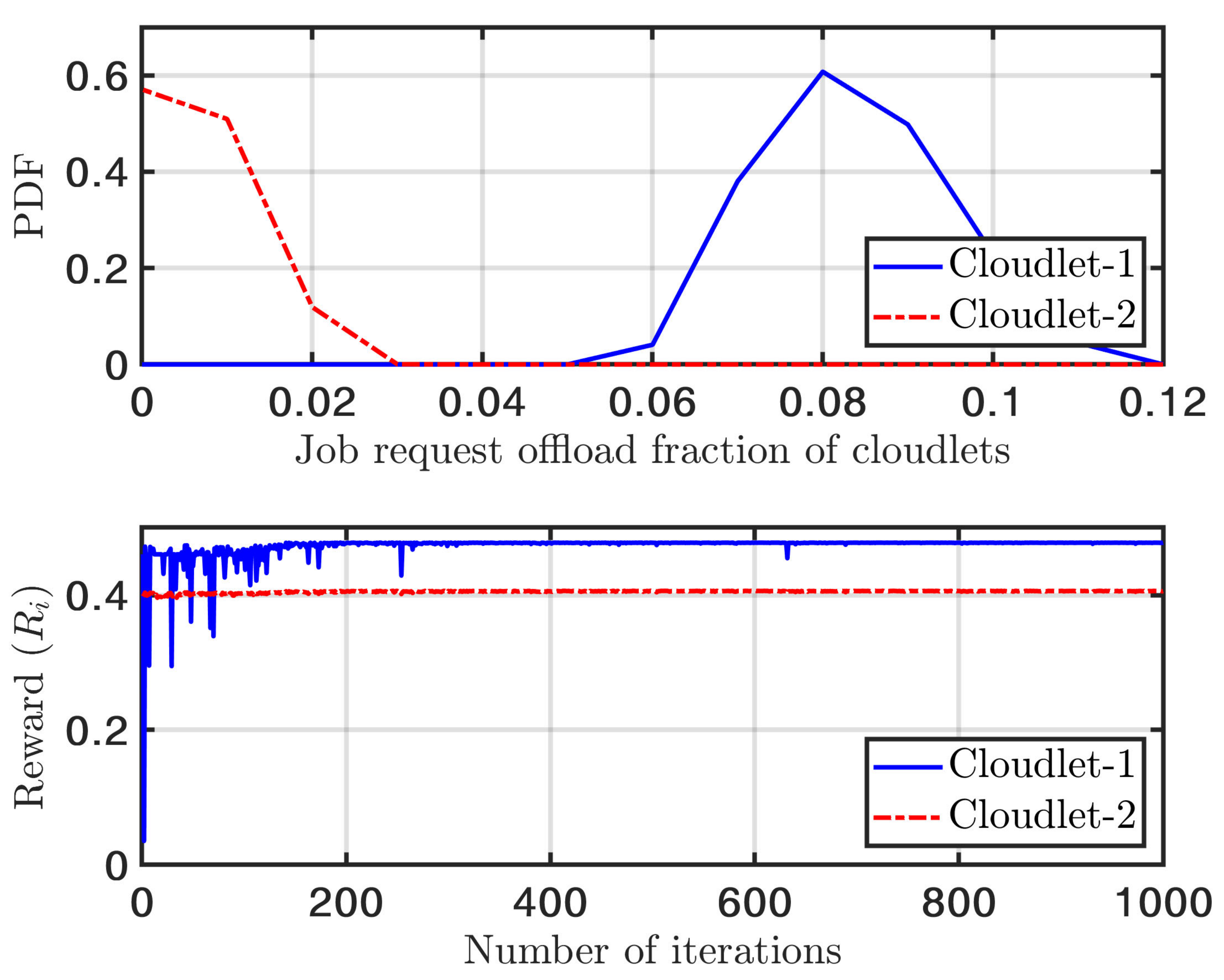}
\caption{Learning NE of the non-cooperative load balancing game with Algorithm \ref{alg2} by two cloudlets with $n_1^m\mu_1^m = n_2^m\mu_2^m = 1000$ jobs/s, $\lambda_1^m$ = 970 jobs/s, $\lambda_2^m$ = 800 jobs/s, and $\Theta = 0.9$, $\sigma = 0.01$.}
\label{converge2}
\end{figure}
\setlength{\textfloatsep}{5pt}
To observe the performance of our proposed centralized load balancing framework with some real-world traces, we consider the cluster usage trace released by Google in 2011 \cite{G2011}. We identify two different classes of incoming job requests with $D_{Q1} = 10$ msec and $D_{Q2} = 20$ msec and fit the PDFs of inter-arrival times with exponential PDFs (refer to Appendix-A). We perform an event-driven simulation on OMNeT++ with these traces among three cloudlets. Each of these cloudlets has $n_i = 10$ processors and different number of processors dedicated to the two different job classes ($n_i^m$) are decided by solving $\mathcal{P}_i^m$.  The job request service rates are $\mu_i^1 = 250$ jobs/s and $\mu_i^2 = 200$ jobs/s and the job request arrival rates ($\lambda_i^m$) for both the classes vary within 0-1500 jobs/s. The intermediate transmission latencies and bandwidths are $t_{12} = t_{21} = 0.5$ msec, $t_{23} = t_{32} = 0.7$ msec, $t_{31} = t_{13} = 0.9$ msec, $B_{12} = B_{21} = 0.6$ Gbps, $B_{23} = B_{32} = 0.8$ Gbps, and $B_{13} = B_{31} = 1$ Gbps, respectively. The duration of each of the utility evaluation time-interval $\tau_x$ is 1 sec. Each cloudlet uses LSTM networks and stochastic gradient descent algorithm to predict their incoming job request arrival rates one timeslot before the beginning of each time interval $\tau_x$ and forecasts them to the mediator. The mediator computes the processor slicing in each cloudlet by solving $\mathcal{P}_i^m$ to compute the NE solution and sends back to the cloudlets. Based on the NE strategies, the overloaded cloudlets randomly offload a fraction of their incoming job requests to their under-loaded neighbors during all timeslots within $\tau_x$ interval. Every time a job request arrives, each $i^{\text{th}}$ overloaded cloudlet randomly offloads the job requests to its $j^{\text{th}}$ neighboring cloudlet according to $\varphi_{ij}^{m*}$ and the job processing simulation is implemented as described in Section \ref{sec3}. In Fig. \ref{predict_compare}, we show three subplots to compare the actual and simulated utilities of each of the cloudlets. Firstly, we plot the theoretically computed utilities with the actual job request arrival rates. Secondly, we plot the simulated utilities with actual job request arrival, but the NE solution computed by using the predicted job request arrival rates. Hence, the NE solution are erroneous and the simulated utilities also deviate from the actual utilities. We observed that the simulated utilities have a mean error of 10\% from the actual utilities. Out of this, nearly 8\% error was due to the prediction error in job request arrival rates and the rest of the error was due to the approximation error in modeling the cloudlets as $M/M/c$ queues.\par
\begin{figure}[!b]
\centering
\includegraphics[width=\columnwidth,height=4.5cm]{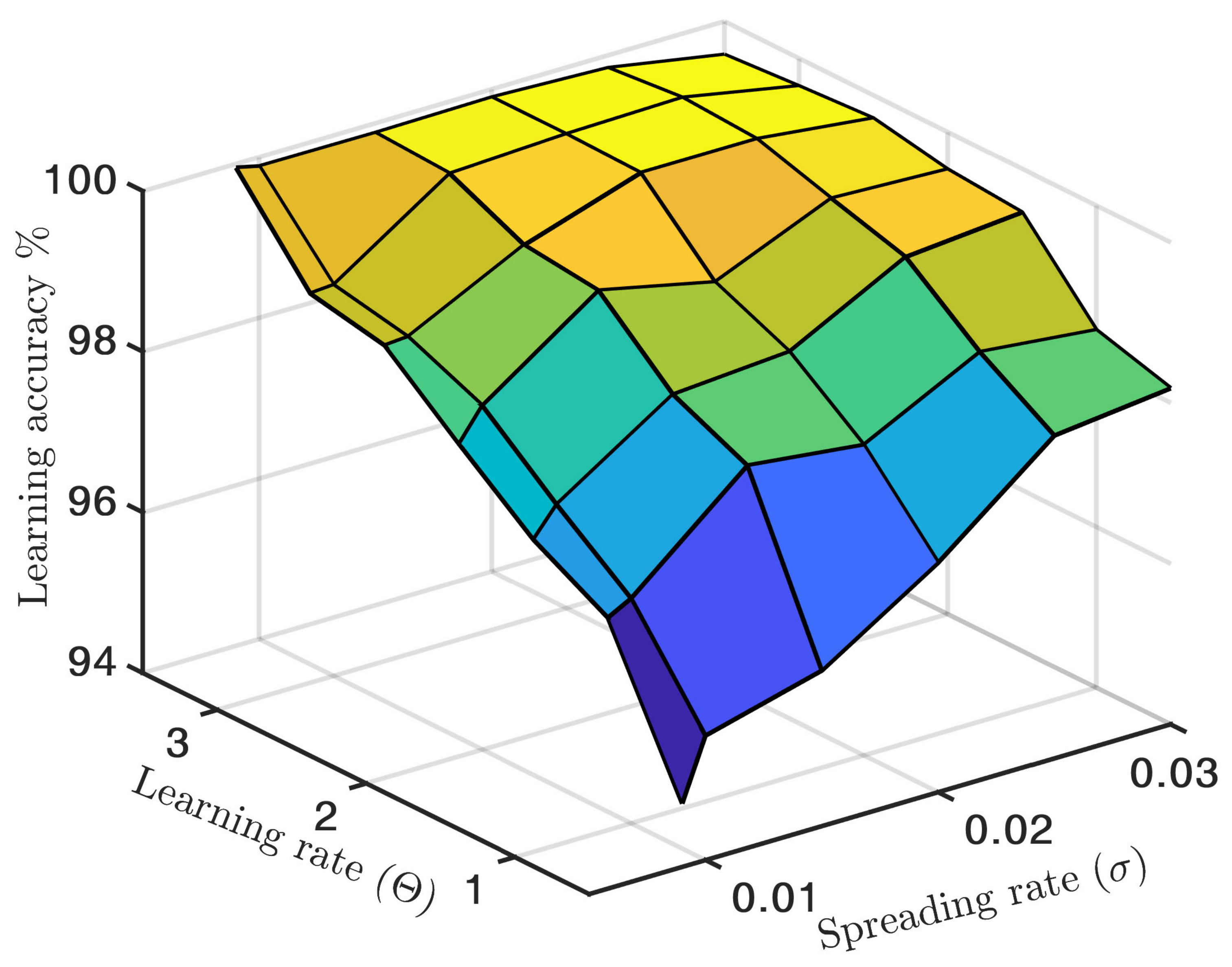}
\caption{Average NE learning accuracy of Algorithm \ref{alg2} against variation of learning rate parameter $\Theta$ and spreading rate parameter $\sigma$.}
\label{accuracy1}
\end{figure}
\setlength{\textfloatsep}{5pt}
In Fig. \ref{converge2}, we observe the convergence properties of the proposed reinforcement learning algorithm.  We consider two neighboring cloudlets, Cloudlet-1 and Cloudlet-2 with intermediate transmission latency $t_{ij}^m = 1$ msec, trying to meet a QoS requirement of $D_{Qm} = 10$ msec.  We also consider that $n_1^m\mu_1^m = n_2^m\mu_2^m = 1000$ jobs/s and $\lambda_1^m = 970$ jobs/s, $\lambda_2^m = 800$ jobs/s, such that Cloudlet-1 is overloaded, but Cloudlet-2 is under-loaded.  Therefore, to meet $D_{Qm}$, Cloudlet-1 needs to offload $0.083\times\lambda_1^m$ job requests to Cloudlet-2, whereas Cloudlet-2 does not need to offload anything (by solving $\mathcal{P}_{\Gamma}^m$). From the PDFs of both the cloudlets also, we observe that the most preferable decision for Cloudlet-2 is not to offload and Cloudlet-1 prefers to offload around 8-9\% of its total incoming job requests.  As we choose the learning and spreading parameters as $\Theta = 0.9$ and $\sigma = 0.01$, respectively, we see that Algorithm \ref{alg2} converges to the expected utility and reward values for both the cloudlets within a few hundred iterations. Note that, instead of searching over the whole strategy space, we considered only the most likely strategies that the cloudlets should possibly consider by using our understanding from the underlying load balancing game. Moreover, by using these techniques, Algorithm \ref{alg2} can perform 100\% accurately when all cloudlets are under-loaded without much exploration.\par
%
It is interesting to note that in the previously considered scenario, even faster convergence to NE is possible by increasing the value of the learning rate parameter $\Theta$, but the learning process may become unstable.  Therefore, to study the performance of Algorithm \ref{alg2}, we plot the average learning accuracy in Fig. \ref{accuracy1} against variation of $\Theta$ and $\sigma$. We observed that a reasonably high degree of accuracy is achievable with a stationarity time of 1 sec as Algorithm \ref{alg2} converges to the NE solution within a few hundred iterations. In this plot, we observe that the NE learning accuracy increases with $\Theta$, but if we simultaneously increase $\sigma$ also, then accuracy performance starts to slightly decrease as exploration increases.\par
\begin{figure}[!t]
\centering
\includegraphics[width=\columnwidth,height=4.5cm]{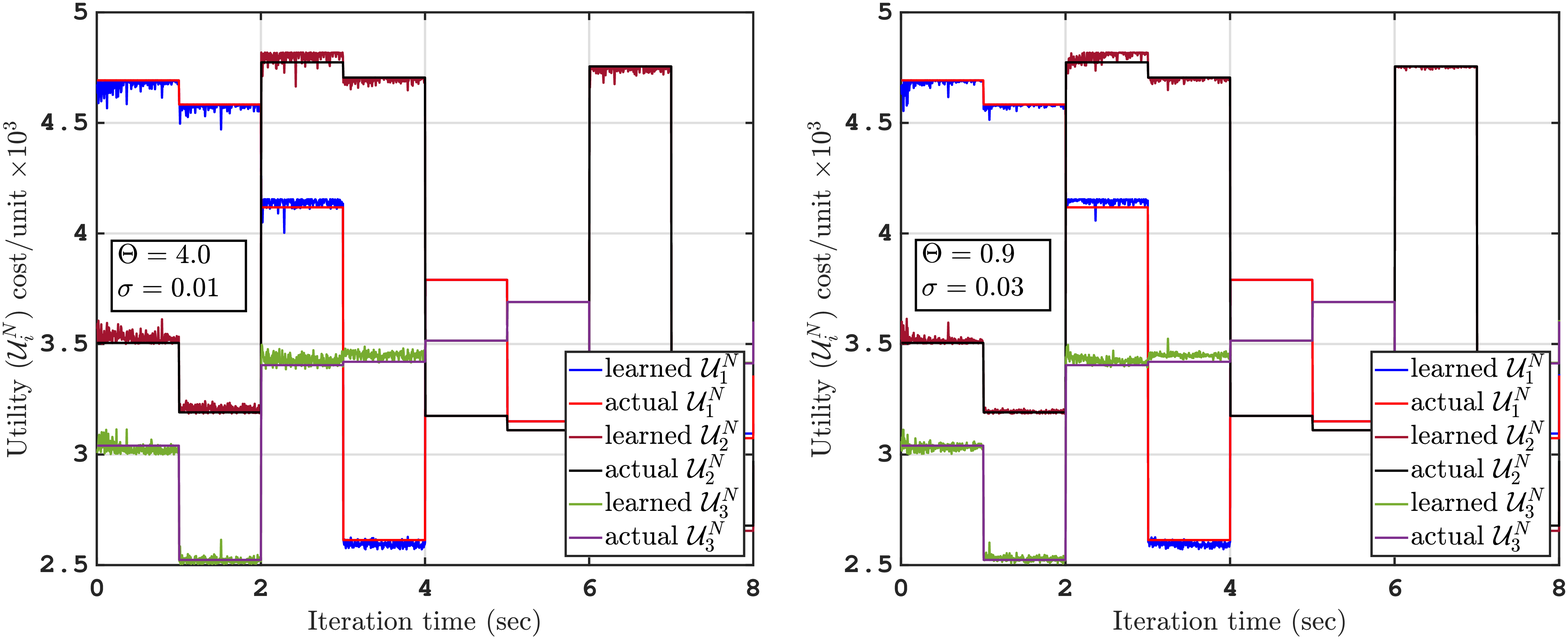}
\caption{Average NE learning accuracy of Algorithm \ref{alg2} for load balancing game among three cloudlets with dynamically varying job request arrival rates ($\lambda_i^m$) to all cloudlets.}
\label{utility}
\end{figure}
\setlength{\textfloatsep}{5pt}
Again, we consider the same scenario of three federated cloudlets with $n_1^m\mu_1^m = n_2^m\mu_2^m = n_3^m\mu_3^m = 1000$ jobs/s and $D_{Qm} = 10$ msec. Moreover, the job request arrival rates $\lambda_i^m$ to each of these cloudlets are dynamically varying and the values are considered from the Google cluster traces. Therefore, the utilities of each $i^{\text{th}}$ cloudlet $\mathcal{U}^N_i$ also changes accordingly. Each cloudlet predicts their incoming job request arrival rates one-timeslot before the beginning of $\tau_x$ time interval with an LSTM network, trained by stochastic sub-gradient algorithm to decide whether it is under-loaded or overloaded. Fig. \ref{utility} presents a comparison between the actual utilities (by solving $\mathcal{P}_{\Gamma}^m$ with true $\lambda_i^m$ values) and the learned utilities through Algorithm \ref{alg2}. The first subplot has a learning rate $\Theta = 0.9$ and spreading rate $\sigma = 0.03$ and Algorithm \ref{alg2} takes some time to learn the actual utility values due to more exploration. The second subplot has $\Theta = 4.0$ and $\sigma = 0.01$ where Algorithm \ref{alg2} learns the actual utility values relatively faster but performance may degrade during sudden changes due to less exploration. Note that if a cloudlet is under-loaded then it decides its NE strategy without any exploration. Also, if an overloaded cannot offload its extra load completely, then the feedback from its under-loaded neighbors help to quickly shift the peak of PDF around the proper NE solution. Overall, we observe that by employing Algorithm \ref{alg2}, the cloudlets are able to achieve fairly accurate utility values.\par
\begin{figure}[t!]
  \centering

  \begin{subfigure}[t]{\columnwidth}
  \centering
  \includegraphics[width=\textwidth,height=4.5cm]{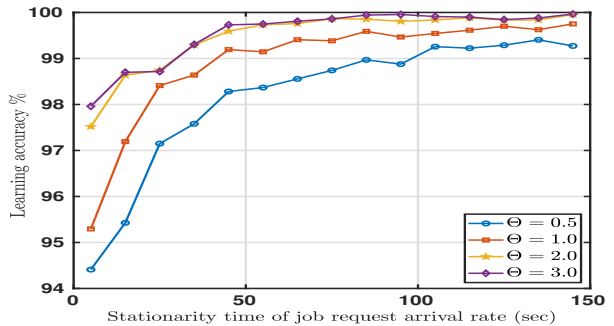}
  \caption{Average NE learning accuracy with $\sigma = 0.01$.}
  \label{accuracy2a}
  \end{subfigure}

  \begin{subfigure}[t]{\columnwidth}
  \centering
  \includegraphics[width=\textwidth,height=4.5cm]{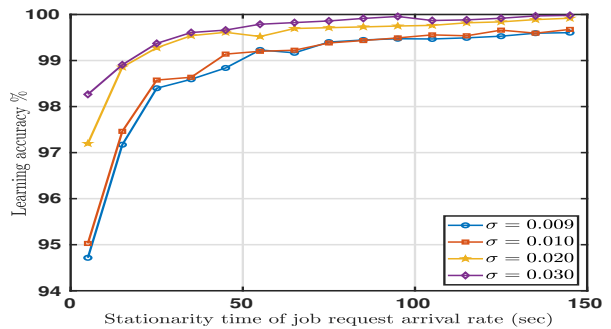}
  \caption{Average NE learning accuracy with $\Theta = 0.9$.} 
  \label{accuracy2b}
  \end{subfigure} 

  \caption{Average NE learning accuracy of Algorithm \ref{alg2} against stationarity time of job request arrival rates ($\lambda_i^m$) to all cloudlets with variations of learning rate $\Theta$ and spreading rate $\sigma$.}
\end{figure}

From the previous results, we found that it is essential to choose the learning rate and spreading rate parameters in such a way so that a proper balance between exploration and exploitation is maintained against the stationarity time of job request arrival rates.  Thus, in Fig. 8 we plot the average NE learning accuracy against the stationarity period of the job request arrival rate.  We vary the stationarity time from 0.5 sec to 150 sec and also tune $\Theta$ from 0.5 to 3 with $\sigma = 0.01$ in Fig. \ref{accuracy2a}, and tune $\sigma$ from 0.009 to 0.030 with $\Theta = 0.9$ in Fig. \ref{accuracy2b}. We observe in general, that the performance of our proposed Algorithm \ref{alg2} increases as the stationarity of the job request arrival rate increases because the algorithm is given more time-slots to exploit and explore the search space.\par
\setlength{\textfloatsep}{5pt}
\section{Conclusions}\label{sec8}
In this paper, we have proposed a novel economic and non-cooperative game-theoretic model for low-latency applications among multiple federated cloudlets. This load balancing framework acts like a non-cooperative game-theoretic model among neighboring cloudlets from different SP and like an optimization model among cloudlets from the same SP. Moreover, this load balancing framework is capable of handling job requests from heterogeneous classes, which is unique over state-of-the-art frameworks. We have proposed a centralized framework where dominant strategy incentive compatible mechanism is imposed on the cloudlets for revealing truthful information. Through numerical evaluations, we have also showed that our proposed framework achieves a better economic utilities under low, medium, and high load conditions compared to some existing frameworks. We have used real job request arrival traces in an event-driven simulation for performance evaluation of the load balancing framework in a realistic scenario and showed that the mean error of the proposed framework is within 10\%. Followed by, we have designed a distributed continuous-action reinforcement learning automata-based algorithm to facilitate the federated cloudlets to learn their NE job request offload strategies independently, with a minimal exchange of control messages with neighboring cloudlets. We have improved the convergence rate of the proposed reinforcement. Through extensive simulation, we have shown that the proposed reinforcement algorithm can achieve nearly $97-99\%$ accuracy with the stationarity time of nearly 10 seconds for the job request arrival process.

\bibliographystyle{IEEEtran}
\bibliography{IEEEabrv,ref_cloudlet_mechanism}
\vspace{-5ex}

\vfill

\end{document}


\title{Appendices for Centralized and Decentralized Non-Cooperative Load-Balancing Games among Federated Cloudlets}
\author{Sourav~Mondal,~\IEEEmembership{Student~Member,~IEEE,}
        Goutam~Das,~\IEEEmembership{Member,~IEEE,}
        and~Elaine~Wong,~\IEEEmembership{Senior~Member,~IEEE}
}

\maketitle

\appendices
\section{Details of Timeslots to guarantee QoS Latency and $M/M/c$ Queuing Model for Cloudlets}
\textbf{(a) Design of timeslots to guarantee QoS latency:} In our proposed load balancing game, each cloudlet uses a \emph{timeslotted model} to ensure the QoS latency target $D_{Qm}$ for the job requests of each job class-$m$. The duration of each fundamental timeslot is chosen to be equal to $D_Q$ and without loss of generality, we assume that $D_{Qm}$ values are integer multiples of $D_Q$, i.e., $D_{Q1} = I_1\times D_Q$, $D_{Q2} = I_2\times D_Q$, $D_{Q3} = I_3\times D_Q$, and so on ($I_1,I_2, I_3,\dots, I_M$ are integer values). Depending on the stationarity of the incoming job request traffic, we choose a bigger time interval $\tau_x$ which is an integer multiple of all of $I_1,I_2, I_3,\dots, I_M$. Each cloudlet uses a workload predictive algorithm a few timeslots before the beginning of each $\tau_x$ time interval to predict the job request arrival rates for all the $M$ job classes. Based on this predicted job request arrival rates, each cloudlet can solve $\mathcal{P}_i^m$ to decide the processor slicing. They can also communicate this information to the mediator in the centralized framework and the mediator brodcasts the Nash equilibrium (NE) load balancing strategies to the cloudlets. This processor slicing and the NE load balancing strategies are applied when the time interval $\tau_x$ actually begins. During each $\tau_x$ interval, the processor slicing and the NE load balancing strategies remain unchanged over all the timeslots.\par 
\begin{figure}[b!]
\centering
\includegraphics[width=\columnwidth]{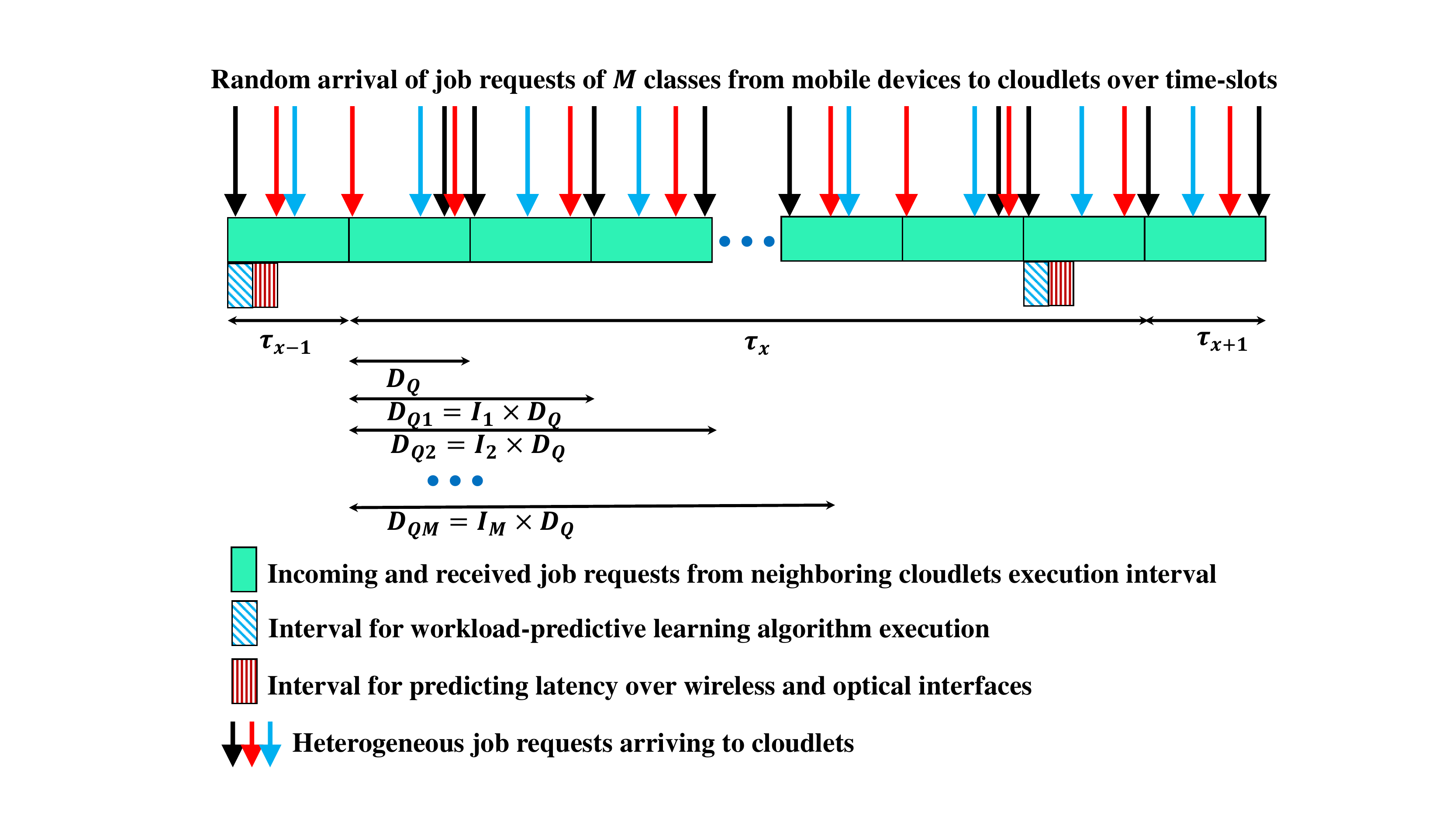}
\caption{A schematic diagram to illustrate the design of timeslots.}
\label{timeslot1}
\end{figure}
\setlength{\textfloatsep}{5pt}
The cloudlets need to use the timeslotted model as shown in Fig. \ref{timeslot1} to ensure the target QoS latency for each job request class. The job requests arrive continuously and randomly within each timeslot and are marked with an integer $\lfloor \frac{D_{Qm}-D_Q}{D_Q}\rfloor = (I_m-1)$. The duration of a timeslot $D_Q$ is subtracted from $D_{Qm}$ to adjust the time lost due to the transmission latencies. If some jobs could not be processed within the timeslot of job request arrival, especially the jobs arriving at the verge of finishing of the timeslot, they are rolled over to the next timeslot by decreasing the marking by 1. This process can continue until those jobs are processed and the jobs can be deleted after the marking becomes 0. This process is followed even if the ongoing $\tau_x$ time interval is over and the next time slot $\tau_{x+1}$ begins. Therefore, the job request arrival queue of the cloudlets can maintain a steady-state unless they are extremely overloaded. As the job request arrival rates are predicted slightly before the actual job request arrival, all these network management operations can be done very efficiently.\par
For more clarity on this timeslotted protocol design, we provide the following numerical example. We consider the values $D_Q = 5$ msec, $D_{Q1} = 10$ msec, $D_{Q2} = 20$ msec, and $\tau_x = 1$ sec. Thus, there are 200 timeslots of $D_Q = 5$ msec within each interval of $\tau_x = 1$ sec. One timeslot before the begining of each $\tau_x$ interval, the processor slicing and NE load balancing strategies are computed based on the one timeslot ahead predicted job request arrival rates. When job requests of class-1 arrive within a particular timeslot, they are marked as 1. If some of these jobs could not be processed within that timeslot, they are rolled over to the next timeslot while updating the marking as 0. If these jobs are still unprocessed at the end of this timeslot, then they are dropped. Similarly, when job requests of class-2 arrive within a particular timeslot, they are marked as 3 and the same process is followed. At the end of each $\tau_x$ interval, the processor slicing and NE load balancing strategies are re-calculated. Therefore, the job request arrival queue of the cloudlets can always remain in the steady-state.\par
\begin{figure*}[t!]
  \centering

  \begin{subfigure}[t]{\textwidth}
  \centering
  \includegraphics[width=\textwidth]{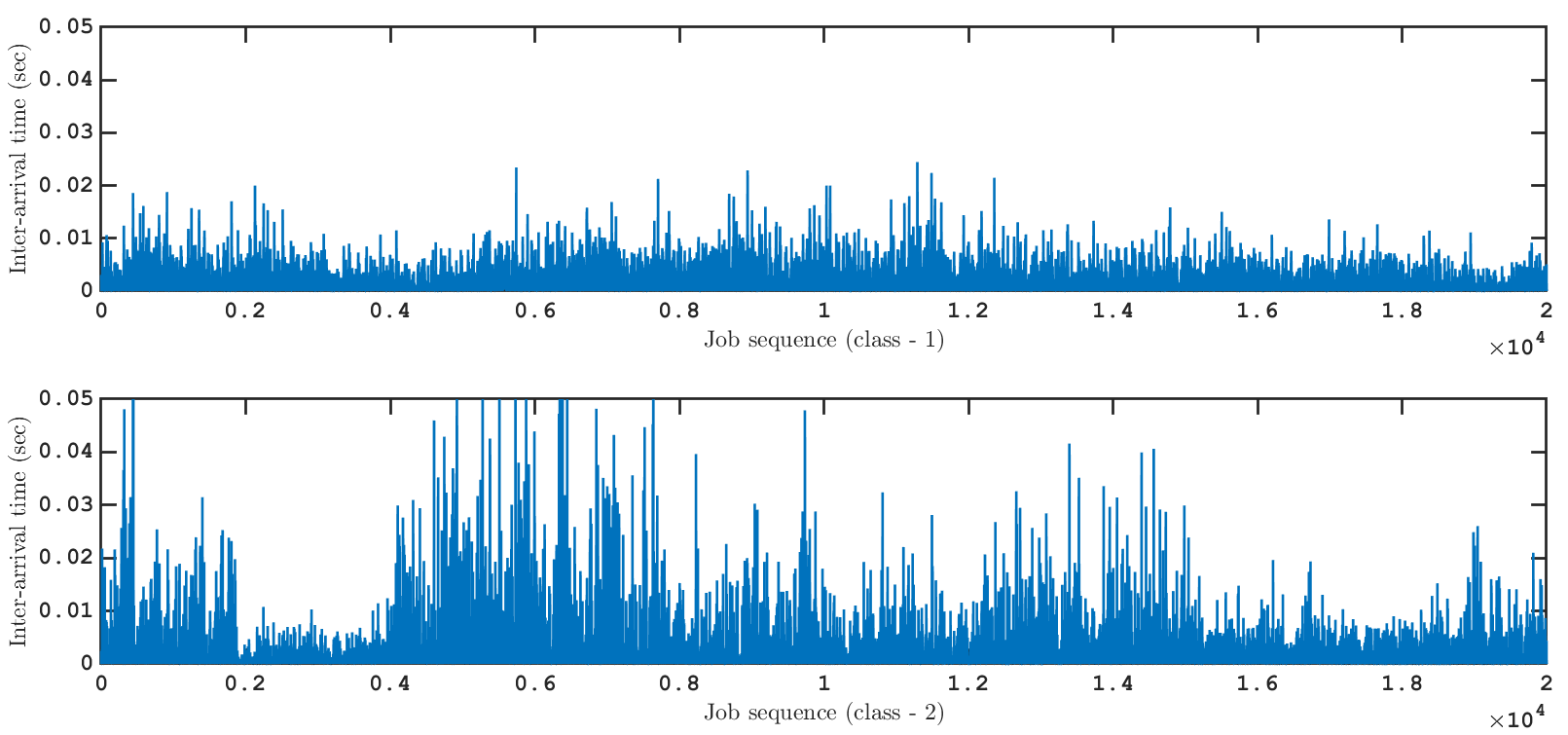}
  \caption{Inter-arrival times of the job requests from three different classes to Google clusters.}
  \label{time_diff}
  \end{subfigure}

  \begin{subfigure}[t]{\textwidth}
  \centering
  \includegraphics[width=\textwidth]{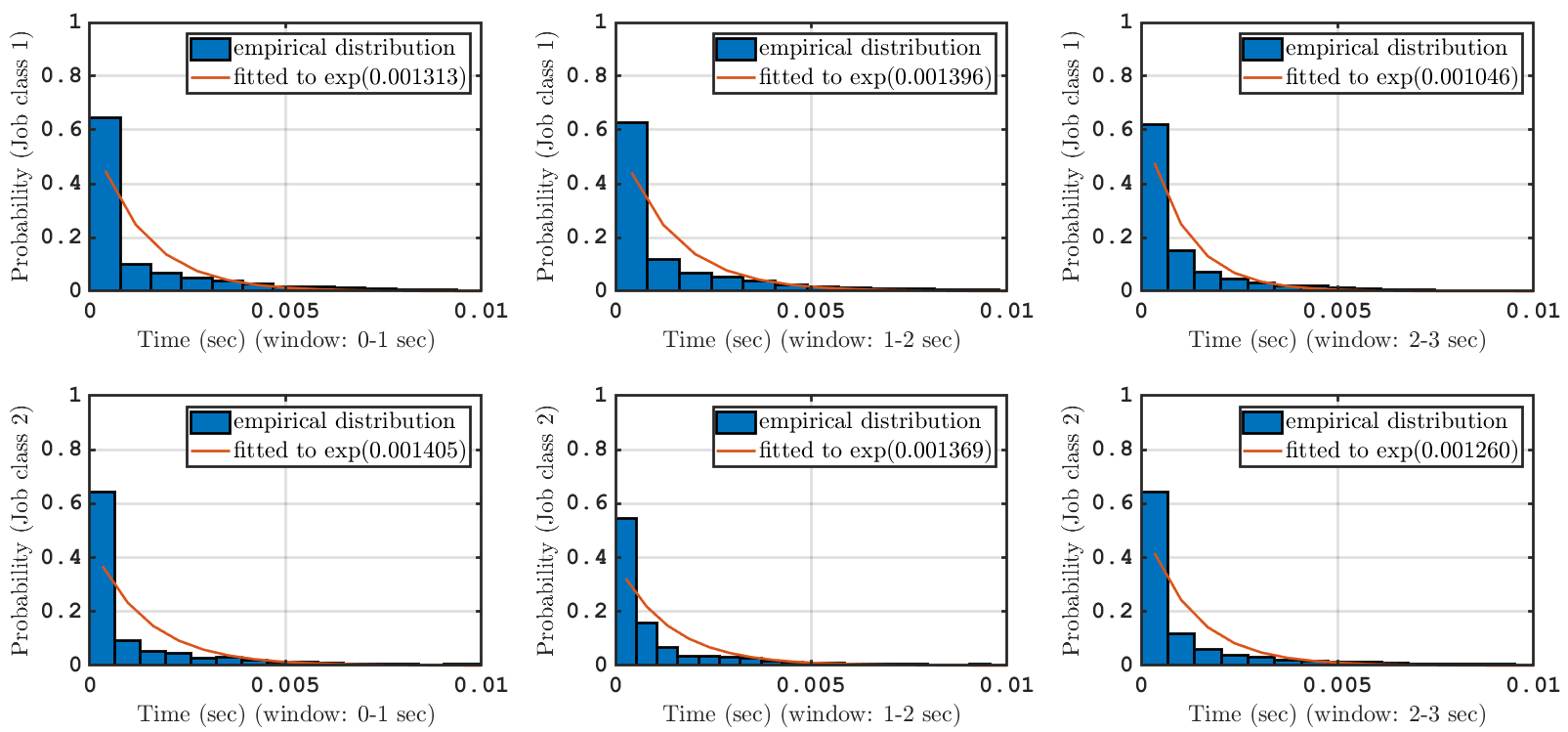}
  \caption{Fitting job request inter-arrival times over 1 sec windows to exponential distribution.}
  \label{time_pdf}
  \end{subfigure}
  
  \caption{Inter-arrival times of the job requests from three different classes to Google clusters and their corresponding distribution fitting.}
\end{figure*}
\setlength{\textfloatsep}{5pt}

\begin{figure*}[t!]
  \centering

  \begin{subfigure}[t]{\textwidth}
  \centering
  \includegraphics[width=\textwidth]{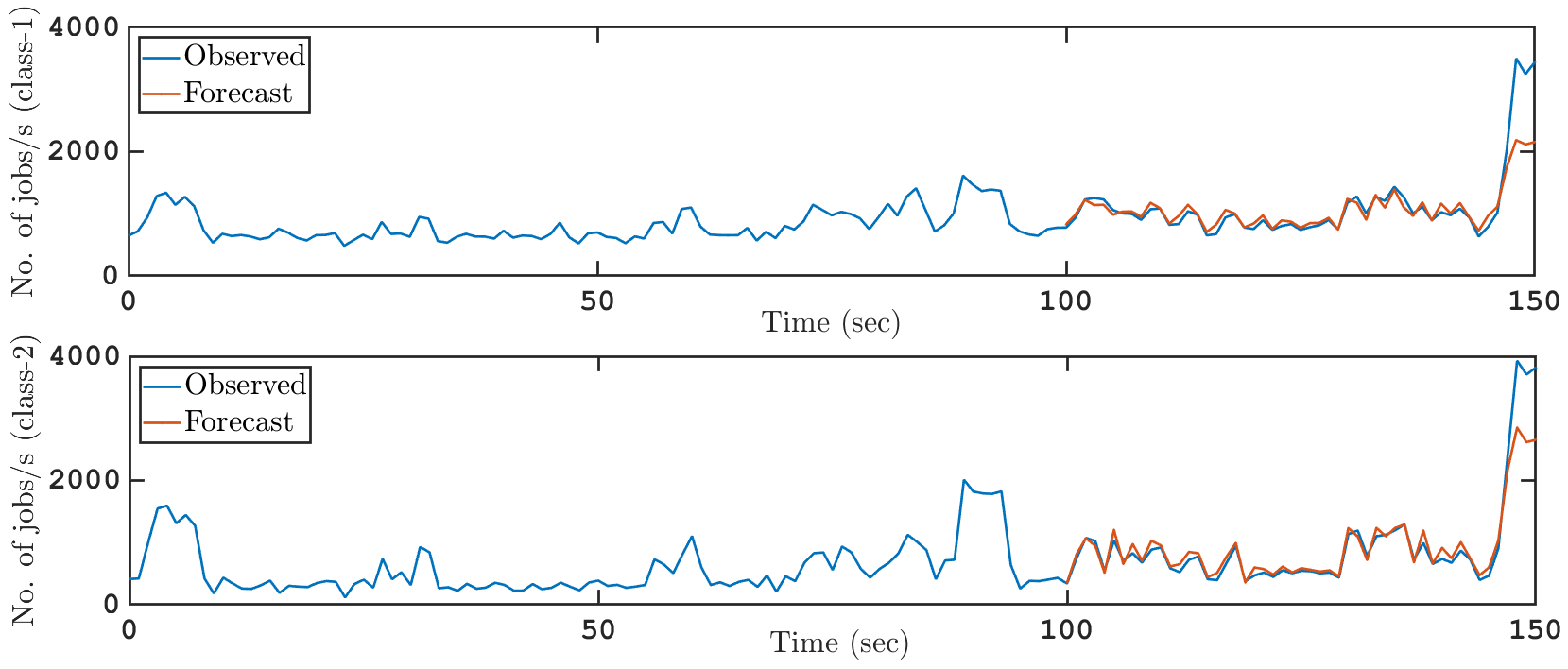}
  \caption{Prediction of job request arrival rates for two different job classes by LSTM algorithm with RMSE values 192 and 543, respectively.}
  \label{lstm}
  \end{subfigure}

  \begin{subfigure}[t]{\textwidth}
  \centering
  \includegraphics[width=\textwidth]{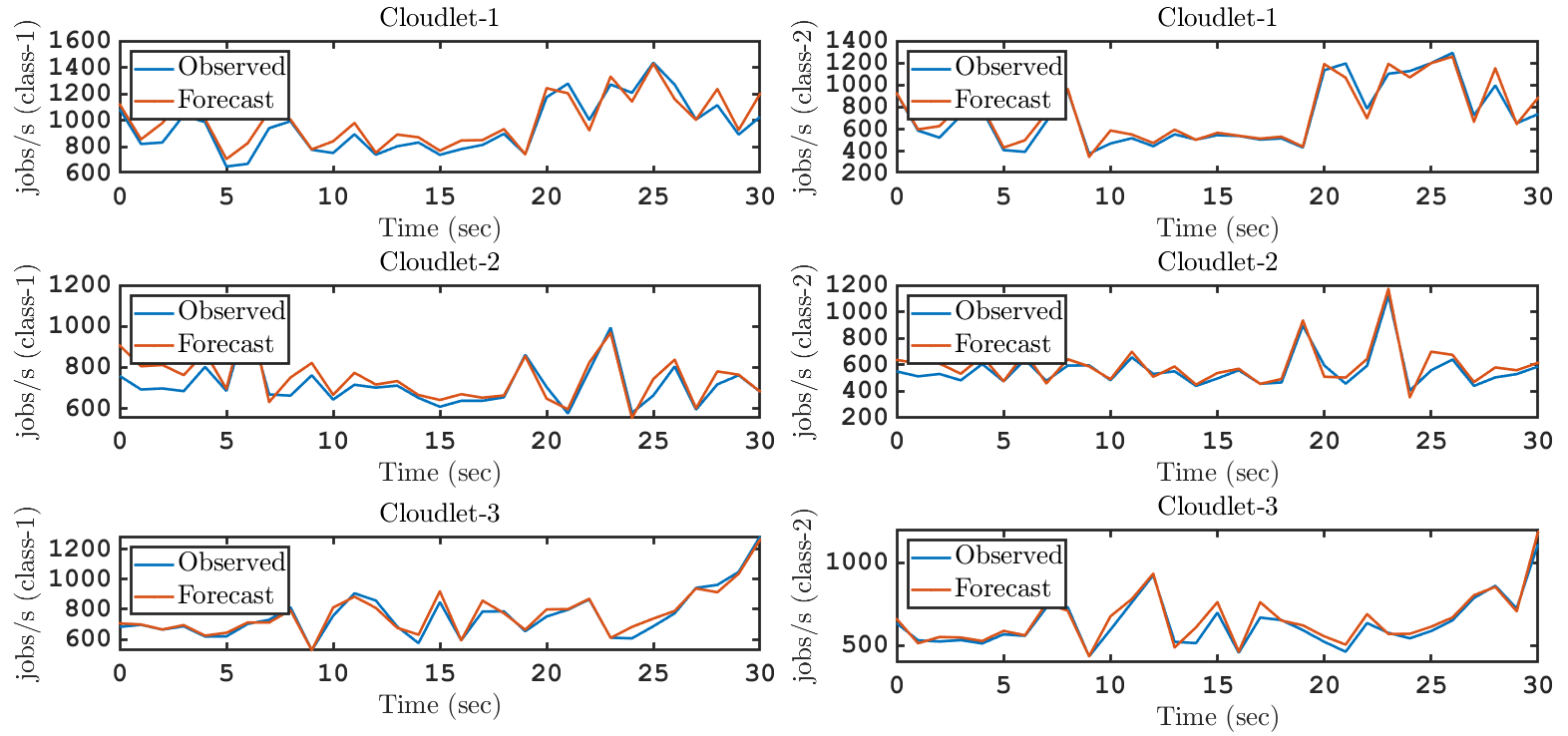}
  \caption{Job request arrival rates of two different classes to three cloudlets used in event-driven simulation with OMNeT++.}
  \label{cloudlet_jobs}
  \end{subfigure}
  
  \caption{Observed and forecasted job request arrival rates of two different classes to cloudlets.}
\end{figure*}
\setlength{\textfloatsep}{5pt}
\textbf{(b) $M/M/c$ queuing model for cloudlets:} We consider that each cloudlet has a finite number of processors, say $c$. In practice, some of the jobs can be completely parallelized and some cannot be parallelized at all. Hence, the actual performance of each cloudlet will lie within the performance bounds provided by $M/M/c$ and $M/M/1$ queuing systems. Recall that the average processing latency of an $M/M/c$ queuing system with job request arrival rate = $\lambda$ jobs/s, service rate = $\mu$ jobs/s and $\rho=\lambda/(c\mu)$ is expressed as follows \cite{Kleinrock}:
\begin{align}
\mathbb{E}(T) = \frac{1}{\mu}+\frac{\mathcal{E}_C(c,\lambda/\mu)}{c\mu-\lambda}, \label{eq01}
\end{align}
where, $\mathcal{E}_C(c,\lambda/\mu)$ is the \emph{Erlang-C formula}, given by:
\begin{align}
\mathcal{E}_C(c,\lambda/\mu) = \frac{1}{1+(1-\rho)\left(\frac{c !}{(c \rho)^{c}}\right)\sum\limits_{k=0}^{c-1} \frac{(c \rho)^k}{k!}}. \label{eq02}
\end{align}
\par Again, the average processing latency of an $M/M/1$ queuing system with the combined service rate of $c\mu$ jobs/s is expressed as follows:
\begin{align}
\mathbb{E}(\hat{T}) = \frac{1}{c\mu-\lambda} = \frac{1}{\mu}+\frac{\rho}{c\mu-\lambda}. \label{eq03}
\end{align}
\par Therefore, under \emph{light load conditions}, we have $\rho \ll 1$, $\frac{\rho}{c\mu-\lambda} \approx 0$, $\frac{\mathcal{E}_C(c,\lambda/\mu)}{c\mu-\lambda} \approx 0$ and hence derive the following:
\begin{align}
\frac{\mathbb{E}(T)}{\mathbb{E}(\hat{T})} = \frac{\frac{1}{\mu}+\frac{\mathcal{E}_C(c,\lambda/\mu)}{c\mu-\lambda}}{\frac{1}{\mu}+\frac{\rho}{c\mu-\lambda}} \approx \frac{\frac{1}{\mu}}{\frac{1}{c\mu}} \approx c. \label{eq04}
\end{align}
\par On the other hand, under \emph{heavy load conditions}, we have $\rho \approx 1$, $\mathcal{E}_C(c,\lambda/\mu) \approx 1$, $\frac{1}{\mu} \ll \frac{\rho}{c\mu-\lambda}$, $\frac{1}{\mu} \ll \frac{\mathcal{E}_C(c,\lambda/\mu)}{c\mu-\lambda}$ and hence derive the following:
\begin{align}
\frac{\mathbb{E}(T)}{\mathbb{E}(\hat{T})} = \frac{\frac{1}{\mu}+\frac{\mathcal{E}_C(c,\lambda/\mu)}{c\mu-\lambda}}{\frac{1}{\mu}+\frac{\rho}{c\mu-\lambda}} \approx \frac{\frac{1}{c\mu-\lambda}}{\frac{1}{c\mu-\lambda}} \approx 1. \label{eq05}
\end{align}
\par From the above analysis, we observe that $\mathbb{E}(T) \geq \mathbb{E}(\hat{T})$ holds true under low as well as high load conditions. Therefore, to guarantee the worst-case performance of the cloudlets in terms of the processing latency, we model the cloudlets as $M/M/c$ queuing systems.\par 
\textbf{(c) Poisson job request arrival to cloudlets:} To validate our assumption that the incoming job requests to the cloudlets follow the Poisson process, we consider the cluster usage trace released by Google in 2011 \cite{G2011}. The Fig. \ref{time_diff} shows the inter-arrival times of the job requests from two different job classes, i.e., the difference between the timestamps of the consecutive job requests. From this plot, we can observe two primary characteristics, viz., \emph{self-similarity} and \emph{non-stationarity}. These properties of the Google cluster traces were studied in more details by the authors of \cite{trace}. Due to self-similarity property, the series of job request inter-arrival times looks the same under any magnification of scales. The non-stationarity property indicates that the distribution and statistical moments of the job request inter-arrival times vary over time. As the job request arrival process is a non-stationary random process, we considered small windows of 1 sec and attempted to fit the distribution of the inter-arrival times with exponential probability distribution in Fig. \ref{time_pdf}. We checked that the distribution fitting satisfies a significance level of 0.05 (verified with Chi-squared goodness-of-fit test). Similar observations were also reported by the authors of \cite{poisson}. Note that we have scaled up the job request arrival rate (jobs per 300 sec are treated as jobs per 1 sec) since we found that the job request arrival rates to the Google clusters are much lower than our expected job request arrival rates to the cloudlets.\par
In Fig. \ref{lstm}, we show job request arrival rates for two different job classes  from the Google cluster traces. As the job request arrival process is self-similar and non-stationary, each cloudlet uses long short-term memory (LSTM) networks and stochastic gradient descent algorithm to predict the incoming job request arrival rates to the cloudlets. In this figure, the job request arrival rates over the first 100 sec interval are used to train the LSTM network. Once the network is trained, it is used to forecast the job request arrival rates 1 sec before the actual job request arrival over the next 50 sec interval. We observe that a very high degree of prediction accuracy is achievable by the LSTM networks because the prediction error is greatly reduced at each iteration by observing the actual job request arrival rates from the previous iteration. Fig. \ref{cloudlet_jobs} shows the job request arrival rates used in the event-driven driven simulation by using OMNeT++.
 
\section{Optimal Processor Slicing in Cloudlets}
\begin{figure}[t!]
\centering
\includegraphics[width=\columnwidth]{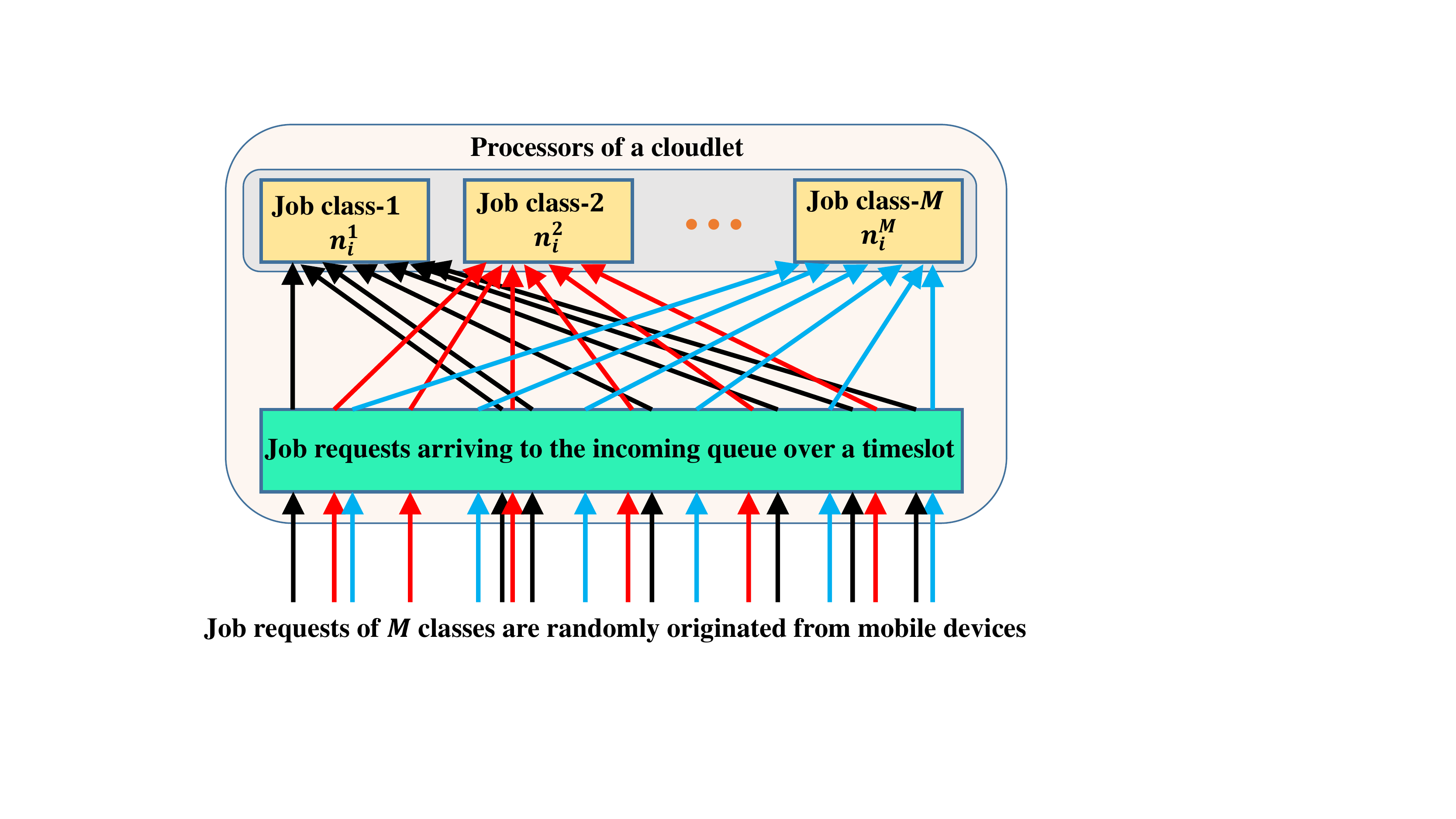}
\caption{A schematic diagram to illustrate the processing of job requests from multiple classes by processor slicing in a cloudlet.}
\label{process}
\end{figure}

In this load balancing problem, each cloudlet uses \emph{processor slicing} technique to handle multiple classes of incoming job requests with heterogeneous QoS target latency requirements, as shown in Fig. \ref{process}. We assume that each cloudlet has multiple processors and denote the total number of processors in $i^{\text{th}}$ cloudlet by $n_i$. If the incoming job requests can be classified in to $M$ classes, then the $i^{\text{th}}$ cloudlet needs to create $M$ slices with $n_i^m$ processors in each slice such that $1\leq n_i^m \leq n_i$. Firstly, each $i^{\text{th}}$ cloudlet observes their individual incoming job request arrival rates for each class and finds the optimal values of $n_i^m$ by solving the following optimization problem:\par
\begin{align}
\mathcal{P}_i^m: \quad \quad \min_{(n_i^1,n_i^2,\dots,n_i^M)} \quad & \max_{m\in\mathcal{J}} \left\{t_{ui}^m+\mathbb{T}_i^m(\mu_i^m,\lambda_i^m;n_i^m) - D_{Qm}\right\} \nonumber\\
\text{subject to}\quad &1 \leq n_i^m \leq n_i, \sum_{m=1}^{M} n_i^m = n_i, \forall m \in \mathcal{J}.  \nonumber
\end{align}
\setlength{\textfloatsep}{5pt}
\par This optimization problem ensures an optimal processor slicing among all the job classes and allows soft processor slicing, i.e., $n_i^m$ can take any real and $\geq 1$ value. We formulate this problem such that a fair number of processors are allocated for all the job classes and avoid situations where the highest number of processors are allocated to a job class whose latency difference from $D_{Qm}$ can be extremely low but high for others. To solve the problem efficiently, we can reformulate the problem as follows:
\begin{align}
\mathcal{R}_i^m: \quad \quad &\min_{(n_i^1,n_i^2,\dots,n_i^M)} \quad Z \nonumber\\
\text{subject to}\quad &1 \leq n_i^m \leq n_i, \sum_{m=1}^{M} n_i^m = n_i, \forall m \in \mathcal{J}, \nonumber\\
& Z \geq \left(t_{ui}^m+\mathbb{T}_i^m(\mu_i^m,\lambda_i^m;n_i^m) - D_{Qm}\right), \forall m \in \mathcal{J}.\nonumber
\end{align}
We can calculate the processing latency values as follows:
\begin{align}
\mathbb{T}_i^m (\mu_i^m,\lambda_i^m;n_i^m) = \frac{1}{\mu_i^m}+ \frac{\mathcal{E}_C(n_i^m,\lambda_i^m/\mu_i^m)}{n_i^m\mu_i^m-\lambda_i^m}, \label{eq06}
\end{align}
where, $\mathcal{E}_C(n_i^m,\lambda_i^m/\mu_i^m)$ is the \emph{Erlang-C formula}, which can be written in an alternative form with $\rho_i^m = \lambda_i^m/(n_i^m \mu_i^m)$ for some advantage in the analysis as follows:
\begin{align}
&\mathcal{E}_C(n_i^m,\lambda_i^m/\mu_i^m) = \frac{1}{1+(1-\rho_i^m)\left(\frac{n_i^m !}{(n_i^m \rho_i^m)^{n_i^m}}\right)\sum\limits_{k=0}^{n_i^m-1} \frac{(n_i^m \rho_i^m)^k}{k!}}\nonumber\\
&= \frac{1}{1+\exp\left(\frac{\lambda_i^m}{\mu_i^m}\right)\left(n_i^m- \frac{\lambda_i^m}{\mu_i^m}\right)\left(\frac{\lambda_i^m}{\mu_i^m}\right)^{-n_i^m}\text{Gamma}\left(n_i^m,\frac{\lambda_i^m}{\mu_i^m}\right)}.\label{eq07}
\end{align}
\par Note that, we are using \emph{incomplete Gamma function} in (\ref{eq07}), which is often used for approximate and exact representation of many Mathematical series \cite{Amore_2005}. The authors of \cite{MMc} proved a conjecture that the average latency of $M/M/c$ queues are convex functions of the number of processors. They considered an expression like (\ref{eq06}) and showed that it is a ratio of two expressions where the numerator is strictly decreasing and the denominator is strictly increasing with $n_i^m$. By further exploiting this property, they proved the convexity of this function. This implies that $\mathcal{R}_i^m$ is a continuous convex optimization problem and hence, it can be solved very efficiently with gradient projection algorithm.

\section{A Detailed Proof of Lemma 4.1}
\begin{IEEEproof}
Recall that in the game $\Gamma$, the utility function of each $i$\textsuperscript{th} cloudlet  $\mathcal{U}^N_i(\bm{\varphi}_i,\bm{\varphi}_{-i}),\forall i\in\mathcal{C}$ is defined as follows:\par
\begin{resizealign} 
&\max_{\bm{\varphi}_i \in \Phi_i}\mathcal{U}^{N}_i(\bm{\varphi}_i,\bm{\varphi}_{-i}) = \sum_{m=1}^M \Omega_{i,1}^m \frac{\lambda_i^m}{n_i^m\mu_i^m} \nonumber\\
& +\sum_{m=1}^M \sum_{j=1,j\neq i}^{N} \Omega_{ji,2}^m \gamma_{ji} \frac{\varphi_{ji}^m\lambda_j^m}{n_i^m\mu_i^m}- \sum_{m=1}^M \sum_{j=1,j\neq i}^{N} \Omega_{ij,2}^m \gamma_{ij}\frac{\varphi_{ij}^m\lambda_i^m}{n_j^m\mu_j^m}\nonumber\\
&-\sum_{m=1}^M \Omega_{i,3}^m \left[\left(1-\sum_{j=1,j\neq i}^{N}\varphi_{ij}^m\right)\frac{\lambda_i^m}{n_i^m \mu_i^m} \max\left\{0,\left(t_{ui}^m+\mathbb{T}_i^m(\mu_i^m,\bar{\lambda}_i^m) -D_{Qm}\right)\right\}\right.\nonumber\\
& \left.+\sum_{j=1,j\neq i}^{N} \frac{\varphi_{ji}^m\lambda_j^m}{n_i^m\mu_i^m} \max\left\{0,\left(t_{uj}^m+\mathbb{T}_i^m(\mu_i^m,\bar{\lambda}_i^m)+t_{ji}^m-D_{Qm} \right)\right\} \right] \label{eq08}
\end{resizealign}\par 
where $\bar{\lambda}_i^m(\bm{\varphi}_i^m,\bm{\varphi}_{-i}^m) = \{(1-\sum_{j\neq i} \varphi_{ij}^m)\lambda_i^m +\sum_{j\neq i} \varphi_{ji}^m\lambda_j^m\}$. From this definition, we can observe that when $i^{\text{th}}$ cloudlet is able to meet the QoS target latency $D_{Qm}$ (under-loaded), or $(t_{uj}^m+ \mathbb{T}_i^m(\mu_i^m,\bar{\lambda}_i^m)+t_{ji}^m-D_{Qm}) \leq 0$, then the utility function can be interpreted as an \emph{affine function} as follows:\par
\begin{resizealign}
\mathcal{U}^N_i&(\bm{\varphi}_i,\bm{\varphi}_{-i}) = \sum_{m=1}^M \Omega_{i,1}^m \frac{\lambda_i^m}{n_i^m\mu_i^m} \nonumber\\
& + \sum_{m=1}^M\sum_{j=1,j\neq i}^{N} \Omega_{ji,2}^m \gamma_{ji} \frac{\varphi_{ji}^m\lambda_j^m}{n_i^m\mu_i^m}- \sum_{m=1}^M \sum_{j=1,j\neq i}^{N} \Omega_{ij,2}^m \gamma_{ij}^m\frac{\varphi_{ij}^m \lambda_i^m}{n_j^m\mu_j^m}. \label{eq09}
\end{resizealign}  
\par Nonetheless, when $i^{\text{th}}$ cloudlet is unable to meet $D_{Qm}$ (overloaded), or $(t_{ui}^m+ \mathbb{T}_i^m(\mu_i^m,\bar{\lambda}_i^m) - D_{Qm}) \geq 0$, then the utility function appears as a \emph{non-linear function} as follows:\par
\begin{resizealign} 
&\max_{\bm{\varphi}_i \in \Phi_i}\mathcal{U}^{N}_i(\bm{\varphi}_i,\bm{\varphi}_{-i}) = \sum_{m=1}^M \Omega_{i,1}^m \frac{\lambda_i^m}{n_i^m\mu_i^m} \nonumber\\
& +\sum_{m=1}^M \sum_{j=1,j\neq i}^{N} \Omega_{ji,2}^m \gamma_{ji} \frac{\varphi_{ji}^m\lambda_j^m}{n_i^m\mu_i^m}- \sum_{m=1}^M \sum_{j=1,j\neq i}^{N} \Omega_{ij,2}^m \gamma_{ij}\frac{\varphi_{ij}^m\lambda_i^m}{n_j^m\mu_j^m}\nonumber\\
&-\sum_{m=1}^M \Omega_{i,3}^m \left[\left(1-\sum_{j=1,j\neq i}^{N}\varphi_{ij}^m\right)\frac{\lambda_i^m}{n_i^m \mu_i^m} \left(t_{ui}^m+\mathbb{T}_i^m(\mu_i^m,\bar{\lambda}_i^m) -D_{Qm}\right)\right.\nonumber\\
& \left.+\sum_{j=1,j\neq i}^{N} \frac{\varphi_{ji}^m\lambda_j^m}{n_i^m\mu_i^m} \left(t_{uj}^m+\mathbb{T}_i^m(\mu_i^m,\bar{\lambda}_i^m)+t_{ji}^m-D_{Qm} \right) \right]. \label{eq10}
\end{resizealign}\par 
\par If (\ref{eq10}) represents a concave function, then the sufficient condition is that its \emph{Hessian matrix} should be a positive semi-definite matrix \cite{Bartsekas}. However, we find that both the diagonal and non-diagonal elements of this matrix are equal, i.e., $\left(\frac{\partial^2 \mathcal{U}^N_i}{\partial (\varphi_{ij}^m)^2}\right) = \left(\frac{\partial^2 \mathcal{U}^N_i}{\partial \varphi_{ij}^m \partial \varphi_{ik}^m}\right),i\neq j,k \in \mathcal{C},m\in\mathcal{J}$. Clearly, this implies that the Hessian of (\ref{eq10}) is not a positive semi-definite matrix and we extend to evaluation of the \emph{bordered Hessian matrix} \cite{Miller}. Note that the $r$\textsuperscript{th} order bordered Hessian matrix of $\mathcal{U}^N_i(\bm{\varphi}_i,\bm{\varphi}_{-i}),\forall i\in\mathcal{C}$, where $r = 1,2, \dots, N$ is written as:\par
\begin{align}
\mathcal{H}_B = \begin{bmatrix}
0 & \frac{\partial \mathcal{U}^N_i}{\partial \varphi_{i1}^m} & \dots & \frac{\partial \mathcal{U}^N_i}{\partial \varphi_{ir}^m}\\ 
\frac{\partial \mathcal{U}^N_i}{\partial \varphi_{i1}^m} & \frac{\partial^2 \mathcal{U}^N_i}{\partial (\varphi_{i1}^m)^2} & \dots & \frac{\partial^2 \mathcal{U}^N_i}{\partial \varphi_{i1}^m \partial \varphi_{ir}^m}\\ 
\vdots & \vdots & \ddots & \vdots\\ 
\frac{\partial \mathcal{U}^N_i}{\partial \varphi_{ir}^m} & \frac{\partial^2 \mathcal{U}^N_i}{\partial \varphi_{ir}^m \partial \varphi_{i1}^m} & \dots & \frac{\partial^2 \mathcal{U}^N_i}{\partial (\varphi_{ir}^m)^2}
\end{bmatrix}, \label{eq11}
\end{align} 
whose bordered elements are the first-order derivatives of $\mathcal{U}^N_i(\bm{\varphi}_i,\bm{\varphi}_{-i})$. Now, we observe that the determinant values of the bordered Hessian matrix (\ref{eq11}), denoted by $D_r$, are negative when $r$ is odd and are positive when $r$ is even, $\forall r = 1,2,\dots,N$:
\begin{align}
&D_1 = 0 - \left(\frac{\partial \mathcal{U}^N_i}{\partial \varphi_{i1}^m}\right)^2 = - \left(\frac{\partial \mathcal{U}^N_i}{\partial \varphi_{i1}^m}\right)^2 \leq 0, \label{eq12}\\
&D_2 = 0 - \left(\frac{\partial \mathcal{U}^N_i}{\partial \varphi_{i1}^m}\right) \left[\left(\frac{\partial \mathcal{U}^N_i}{\partial \varphi_{i1}^m}\right) \left(\frac{\partial^2 \mathcal{U}^N_i}{\partial (\varphi_{i2}^m)^2}\right)\right.\nonumber \\
&\left. \quad \quad \quad \quad \quad \quad \quad \quad \quad \quad \quad \quad \quad \quad- \left(\frac{\partial^2 \mathcal{U}^N_i}{\partial \varphi_{i1}^m\partial \varphi_{i2}^m}\right)\left(\frac{\partial \mathcal{U}^N_i}{\partial \varphi_{i2}^m}\right) \right]\nonumber\\
&+\left(\frac{\partial \mathcal{U}^N_i}{\partial \varphi_{i2}^m}\right) \left[\left(\frac{\partial \mathcal{U}^N_i}{\partial \varphi_{i1}^m}\right)\left(\frac{\partial^2 \mathcal{U}^N_i}{\partial \varphi_{i2}^m\partial \varphi_{i1}^m}\right)- \left(\frac{\partial^2 \mathcal{U}^N_i}{\partial (\varphi_{i1}^m)^2}\right)\left(\frac{\partial \mathcal{U}^N_i}{\partial \varphi_{i2}^m}\right) \right]\nonumber\\
&= - \left(\frac{\partial^2 \mathcal{U}^N_i}{\partial \varphi_{i1}^m \partial \varphi_{i2}^m}\right)\left[\left(\frac{\partial \mathcal{U}^N_i}{\partial \varphi_{i1}^m}\right) - \left(\frac{\partial \mathcal{U}^N_i}{\partial \varphi_{i2}^m}\right)\right]^2 \geq 0, \label{eq13}
\end{align} 
and so on.  Note that, the inequalities in (\ref{eq12})-(\ref{eq13}) hold true, because we can observe that the second-order derivatives are all equal, i.e., $\left(\frac{\partial^2 \mathcal{U}^N_i}{\partial (\varphi_{i1}^m)^2}\right) = \left(\frac{\partial^2 \mathcal{U}^N_i}{\partial (\varphi_{i2}^m)^2}\right) = \left(\frac{\partial^2 \mathcal{U}^N_i}{\partial \varphi_{i1}^m \partial \varphi_{i2}^m}\right) = \left(\frac{\partial^2 \mathcal{U}^N_i}{\partial \varphi_{i2}^m \partial \varphi_{i1}^m}\right) = 0$, when the condition of stable operation $\{(1-\sum_{j\neq i}\varphi_{ij}^m)\lambda_i^m+\sum_{j\neq i}\varphi_{ji}^m \lambda_j^m\}/(n_i^m \mu_i^m) < 1, \forall i,j\neq i \in\mathcal{C}, \forall m \in \mathcal{J}$ holds. Therefore, we can conclude that (\ref{eq10}) is a \emph{quasi-concave function} of $\bm{\varphi}_i^m$ \cite{Miller}. In general, we can conclude that the utility functions of each $i$\textsuperscript{th} cloudlet $\mathcal{U}^N_i(\bm{\varphi}_i,\bm{\varphi}_{-i}),\forall i\in\mathcal{C}$ are affine functions of $\bm{\varphi}_i$ when they are under-loaded and quasi-concave functions of $\bm{\varphi}_i$ when they are overloaded. Hence proved.
\end{IEEEproof}

\section{Proof of Theorem 4.3}
\begin{IEEEproof}
We observe that the objective of the problem $\mathcal{P}_{\Gamma}^m$ is defined as the sum of complementary slackness conditions of each of the cloudlets. Therefore, the value of the objective is non-negative with any feasible solution. If we can find a vector $\bm{\varphi}^* \in \Phi$ as an NE point of the game $\Gamma$, then there exist a set of non-negative values of the variables $\bm{\alpha}_i^{m*}\in\mathbb{R}^{N-1}$, $\bm{\beta}_i^{m*}\in\mathbb{R}^{N-1}$, and $\bm{\xi}_i^{m*}\in\mathbb{R}^{N-1}$ such that all the necessary first-order and complementary slackness conditions of $\Gamma$ are satisfied \cite{game_opt}. Clearly, with these values of the variables, the objective function of $\mathcal{P}_{\Gamma}^m$ becomes 0 and hence, the non-negative optimal value of the objective function is reached. On the other hand, due to the convexity of the constraint functions, the KKT conditions are sufficient for $\bm{\varphi}_i^*$ to be the maxima of utilities of each cloudlet $\mathcal{U}^N_i(\bm{\varphi}_i,\bm{\varphi}_{-i})$.
\end{IEEEproof}
\begin{figure}[t!]
\centering
\includegraphics[width=\columnwidth]{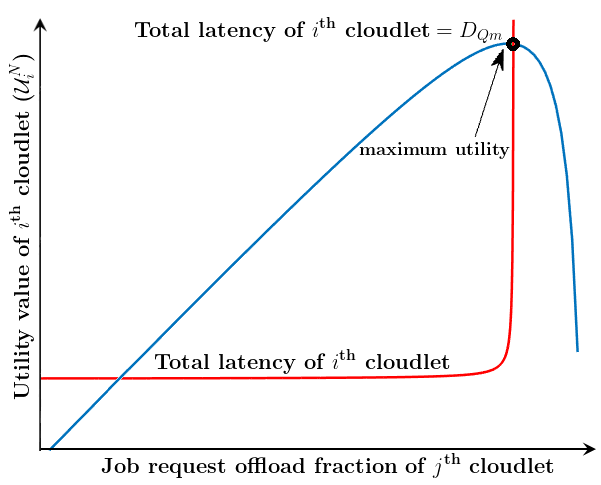}
\caption{A sample utility function of $i^{\text{th}}$ cloudlet against job request offload fraction of neighboring $j^{\text{th}}$ cloudlet $\varphi_{ji}^m$.}
\label{utility_cloudlet}
\end{figure}
\setlength{\textfloatsep}{5pt}
\par It is interesting to note that in this load balancing game, each competing cloudlet is interested in \emph{maximizing their individual utilities rather than strictly minimizing the average end-to-end latency} as most of the existing works.  Hence, the cloudlets are always interested in receiving some job requests from neighboring cloudlets as long as the QoS latency requirement $D_{Qm}$ is met and some extra incentive is gained.  Fig. \ref{utility_cloudlet} shows that with a sufficient amount of job requests and a set of properly chosen parameters $\Omega_{i,1}^m$, $\Omega_{ij,2}^m$, and $\Omega_{i,3}^m$, the utility function $i^{\text{th}}$ cloudlet monotonically increases as more job requests are offloaded by the neighboring $j^{\text{th}}$ cloudlet until the total end-to-end latency is equal to the target QoS latency value $D_Q$.  The maximum utility is achieved at the point where the total end-to-end latency is equal to $D_{Qm}$ and the utility starts to decrease beyond this point. Therefore, the overloaded cloudlets can only offload to its under-loaded neighboring cloudlets until their overall latency is equal to $D_{Qm}$. Thus, the fraction of incoming job requests offloaded by an overloaded cloudlet is controlled by the overloaded cloudlet as well as its under-loaded neighboring cloudlets.\par
Furthermore, note that in our game formulation, there is no explicit latency constraints on the federated cloudlets, even highly over-loaded cloudlets can participate in the game and can offload some of the job requests to the relatively under-loaded neighboring cloudlets. This makes their utility higher than the utility by not participating in the game.  Nonetheless, under such conditions the game formulation in \cite{TONload} that has explicit delay bound on participating cloudlets becomes infeasible and hence, a valid NE solution can not be obtained. As we compute the NE solution of this load balancing game, we can observe a very important property of the solution. The authors of \cite{MMc2} proved that the average latency expression of $M/M/c$ queuing systems is monotonically increasing and convex against load $\rho$. Intuitively, this implies that as the class-$m$ job request arrival rate ($\lambda_i^m$) increases to $i^{\text{th}}$ cloudlet, it needs to offload a higher fraction of job requests to its neighboring cloudlets for maintaining the stable condition of operation. Mathematically, if $i^{\text{th}}$ cloudlet is overloaded such that $\hat{\lambda}_i^m > \lambda_i^m > 0$, then $\hat{\varphi}_{ij}^m > \varphi_{ij}^m > 0$.\par
%

\section{Proof of Proposition 5.1}
\begin{IEEEproof}
From the NE solution of the proposed non-cooperative load balancing game, we know that any overloaded cloudlet offloads only a small fraction of its total job request arrival rate, i.e., $\varphi_{ij}^{m*} \approx 0$, $\varphi_{ij}^{m*} \lambda_i^m \ll \lambda_i^m$. Now, for the proposed incentive mechanism to work perfectly, we need to show that $\Omega_{ij,2}^m \geq \Omega_{i,3}^m [t_{ui}^m+\mathbb{T}_i^m(\mu_i^m,\lambda_i^{m,max})] + \Omega_{i,4}^m$ and $n_i \Omega_{ij,2}^m \leq (\Omega_{i,3}^m+\Omega_{i,4}^m), \forall i,j \in\mathcal{C},m \in \mathcal{J}$ are necessary conditions among the primary networks parameters such that $(\hat{\mathcal{U}}_i^N - \mathcal{U}_i^N) \leq 0$ holds for the below expression for \textbf{Sub-case-2A(i)}:\par
\begin{resizealign} 
&(\hat{\mathcal{U}}_i^N -\mathcal{U}_i^N) = -\sum_{m=1}^M\sum_{j=1,j\neq i}^N \Omega_{ij,2}^m \frac{\gamma_{ij}}{n_j^m\mu_j^m}(\hat{\varphi}_{ij}^{m*} \hat{\lambda}_i^m - \varphi_{ij}^{m*} \lambda_i^m)\nonumber\\
&- \sum_{m=1}^M \left(\frac{\lambda_i^m- \sum_{j\neq i}\hat{\varphi}_{ij}^{m*} \hat{\lambda}_i^m}{n_i^m\mu_i^m}\right) \left\{\Omega_{i,3}^m \left[t_{ui}^m+ \mathbb{T}_i^m(\mu_i^m,(\lambda_i^m- \sum_{j\neq i}\hat{\varphi}_{ij}^{m*}\hat{\lambda}_i^m))-D_{Qm} \right]\right\}\nonumber\\
&+\sum_{m=1}^M \left(\frac{\lambda_i^m- \sum_{j\neq i}\varphi_{ij}^{m*} \lambda_i^m}{n_i^m\mu_i^m}\right) \left\{\Omega_{i,3}^m \left[t_{ui}^m+ \mathbb{T}_i^m(\mu_i^m,(\lambda_i^m- \sum_{j\neq i}\varphi_{ij}^{m*}\lambda_i^m))-D_{Qm} \right]\right\} \nonumber\\
&- \sum_{m=1}^M \Omega_{i,4}^m \left(\frac{\hat{\lambda}_i^m}{n_i^m\mu_i^m}\right) + \sum_{m=1}^M \Omega_{i,4}^m \left(\frac{\lambda_i^m}{n_i^m\mu_i^m}\right) \leq 0. \label{eq14}
\end{resizealign}
\par In this case, we observe that $\lambda_i^{m,max}\geq \hat{\lambda}_i^m \geq \lambda_i^m$ and $\hat{\varphi}_{ij}^{m*}\hat{\lambda}_i^m \geq \varphi_{ij}^{m*}\lambda_i^m$. Thus, to show that (\ref{eq14}) holds true, firstly we verify that,
\begin{align}
\Omega_{i,3}^m &\left[t_{ui}^m+\mathbb{T}_i^m(\mu_i^m,\lambda_i^{m,max}) \right] \nonumber\\
&\geq \Omega_{i,3}^m \left[t_{ui}^m+ \mathbb{T}_i^m(\mu_i^m,(\lambda_i^m- \sum_{j\neq i}\hat{\varphi}_{ij}^{m*}\hat{\lambda}_i^m))-D_{Qm} \right], \label{eq15}\\
\Omega_{i,3}^m &\left[t_{ui}^m+\mathbb{T}_i^m(\mu_i^m,\lambda_i^{m,max}) \right] \nonumber\\
&\geq \Omega_{i,3}^m \left[t_{ui}^m+ \mathbb{T}_i^m(\mu_i^m,(\lambda_i^m- \sum_{j\neq i}\varphi_{ij}^{m*}\lambda_i^m))-D_{Qm} \right]. \label{eq16}
\end{align}
\par By using (\ref{eq15})-(\ref{eq16}) and assuming that $(\frac{\lambda_i^m-\sum_{j\neq i}\hat{\varphi}_{ij}^{m*} \hat{\lambda}_i^m}{n_i^m\mu_i^m}) \geq 0$ and $(\frac{\lambda_i^m-\sum_{j\neq i}\varphi_{ij}^{m*} \lambda_i^m}{n_i^m\mu_i^m}) \geq 0$, we can further deduce that,\par
\begin{resizealign}
\sum_{m=1}^M &\left(\frac{\lambda_i^m-\sum_{j\neq i}\hat{\varphi}_{ij}^{m*} \hat{\lambda}_i^m}{n_i^m\mu_i^m}\right) \Omega_{i,3}^m \left[t_{ui}^m+\mathbb{T}_i^m(\mu_i^m,\lambda_i^{m,max}) \right] \nonumber\\
&\geq \sum_{m=1}^M \left(\frac{\lambda_i^m- \sum_{j\neq i}\hat{\varphi}_{ij}^{m*} \hat{\lambda}_i^m}{n_i^m\mu_i^m}\right) \Omega_{i,3}^m \left[t_{ui}^m+ \mathbb{T}_i^m(\mu_i^m,(\lambda_i^m- \sum_{j\neq i}\hat{\varphi}_{ij}^{m*}\hat{\lambda}_i^m))-D_{Qm} \right], \label{eq17}\\
\sum_{m=1}^M &\left(\frac{\lambda_i^m- \sum_{j\neq i}\varphi_{ij}^{m*} \lambda_i^m}{n_i^m\mu_i^m}\right) \Omega_{i,3}^m \left[t_{ui}^m+\mathbb{T}_i^m(\mu_i^m,\lambda_i^{m,max}) \right] \nonumber\\
&\geq \sum_{m=1}^M \left(\frac{\lambda_i^m- \sum_{j\neq i}\varphi_{ij}^{m*} \lambda_i^m}{n_i^m\mu_i^m}\right) \Omega_{i,3}^m \left[t_{ui}^m+ \mathbb{T}_i^m(\mu_i^m,(\lambda_i^m- \sum_{j\neq i}\varphi_{ij}^{m*}\lambda_i^m))-D_{Qm} \right]. \label{eq18}
\end{resizealign}
\par If $(\frac{\lambda_i^m-\sum_{j\neq i}\hat{\varphi}_{ij}^{m*} \hat{\lambda}_i^m}{n_i^m\mu_i^m}) < 0$ or $(\frac{\lambda_i^m-\sum_{j\neq i}\varphi_{ij}^{m*} \lambda_i^m}{n_i^m\mu_i^m}) < 0$, then the inequalities in (\ref{eq17}) and (\ref{eq18}) do not hold. However, in general, $\lambda_i^m \geq \sum_{j\neq i}\varphi_{ij}^{m*} \lambda_i^m$, and as the overloaded cloudlets cannot share the load completely in this case, hence $\lambda_i^m \geq \sum_{j\neq i} \hat\varphi_{ij}^{m*} \hat\lambda_i^m$. Therefore, we can safely continue with (\ref{eq17})-(\ref{eq18}) and by subtracting (\ref{eq17}) from (\ref{eq18}), we get,\par
\begin{resizealign}
\sum_{m=1}^M &\left(\frac{\sum_{j\neq i}\hat{\varphi}_{ij}^{m*} \hat{\lambda}_i^m - \sum_{j\neq i}\varphi_{ij}^{m*} \lambda_i^m}{n_i^m\mu_i^m}\right) \Omega_{i,3}^m \left[t_{ui}^m+\mathbb{T}_i^m(\mu_i^m,\lambda_i^{m,max}) \right] \nonumber\\ 
& \geq - \sum_{m=1}^M \left(\frac{\lambda_i^m-\sum_{j\neq i}\hat{\varphi}_{ij}^{m*} \hat{\lambda}_i^m}{n_i^m\mu_i^m}\right) \Omega_{i,3}^m \left[t_{ui}^m+ \mathbb{T}_i^m(\mu_i^m,(\lambda_i^m- \sum_{j\neq i}\hat{\varphi}_{ij}^{m*}\hat{\lambda}_i^m))-D_{Qm} \right] \nonumber\\
& \quad+ \sum_{m=1}^M \left(\frac{\lambda_i^m- \sum_{j\neq i}\varphi_{ij}^{m*} \lambda_i^m}{n_i^m\mu_i^m}\right) \Omega_{i,3}^m \left[t_{ui}^m+ \mathbb{T}_i^m(\mu_i^m,(\lambda_i^m- \sum_{j\neq i}\varphi_{ij}^{m*}\lambda_i^m))-D_{Qm} \right]. \label{eq19}
\end{resizealign}
\par Now, from our considered relation $\Omega_{ij,2}^m \geq \Omega_{i,3}^m [t_{ui}^m+\mathbb{T}_i^m(\mu_i^m,\lambda_i^{m,max})]+\Omega_{i,4}^m, \forall i,j \in\mathcal{C},m \in \mathcal{J}$, we can conclude,\par
\begin{resizealign}
\sum_{m=1}^M &\left(\frac{\sum_{j\neq i}\hat{\varphi}_{ij}^{m*} \hat{\lambda}_i^m - \sum_{j\neq i}\varphi_{ij}^{m*} \lambda_i^m}{n_i^m\mu_i^m}\right) \left(\Omega_{ij,2}^m -\Omega_{i,4}^m \right) \nonumber\\
&\geq \sum_{m=1}^M \left(\frac{\sum_{j\neq i}\hat{\varphi}_{ij}^{m*} \hat{\lambda}_i^m - \sum_{j\neq i}\varphi_{ij}^{m*} \lambda_i^m}{n_i^m\mu_i^m}\right)\Omega_{i,3}^m \left[t_{ui}^m+\mathbb{T}_i^m(\mu_i^m,\lambda_i^{m,max}) \right]. \label{eq20}
\end{resizealign}
\par After some algebraic manipulations along with the inequalities $\hat{\lambda}_i^m \geq \sum_{j\neq i} \hat{\varphi}_{ij}^{m*}\hat{\lambda}_i^m$ and $\lambda_i^m \geq \sum_{j\neq i} \varphi_{ij}^{m*}\lambda_i^m$, we obtain the following expression,\par
\begin{resizealign}
&\sum_{m=1}^M \sum_{j=1,j\neq i}^N \Omega_{ij,2}^m \frac{\gamma_{ij}}{n_j^m\mu_j^m}(\hat{\varphi}_{ij}^{m*} \hat{\lambda}_i^m - \varphi_{ij}^{m*} \lambda_i^m) +\sum_{m=1}^M \Omega_{i,4}^m \left(\frac{\hat{\lambda}_i^m}{n_i^m\mu_i^m}\right) - \sum_{m=1}^M \Omega_{i,4}^m \left(\frac{\lambda_i^m}{n_i^m\mu_i^m}\right) \nonumber\\
& \geq \sum_{m=1}^M \left(\frac{\sum_{j\neq i}\hat{\varphi}_{ij}^{m*} \hat{\lambda}_i^m - \sum_{j\neq i}\varphi_{ij}^{m*} \lambda_i^m}{n_i^m\mu_i^m}\right)\Omega_{i,3}^m \left[t_{ui}^m+\mathbb{T}_i^m(\mu_i^m,\lambda_i^{m,max}) \right]. \label{eq21}
\end{resizealign}
\par Therefore, by using (\ref{eq19}) and (\ref{eq21}), we can show that the inequality in (\ref{eq14}) holds true. Similarly, it is very straightforward to show by extending our analysis from the necessary conditions that $(\hat{\mathcal{U}}_i^N - \mathcal{U}_i^N) \leq 0$ holds for \textbf{Sub-case-2A(ii)}:\par
\begin{resizealign} 
&(\hat{\mathcal{U}}_i^N -\mathcal{U}_i^N) = -\sum_{m=1}^M\sum_{j=1,j\neq i}^N \Omega_{ij,2}^m \frac{\gamma_{ij}}{n_j^m\mu_j^m}(\hat{\varphi}_{ij}^{m*} \hat{\lambda}_i^m - \varphi_{ij}^{m*} \lambda_i^m) - \sum_{m=1}^M \Omega_{i,4}^m \left(\frac{\hat{\lambda}_i^m}{n_i^m\mu_i^m}\right)\nonumber\\
& - \sum_{m=1}^M \left(\frac{\lambda_i^m- \sum_{j\neq i}\hat{\varphi}_{ij}^{m*} \hat{\lambda}_i^m}{n_i^m\mu_i^m}\right) \left\{\Omega_{i,3}^m \left[t_{ui}^m+ \mathbb{T}_i^m(\mu_i^m,(\lambda_i^m- \sum_{j\neq i}\hat{\varphi}_{ij}^{m*}\hat{\lambda}_i^m))-D_{Qm} \right]\right\} \leq 0. \label{eq22}
\end{resizealign} 
\par In this case, $\lambda_i^{m,max} \geq \lambda_i^m \geq \hat\lambda_i^m$ and $\varphi_{ij}^{m*}\lambda_i^m \geq \hat{\varphi}_{ij}^{m*}\hat{\lambda}_i^m$ i.e., $(\varphi_{ij}^{m*}\lambda_i^m - \hat{\varphi}_{ij}^{m*}\hat{\lambda}_i^m) \geq 0$ or, $-(\hat{\varphi}_{ij}^{m*}\hat{\lambda}_i^m - \varphi_{ij}^{m*}\lambda_i^m) \geq 0$. Also, the overloaded cloudlet could offload its extra load to the under-loaded neighbors completely if it had revealed the true $\lambda_i^m$ value. Now, we use the inequalities $n_i\Omega_{ij,2}^m \leq (\Omega_{i,3}^m+\Omega_{i,4}^m)$ such that $n_i \geq n_i^m, \forall m\in \mathcal{J}$ to proceed as follows:\par
\begin{resizealign}
&-\sum_{m=1}^M\sum_{j=1,j\neq i}^N \Omega_{ij,2}^m \frac{\gamma_{ij}}{n_j^m\mu_j^m}(\hat{\varphi}_{ij}^{m*} \hat{\lambda}_i^m- \varphi_{ij}^{m*} \lambda_i^m) \nonumber\\
&\leq - \frac{1}{n_i}\sum_{m=1}^M\sum_{j=1,j\neq i}^N \frac{\Omega_{i,3}^m \gamma_{ij}}{n_j^m \mu_j^m}(\hat{\varphi}_{ij}^{m*} \hat{\lambda}_i^m- \varphi_{ij}^{m*} \lambda_i^m) \nonumber\\
&\quad \quad -\frac{1}{n_i}\sum_{m=1}^M\sum_{j=1,j\neq i}^N \frac{\Omega_{i,4}^m\gamma_{ij}}{n_j^m \mu_j^m}(\hat{\varphi}_{ij}^{m*} \hat{\lambda}_i^m- \varphi_{ij}^{m*} \lambda_i^m)  \nonumber \\
&\leq \sum_{m=1}^M \Omega_{i,3}^m \gamma_{ij}\frac{(\sum_{j\neq i}\varphi_{ij}^{m*} \lambda_i^m - \sum_{j\neq i}\hat{\varphi}_{ij}^{m*} \hat{\lambda}_i^m)}{n_i^m\mu_i^m} \nonumber\\
&\quad \quad +\sum_{m=1}^M \Omega_{i,4}^m\gamma_{ij}\frac{(\sum_{j\neq i}\varphi_{ij}^{m*} \lambda_i^m - \sum_{j\neq i}\hat{\varphi}_{ij}^{m*} \hat{\lambda}_i^m)}{n_i^m\mu_i^m} \nonumber \\
& \leq \sum_{m=1}^M \left(\frac{\lambda_i^m- \sum_{j\neq i}\hat{\varphi}_{ij}^{m*} \hat{\lambda}_i^m}{n_i^m\mu_i^m}\right) \left\{\Omega_{i,3}^m \left[t_{ui}^m+ \mathbb{T}_i^m(\mu_i^m,(\lambda_i^m- \sum_{j\neq i}\hat{\varphi}_{ij}^{m*}\hat{\lambda}_i^m))-D_{Qm} \right]\right\} \nonumber\\
& \quad + \sum_{m=1}^M \Omega_{i,4}^m \left(\frac{\hat{\lambda}_i^m}{n_i^m\mu_i^m}\right). \label{eq23}
\end{resizealign}
\par Therefore, we can see that the expressions (\ref{eq23}) and (\ref{eq24}) are equivalent. Next, we consider the expression for \textbf{Sub-case-2A(iii)}:\par
\begin{resizealign}
&(\hat{\mathcal{U}}_i^N - \mathcal{U}_i^N) = \sum_{m=1}^M \sum_{j=1,j\neq i}^N \Omega_{ji,2}^m \gamma_{ji} \frac{\hat{\varphi}_{ji}^{m*}\hat{\lambda}_j^m}{n_i^m\mu_i^m} \nonumber\\
&+ \sum_{m=1}^M \sum_{j=1,j\neq i}^N \Omega_{ij,2}^m \gamma_{ij}  \frac{\varphi_{ij}^{m*} \lambda_i^m}{n_j^m\mu_j^m} - \sum_{m=1}^M \Omega_{i,4}^m \left\{\frac{\hat{\lambda}_i^m+\sum_{j\neq i}\hat{\varphi}_{ji}^{m*} \hat{\lambda}_j^m+\aleph_i^m}{n_i^m\mu_i^m} \right\} \nonumber\\
& - \sum_{m=1}^M \Omega_{i,3}^m \left[\frac{\lambda_i^m}{n_i^m\mu_i^m}\left(t_{ui}^m+ \mathbb{T}_i^m(\mu_i^m,\lambda_i^m+\sum_{j\neq i} \hat{\varphi}_{ji}^{m*} \hat{\lambda}_j^m+\aleph_i^m) -D_{Qm} \right) \right.\nonumber\\
& \quad\left. +\sum_{j=1,j\neq i}^{N}\frac{\hat{\varphi}_{ji}^{m*} \hat{\lambda}_j^m}{n_i^m\mu_i^m} \left(t_{uj}^m+\mathbb{T}_i^m(\mu_i^m,\lambda_i^m+\sum_{j\neq i} \hat{\varphi}_{ji}^{m*} \hat{\lambda}_j^m+\aleph_i^m)+t_{ji}^m-D_{Qm} \right)\right] \leq 0. \label{eq24}
\end{resizealign} 
\par In this case also, we have $\lambda_i^{m,max} \geq \lambda_i^m \geq \hat\lambda_i^m$ and $\varphi_{ij}^{m*}\lambda_i^m \geq \hat{\varphi}_{ij}^{m*}\hat{\lambda}_i^m$ and we use our considered relation $n_i\Omega_{ij,2}^m \leq (\Omega_{i,3}^m+\Omega_{i,4}^m)$ to proceed as follows:\par
\begin{resizealign}
\sum_{m=1}^M& \sum_{j=1,j\neq i}^N \Omega_{ji,2}^m \gamma_{ji} \frac{\hat{\varphi}_{ji}^{m*}\hat{\lambda}_j^m}{n_i^m\mu_i^m} + \sum_{m=1}^M \sum_{j=1,j\neq i}^N \Omega_{ij,2}^m \gamma_{ij}  \frac{\varphi_{ij}^{m*} \lambda_i^m}{n_j^m\mu_j^m} \nonumber\\
&\leq \frac{1}{n_i} \sum_{m=1}^M \sum_{j=1,j\neq i}^N \gamma_{ji} \frac{\hat{\varphi}_{ji}^{m*}\hat{\lambda}_j^m}{n_i^m \mu_i^m}(\Omega_{i,3}^m + \Omega_{i,4}^m) \nonumber\\
&\quad + \frac{1}{n_i} \sum_{m=1}^M \sum_{j=1,j\neq i}^N \gamma_{ij}  \frac{\varphi_{ij}^{m*} \lambda_i^m}{n_j^m \mu_j^m} (\Omega_{i,3}^m + \Omega_{i,4}^m) \nonumber\\
& \leq \sum_{m=1}^M \sum_{j=1,j\neq i}^N \gamma_{ji} \frac{\hat{\varphi}_{ji}^{m*}\hat{\lambda}_j^m}{n_i^m\mu_i^m}(\Omega_{i,3}^m + \Omega_{i,4}^m) + \sum_{m=1}^M \sum_{j=1,j\neq i}^N \gamma_{ij}  \frac{\varphi_{ij}^{m*} \lambda_i^m}{n_i^m\mu_i^m} (\Omega_{i,3}^m + \Omega_{i,4}^m) \nonumber\\
& \leq \sum_{m=1}^M  \frac{\Omega_{i,3}^m\gamma_{ji}}{n_i^m\mu_i^m}\sum_{j=1,j\neq i}^N (\hat{\varphi}_{ji}^{m*}\hat{\lambda}_j^m+ {\varphi}_{ji}^{m*}{\lambda}_j^m) + \sum_{m=1}^M  \frac{\Omega_{i,4}^m\gamma_{ji}}{n_i^m\mu_i^m}\sum_{j=1,j\neq i}^N (\hat{\varphi}_{ji}^{m*}\hat{\lambda}_j^m+ {\varphi}_{ji}^{m*}{\lambda}_j^m) \nonumber\\
& \leq \sum_{m=1}^M \Omega_{i,4}^m \left\{\frac{\hat{\lambda}_i^m+\sum_{j\neq i}\hat{\varphi}_{ji}^{m*} \hat{\lambda}_j^m+\aleph_i^m}{n_i^m\mu_i^m} \right\} \nonumber\\
& \quad + \sum_{m=1}^M \Omega_{i,3}^m \left[\frac{\lambda_i^m}{n_i^m\mu_i^m}\left(t_{ui}^m+ \mathbb{T}_i^m(\mu_i^m,\lambda_i^m+\sum_{j\neq i} \hat{\varphi}_{ji}^{m*} \hat{\lambda}_j^m+\aleph_i^m) -D_{Qm} \right) \right.\nonumber\\
& \quad \quad \left. +\sum_{j=1,j\neq i}^{N}\frac{\hat{\varphi}_{ji}^{m*} \hat{\lambda}_j^m}{n_i^m\mu_i^m} \left(t_{uj}^m+\mathbb{T}_i^m(\mu_i^m,\lambda_i^m+\sum_{j\neq i} \hat{\varphi}_{ji}^{m*} \hat{\lambda}_j^m+\aleph_i^m)+t_{ji}^m-D_{Qm} \right)\right]. \label{eq25}
\end{resizealign}
\par Hence, from the expression in (\ref{eq25}), we can derive the expression in (\ref{eq24}). Next, we consider the expression for \textbf{Sub-case-2B(i)/(iii)}:\par
\begin{resizealign}
&(\hat{\mathcal{U}}_i^N - \mathcal{U}_i^N) = \sum_{m=1}^M \sum_{j=1,j\neq i}^N \Omega_{ji,2}^m \frac{\gamma_{ji}}{n_i^m\mu_i^m}\left(\hat{\varphi}_{ji}^{m*}\hat{\lambda}_j^m-\varphi_{ji}^{m*}\lambda_j^m\right) \nonumber\\
& - \sum_{m=1}^M \Omega_{i,4}^m \left\{\frac{\hat{\lambda}_i^m+\sum_{j\neq i}\hat{\varphi}_{ji}^{m*} \hat{\lambda}_j^m+\aleph_i^m}{n_i^m\mu_i^m} \right\} \nonumber\\
& - \sum_{m=1}^M \Omega_{i,3}^m \left[\frac{\lambda_i^m}{n_i^m\mu_i^m}\left(t_{ui}^m+ \mathbb{T}_i^m(\mu_i^m,\lambda_i^m+\sum_{j\neq i} \hat{\varphi}_{ji}^{m*} \hat{\lambda}_j^m+\aleph_i^m) -D_{Qm} \right) \right.\nonumber\\
& \left. +\sum_{j=1,j\neq i}^{N}\frac{\varphi_{ji}^{m*} \lambda_j^m}{n_i^m\mu_i^m} \left(t_{uj}^m+\mathbb{T}_i^m(\mu_i^m,\lambda_i^m+\sum_{j\neq i} \hat{\varphi}_{ji}^{m*} \hat{\lambda}_j^m+\aleph_i^m)+t_{ji}^m-D_{Qm} \right)\right] \leq 0. \label{eq26}
\end{resizealign}  
\par In this case, again we have $\lambda_i^{m,max} \geq \lambda_i^m \geq \hat\lambda_i^m$ and $\varphi_{ij}^{m*}\lambda_i^m \geq \hat{\varphi}_{ij}^{m*}\hat{\lambda}_i^m$. Thus, we use our considered relation $n_i \Omega_{ij,2}^m \leq (\Omega_{i,3}^m+\Omega_{i,4}^m)$ to proceed as follows:\par
\begin{resizealign}
\sum_{m=1}^M & \sum_{j=1,j\neq i}^N \Omega_{ji,2}^m \frac{\gamma_{ji}}{n_i^m\mu_i^m}\left(\hat{\varphi}_{ji}^{m*}\hat{\lambda}_j^m-\varphi_{ji}^{m*}\lambda_j^m\right) \nonumber\\
& \leq \sum_{m=1}^M \sum_{j=1,j\neq i}^N \frac{\gamma_{ji}}{n_i^m \mu_i^m}\left(\hat{\varphi}_{ji}^{m*}\hat{\lambda}_j^m-\varphi_{ji}^{m*}\lambda_j^m\right) (\Omega_{i,3}^m + \Omega_{i,4}^m) \nonumber\\
& \leq \sum_{m=1}^M \Omega_{i,4}^m \left\{\frac{\hat{\lambda}_i^m+\sum_{j\neq i}\hat{\varphi}_{ji}^{m*} \hat{\lambda}_j^m+\aleph_i^m}{n_i^m\mu_i^m} \right\} \nonumber\\
& \quad +\sum_{m=1}^M \sum_{j=1,j\neq i}^N \frac{\gamma_{ji}}{n_i^m\mu_i^m}\left(\hat{\varphi}_{ji}^{m*}\hat{\lambda}_j^m-\varphi_{ji}^{m*}\lambda_j^m\right) \Omega_{i,3}^m \nonumber\\
& \leq \sum_{m=1}^M \Omega_{i,4}^m \left\{\frac{\hat{\lambda}_i^m+\sum_{j\neq i}\hat{\varphi}_{ji}^{m*} \hat{\lambda}_j^m+\aleph_i^m}{n_i^m\mu_i^m} \right\} \nonumber\\
& \quad+ \sum_{m=1}^M \Omega_{i,3}^m \left[\frac{\lambda_i^m}{n_i^m\mu_i^m}\left(t_{ui}^m+ \mathbb{T}_i^m(\mu_i^m,\lambda_i^m+\sum_{j\neq i} \hat{\varphi}_{ji}^{m*} \hat{\lambda}_j^m+\aleph_i^m) -D_{Qm} \right) \right.\nonumber\\
& \quad \left. +\sum_{j=1,j\neq i}^{N}\frac{\varphi_{ji}^{m*} \lambda_j^m}{n_i^m\mu_i} \left(t_{uj}^m+\mathbb{T}_i^m(\mu_i^m,\lambda_i^m+\sum_{j\neq i} \hat{\varphi}_{ji}^{m*} \hat{\lambda}_j^m+\aleph_i^m)+t_{ji}^m-D_{Qm} \right)\right]. \label{eq27}
\end{resizealign}
\par Therefore, we can compare the expressions in (\ref{eq26}) and (\ref{eq27}) to conclude that these are equivalent expressions. Lastly, we investigate the \textbf{Sub-case-3B}:\par
\begin{resizealign}
&(\hat{\mathcal{U}}_i^N - \mathcal{U}_i^N) = \sum_{m=1}^M \sum_{j=1,j\neq i}^N \Omega_{ji,2}^m \gamma_{ji} \frac{\hat{\varphi}_{ji}^{m*} \hat{\lambda}_j^m}{n_i^m\mu_i^m}- \sum_{m=1}^M \Omega_{i,4}^m \left\{\frac{\hat{\lambda}_i^m+\sum_{j\neq i}\hat{\varphi}_{ji}^{m*} \hat{\lambda}_j^m+\aleph_i^m}{n_i^m\mu_i^m} \right\}\nonumber\\
& + \sum_{m=1}^M \Omega_{i,3}^m \frac{\lambda_i^m}{n_i^m\mu_i}\left[t_{ui}^m+ \mathbb{T}_i^m(\mu_i^m,\lambda_i^m)-D_{Qm} \right]\nonumber\\
& - \sum_{m=1}^M \Omega_{i,3}^m \frac{\lambda_i^m}{n_i^m\mu_i^m}\left[t_{ui}^m+ \mathbb{T}_i^m(\mu_i^m,\lambda_i^m+\sum_{j\neq i} \hat{\varphi}_{ji}^{m*} \hat{\lambda}_j^m+\aleph_i^m)-D_{Qm} \right]\nonumber\\
& - \sum_{m=1}^M \sum_{j=1,j\neq i}^N \Omega_{i,3} \frac{\hat{\varphi}_{ji}^{m*}\hat{\lambda}_j^m}{n_i^m\mu_i^m}\left[t_{uj}^m+ \mathbb{T}_i^m(\mu_i^m,\lambda_i^m+\sum_{j\neq i} \hat{\varphi}_{ji}^{m*} \hat{\lambda}_j^m+\aleph_i^m)+t_{ji}^m-D_{Qm} \right] \leq 0. \label{eq28}
\end{resizealign}
\par In this case, $\lambda_i^{m,max} \geq \lambda_i^m \geq \hat\lambda_i^m$ and $\varphi_{ij}^{m*}\lambda_i^m \geq \hat{\varphi}_{ij}^{m*}\hat{\lambda}_i^m$. Now, we use our considered relation $n_i \Omega_{ij,2}^m \leq (\Omega_{i,3}^m+\Omega_{i,4}^m)$ to proceed as follows:\par
\begin{resizealign}
&\sum_{m=1}^M \sum_{j=1,j\neq i}^N \frac{\hat{\varphi}_{ji}^{m*}\hat{\lambda}_j^m}{n_i^m\mu_i^m} \Omega_{ij,2}^m \leq \sum_{m=1}^M \sum_{j=1,j\neq i}^N \frac{\hat{\varphi}_{ji}^{m*}\hat{\lambda}_j^m}{n_i^m\mu_i^m}(\Omega_{i,3}^m+\Omega_{i,4}^m) \nonumber\\
& \leq \sum_{m=1}^M \sum_{j=1,j\neq i}^N \frac{\hat{\varphi}_{ji}^{m*}\hat{\lambda}_j^m}{n_i^m\mu_i^m}\Omega_{i,3}^m + \sum_{m=1}^M \Omega_{i,4}^m \left\{\frac{\hat{\lambda}_i^m+\sum_{j\neq i}\hat{\varphi}_{ji}^{m*} \hat{\lambda}_j^m+\aleph_i^m}{n_i^m\mu_i^m} \right\} \nonumber\\
& \leq \sum_{m=1}^M \Omega_{i,4}^m \left\{\frac{\hat{\lambda}_i^m+\sum_{j\neq i}\hat{\varphi}_{ji}^m \hat{\lambda}_j^m+\aleph_i^m}{n_i^m\mu_i^m} \right\} \nonumber\\
&\quad +\sum_{m=1}^M \sum_{j=1,j\neq i}^N \Omega_{i,3}^m \frac{\hat{\varphi}_{ji}^{m*}\hat{\lambda}_j^m}{n_i^m\mu_i^m}\left[t_{uj}^m+ \mathbb{T}_i^m(\mu_i^m,\lambda_i^m+\sum_{j\neq i} \hat{\varphi}_{ji}^m \hat{\lambda}_j^m+\aleph_i^m)+t_{ji}^m-D_{Qm} \right] \nonumber\\
& \leq \sum_{m=1}^M \Omega_{i,4}^m \left\{\frac{\hat{\lambda}_i^m+\sum_{j\neq i}\hat{\varphi}_{ji}^m \hat{\lambda}_j^m+\aleph_i^m}{n_i^m\mu_i^m} \right\} \nonumber\\
&\quad +\sum_{m=1}^M \sum_{j=1,j\neq i}^N \Omega_{i,3}^m \frac{\hat{\varphi}_{ji}^{m*}\hat{\lambda}_j^m}{n_i^m\mu_i^m}\left[t_{uj}^m+ \mathbb{T}_i^m(\mu_i^m,\lambda_i^m+\sum_{j\neq i} \hat{\varphi}_{ji}^m \hat{\lambda}_j^m+\aleph_i^m)+t_{ji}^m-D_{Qm} \right] \nonumber\\
& \quad +\sum_{m=1}^M \Omega_{i,3}^m \frac{\lambda_i^m}{n_i^m\mu_i^m}\left[t_{ui}^m+ \mathbb{T}_i^m(\mu_i^m,\lambda_i^m+\sum_{j\neq i} \hat{\varphi}_{ji}^m \hat{\lambda}_j^m+\aleph_i^m)-D_{Qm} \right]  \nonumber\\
& \quad - \sum_{m=1}^M \Omega_{i,3}^m \frac{\lambda_i^m}{n_i^m\mu_i}\left[t_{ui}^m+ \mathbb{T}_i^m(\mu_i^m,\lambda_i^m)-D_{Qm} \right]. \label{eq29}
\end{resizealign}
\par Clearly, (\ref{eq28}) and (\ref{eq29}) are equivalent expressions and hence, $\Omega_{ij,2}^m \geq \Omega_{i,3}^m [t_{ui}^m+\mathbb{T}_i^m(\mu_i^m,\lambda_i^{m,max})]+ \Omega_{i,4}^m$ and $n_i \Omega_{ij,2}^m \leq (\Omega_{i,3}^m+\Omega_{i,4}^m), \forall i,j \in\mathcal{C},m \in \mathcal{J}$ are the necessary conditions for the successful implementation of the incentive compatible load balancing mechanism. 
\end{IEEEproof}

\section{Proof of Theorem 6.1}
\begin{IEEEproof}
We considered that the strategy space $\Phi_i^m$ is a compact set in the load balancing game $\Gamma$ and the stochastic reward values are normalized, i.e., $\mathcal{R}_i^m: \mathbb{R}\to [0,1], \forall i \in \mathcal{C},m\in\mathcal{J}$.  The authors of \cite{carla} showed that with such conditions, a continuous action reinforcement learning automata-based PDF update rule,\par
\begin{resizealign}
\bm{f}_i^{(n+1)}&(\bm{\varphi}_i) = \chi^{(n)} \left[\bm{f}_i^{(n)}(\bm{\varphi}_i) +\Theta^{(n)} \left(\mathcal{R}_i^{(n)}-\mathcal{R}_i^{(n-1)}\right) \exp\left\{-\frac{1}{2}\left(\frac{\bm{\varphi}_i -\bm{\varphi}_i^{(n)}}{\sigma^{(n)}}\right )^2\right\} \right], \label{eq31}
\end{resizealign}  
\par can be guaranteed to converge to a local optimal NE, if the following necessary restrictions are imposed:\par
\begin{enumerate}[(i)]
\item We choose a sufficiently small value of $\Theta^m$ such that each $i^{\text{th}}$ cloudlet can match their expected strategy through iterations.
\item We choose a value of $\sigma^m$ such that the equilibrium point of $\bm{f}_i(\bm{\varphi}_i^m)$ has an upper bound $\frac{1}{\sigma^m\sqrt{2\pi}}$.
\end{enumerate}
\par Thus, if the above restrictions are satisfied, then the continuous action reinforcement learning automata-based algorithm will converge to at least one of the existing NE of the underlying continuous-kernel game.
\end{IEEEproof}

\bibliographystyle{IEEEtran}
\bibliography{IEEEabrv,ref_cloudlet_mechanism}
\vspace{-5ex}
